\documentclass[a4paper,fleqn,usenatbib]{mnras}

\usepackage{newtxtext,newtxmath}

\usepackage[T1]{fontenc}
\usepackage{ae,aecompl}
\usepackage{hyperref}

\usepackage{graphicx}	
\usepackage{amsmath}	
\usepackage{amssymb}	
\usepackage{adjustbox}
\usepackage{subfigure}

\newcommand{\vsini}{\ensuremath{{\upsilon}\sin i}}
\newcommand{\kms}{\,km\,s$^{-1}$}

\title[New and improved rotational periods of mCP stars]{New and improved rotational periods of magnetic CP stars from ASAS-3, KELT and MASCARA data}

\author[Bernhard et al.]{
Klaus Bernhard,$^{1,2}$\thanks{E-mail: klaus.bernhard@liwest.at}
Stefan H{\"u}mmerich,$^{1,2}$
Ernst Paunzen$^{3}$
\\
$^{1}$Bundesdeutsche Arbeitsgemeinschaft f{\"u}r Ver{\"a}nderliche Sterne e.V. (BAV), Berlin, Germany\\
$^{2}$American Association of Variable Star Observers (AAVSO), Cambridge, USA\\
$^{3}$Department of Theoretical Physics and Astrophysics, Masaryk University, Kotl\'a\v{r}sk\'a 2, 611 37 Brno, Czech Republic\\
}

\date{Accepted XXX. Received YYY; in original form ZZZ}

\pubyear{2019}

\begin{document}

\def\teff{${T}_{\rm eff}$}
\def\kms{{km\,s}$^{-1}$}
\def\logg{$\log g$}
\def\micro{$\xi_{\rm t}$}
\def\macro{$\zeta_{\rm RT}$}
\def\rad{$v_{\rm r}$}
\def\vsini{$v\sin i$}
\def\ebv{$E(B-V)$}

\label{firstpage}
\pagerange{\pageref{firstpage}--\pageref{lastpage}}
\maketitle

\begin{abstract} 
Magnetic chemically peculiar (mCP) stars allow the investigation of such diverse phenomena as atomic diffusion, magnetic fields and stellar rotation. The aim of the present investigation is to enhance our knowledge of the rotational properties of mCP stars by increasing the sample of objects with accurately determined rotational periods. To this end, archival photometric time-series data from the ASAS-3, KELT and MASCARA surveys were employed to improve existing rotational period information and derive rotational periods for mCP stars hitherto not known to be photometric variables. Our final sample consists of 294 mCP stars, a considerable amount of which (more than 40\%) are presented here as photometric variables for the first time. In addition, we identified 24 mCP star candidates that show light variability in agreement with rotational modulation but lack spectroscopic confirmation. The rotational period distribution of our sample agrees well with the literature. Most stars are between 100\,Myr and 1\,Gyr old, with an apparent lack of very young stars. No objects were found on the zero age main sequence; several stars seem to have evolved to the subgiant stage, albeit well before the first dredge-up. We identified four eclipsing binaries (HD\,244391, HD\,247441, HD\,248784 and HD\,252519), which potentially host an mCP star. This is of great interest because mCP stars are very rarely found in close binary systems, particularly eclipsing ones. Using archival spectra, we find strong evidence that the HD\,252519 system indeed harbours an mCP star component.
\end{abstract}




\section{Introduction} \label{introduction}

Chemically peculiar (CP) stars form a significant fraction (about 10\,\%) of upper main-sequence stars and are mostly found between spectral types early B to early F. The magnetic Of?p stars extend this range to spectral types of early O \citep{wade11,wade11_2}; they are, however, rare objects. The defining characteristic of CP stars is the presence of spectral peculiarities, which indicate unusual elemental abundance patterns. Several groups of CP stars have been defined, such as the metallic-line or Am (CP1) stars, the magnetic Bp/Ap (CP2) stars, the Mercury-Manganese (HgMn/CP3) stars, the He-weak (CP4) stars or the $\lambda$ Bootis stars \citep{preston74,maitzen84,paunzen04,ghazaryan18}.

Relevant to the present investigation is the group of the magnetic CP (mCP) stars that traditionally consists of the CP2 stars and the He-peculiar (CP4 and He-rich) stars. These stars possess strong (up to several tens of kG; \citealt{auriere07}) and organized magnetic fields and exhibit characteristic abundance anomalies that are generally thought to be the result of the competition between radiative levitation and gravitational settling (atomic diffusion; \citealt{michaud70,richer00}). The class-defining characteristics of the He-peculiar stars are their strong or weak lines of neutral He. In addition, certain other peculiarities, such as photospheric overabundances of Si, P, Ga, Sr or Ti in the He-weak stars, may be present. The CP2 stars show greatly enhanced abundances of certain elements like Si, Fe, Sr, Cr, and the rare-earth elements, whose absorption lines in the blue-violet spectral region are traditionally used for a classification of these objects \citep{preston74}. CP2 stars also show significant, though scattered, underabundances of He and quite often peculiarly weak or strong Ca II K lines, which renders spectral classification difficult \citep{gray09,ghazaryan18}.

Likely due to the influence of the magnetic field on the diffusion processes, mCP stars show a non-uniform distribution of chemical elements (chemical spots or belts) on their surfaces. Flux is redistributed in the abundance patches (line and continuum blanketing; for details cf. e.g. \citealt{lanz96,shulyak10}) and mCP stars show strictly periodic light, spectral, and magnetic variations with the rotation period, which are well understood in terms of the oblique-rotator model in which the magnetic axis is oblique to the rotation axis \citep{stibbs50}. According to convention, photometrically variable mCP stars are also referred to, after their bright prototype, as $\alpha^{2}$ Canum Venaticorum (ACV) variables.

The observed amplitudes of the light variations in ACV variables are generally of the order of only several hundredth of a magnitude in the optical region, reaching up to more than 0.1\,mag ($V$) in rare instances. However, analyzing photometric time-series data is an efficient means of deriving information on the rotational period, in particular as they are much easier acquired than spectroscopic time series. Many studies have successfully exploited the wealth of archival photometric observations provided by numerous past and on-going sky surveys \citep{paunzen98,wraight12,bernhard15b,bernhard15a,huemmerich16,bowman18,huemmerich18,david_uraz19,sikora19}, which has greatly enhanced our knowledge of the rotational properties of mCP stars \citep[e.g.][]{netopil17}.

The present investigation's aim is to add to our knowledge of the rotational properties of mCP stars by increasing the sample of these objects with accurately determined rotational periods, thereby continuing our efforts in this respect \citep{paunzen98,bernhard15b,bernhard15a,huemmerich16,huemmerich18}. To this end, archival photometric time-series data were employed to (i) improve existing rotational period information and (ii) derive rotational periods for stars hitherto not known to be photometric variables / ACV variables. Our final sample consists of 318 stars: 294 mCP stars, a considerable amount of which (more than 40\%) are presented here as photometric variables for the first time, and 24 mCP star candidates.

We have identified four eclipsing binary systems among our targets, which are potentially of great interest because mCP stars are very rarely found in close binary systems, particularly eclipsing ones \citep{RM09,niemczura17,kochukhov18}. Using archival spectra, we find strong evidence that the system of HD\,252519 indeed harbours an mCP star component; all other systems should be investigated with time-resolved spectroscopy for confirmation.

The employed photometric time-series data, the process of target selection and the methods of data analysis and classification are described in Section \ref{observ}. Results are presented in Section \ref{result}, which also discusses the eclipsing binary systems in more detail. We conclude in Section \ref{conclu}.

\section{Observations, target selection and data analysis} \label{observ}

This section contains a short description of the photometric survey sources employed in the present study. Further detailed information can be gleaned from the references given. In addition, we detail the processes of target selection, data analysis and classification that led to the here presented sample of mCP stars and candidates.

\subsection{Observations} \label{observations} 

\subsubsection{The All Sky Automated Survey (ASAS)} \label{surveys_ASAS3}
The All Sky Automated Survey (ASAS)\footnote{http://www.astrouw.edu.pl/asas/} constantly monitored the entire southern sky and part of the northern sky to about $\delta$\,$<$\,+28\degr, with the goal of detecting and analyzing any kind of photometric variability. In the present study, data from the third phase of the project (ASAS-3) have been employed, which lasted from 2000 until 2009 \citep{ASAS1}. Observations were procured at the 10-inch astrograph dome of the Las Campanas Observatory, Chile with two wide-field telescopes equipped with f/2.8 200\,mm Minolta lenses and 2048 x 2048 AP 10 Apogee CCD cameras (field of view of 8\fdg8\,x\,8\fdg8). Observations were acquired in Johnson ($V$) and reasonable photometry is available for stars in the magnitude range from 7 to 14. The strict observing cadence of the ASAS survey (typically one observation per day; \citealt{ASAS2}) introduces strong daily aliasing in the resulting Fourier amplitude spectra. More information on the ASAS survey is found in \citet{ASAS1}.

\subsubsection{The Kilodegree Extremely Little Telescope (KELT)} \label{surveys_KELT}
The Kilodegree Extremely Little Telescope (KELT) survey\footnote{https://keltsurvey.org/} is an ongoing project that aims at the detection of transiting exoplanets and employs two robotic wide-field telescopes situated at the Winer Observatory in Arizona (KELT-N) and the South African Astronomical Observatory near Sutherland (KELT-S) \citep{KELT1,KELT2}. The KELT systems consist of a 4096 x 4096 AP16E Apogee CCD camera (KELT-N) and a 4096 x 4096 Alta U16M Apogee CCD camera (KELT-S) that are used with a Kodak Wratten \#8 red-pass filter and either f/1.9 42\,mm (wide-angle survey mode; field of view of 26\fdg0\,x\,26\fdg0) or f/2.8 200\,mm (narrow-angle campaign mode; field of view of 10\fdg8\,x\,10\fdg8) Mamiya lenses. The observing cadence is 10 to 30\,min. The obtained magnitudes are comparable to standard $R$ band photometry and useful for stars in the magnitude range from 7 to 13. The time baseline of the KELT data reaches up to 10 years but differs widely from object to object. More information on the KELT survey is found in \citet{KELT1,KELT2}.

\subsubsection{The Multi-site All-Sky CAmeRA (MASCARA)} \label{surveys_MASCARA}
The Multi-site All-Sky CAmeRA (MASCARA)\footnote{http://mascara1.strw.leidenuniv.nl/} is an ongoing survey run by Leiden University that is dedicated to the search for exoplanets around bright stars in the $V$ magnitude range of 4 to 8 \citep{MASCARA2}. MASCARA operates two stations located in the northern hemisphere at the Roque de los Muchachos Observatory, Canary Islands and in the southern hemisphere at La Silla Observatory, Chile, which have been operational since early 2015 and October 2016, respectively. The stations consist of five 4008 x 2672 CCD cameras (northern station: ATIK 11000M; southern station: FLI6 ML11002) equipped with Kodak KAI-11002 interline CCD sensors and f/1.4 24\,mm Canon lenses with a 17\,mm aperture (field of view per camera of 53\fdg0\,x\,74\fdg0). No filters are used, and data are acquired at a cadence of 6.4\,s. The currently publicly available data have been binned by 50 points to a cadence of 320\,s \citep{Burggraaff18}. More information on the MASCARA survey can be gleaned from \citet{MASCARA1} and \citet{MASCARA2}.

\subsection{Target stars} \label{target}


For the investigation with ASAS-3 and KELT data, a list of mCP stars and mCP star candidates was selected from the most recent version of the Catalogue of Ap, HgMn, and Am stars \citep[][RM09 hereafter]{RM09}. Here, the emphasis was on deriving rotational periods for objects that hitherto lacked this information; therefore, we excluded known ACV variables with well determined rotational periods from further analysis. To this end, the General Catalogue of Variable Stars \citep{GCVS}, the International Variable Star Index of the AAVSO (VSX; \citealt{VSX}) and the SIMBAD \citep{SIMBAD} and VizieR \citep{VizieR} databases were queried for classificatory and rotational period information. Suspected or misclassified variables and variables of undetermined type (catalogue-types 'VAR', 'MISC' etc.) were kept. These stars are identified in Table \ref{table_ASASKELT} via footnotes that provide essential information from the VSX. 112 mCP stars having ASAS-3 data and 38 mCP stars with KELT data were selected via this approach.

At the time of this writing, MASCARA data were only publicly available for the stars studied by \citet{Burggraaff18}. We crossmatched the light curves published by \citet{Burggraaff18} with the RM09 catalogue and retrieved data for 144 mCP stars. All of these bright objects are already known as photometric variables with period information of differing quality. Therefore, because of their high quality and cadence, MASCARA data were employed to verify or improve the published periods.

In order to search for rotational variables that might be spectroscopically misclassified mCP stars, we subsequently investigated the remaining objects from the RM09 catalogue (i.e. objects without mCP stars classifications) for which KELT and MASCARA data were available (mainly CP1 and CP3 stars).\footnote{Because of the ASAS survey's extensive sky coverage, an investigation of non-mCP stars from the RM09 list using ASAS-3 data is a major project that is beyond the scope of the present study.} In this way, 24 stars that show light variability in agreement with rotational modulation were selected and included into the sample as ACV candidates.

\subsection{Data analysis and classification} \label{dataan}

The following sections give details on the data processing, the methods used in period analysis and the process of classification.

\subsubsection{Data processing and period analysis}  \label{datapr}

The light curves of our sample stars were downloaded from the ASAS-3 website, the KELT time series search page on the NASA exoplanet archive\footnote{https://exoplanetarchive.ipac.caltech.edu/cgi-bin/TblSearch/nph-tblSearchInit?app=ExoTbls\&config=kelttimeseries} and the VizieR catalogue of light curves published by \citet{Burggraaff18}.\footnote{http://vizier.u-strasbg.fr/viz-bin/VizieR?-source=J/A+A/617/A32}

Only data sets containing at least 100 measurements were included into the analysis. In the case of ASAS-3 data, no stars brighter than $V$\,=\,7\,mag were investigated to avoid saturation issues. No such issues were identified in the KELT and MASCARA data sets; problematic data sets seem to have been sorted out prior to publication.

In a first step, all data sets were cleaned of outliers by a 3$\sigma$ clipping and searched for periodic signals in the frequency range of 0\,$<$\,$f$(d$^{-1}$)\,$<$\,20 using \textsc{Period04} \citep{period04}. Subsequently, remaining outliers were carefully removed by visual inspection and left-over data points with a quality flag of 'D' (='worst data, probably useless') were removed from the ASAS-3 data sets. No significant instrumental trends were identified in the data for our sample stars.

For a frequency to be deemed significant, it had to show a signal-to-noise (S/N) ratio greater than four (SNR\,$>$\,4). SNR values were determined using \textsc{Period04}. In addition, to avoid spurious detections, all objects with 4\,$<$\,SNR\,$<$\,6 were required to show a False Alarm Probability (FAP) less than 1\,\% (FAP\,$<$\,0.01). The FAP was calculated using the ANOVA method within the program package PERANSO \citep{PERANSO}. Objects with FAP\,$\ge$\,0.01 were discarded, except for the star HD 273763. This object shows only weak variability in ASAS-3 data (FAP\,$\sim$\,0.1); however, its light curve is much better defined in data from the All Sky Automated Survey for SuperNovae (ASAS-SN; \citealt{shappee14}), which prove the reality of the light variations.

For the final frequency analysis, the pretreated datasets were again searched for periodic signals in the frequency domain of 0\,$<$\,$f$(d$^{-1}$)\,$<$\,5 with \textsc{Period04}. The resulting phase plots, folded with the most significant frequency, were visually inspected and objects exhibiting convincing phase plots were kept. In all objects, only one significant frequency, its corresponding alias peaks, and, in many cases, harmonics were detected.

Although recent studies based on highly-precise satellite photometry \citep[e.g.][]{huemmerich18} have shown that a considerable fraction of mCP star light curves is more complex than initially thought, most mCP star light curves can be well described by a sine wave and its first harmonic (e.g. \citealt{north84,mathys85,bernhard15b}), at least when dealing with ground-based photometry. Therefore, a least-squares fit to the data of all mCP stars and candidates was performed using \textsc{Period04} and each light curve was fitted using a Fourier series consisting of the fundamental sine wave and its first harmonic. From this procedure, the semi-amplitudes $A_{\mathrm 1}$, $A_{\mathrm 2}$ and the corresponding phases $\phi_{\mathrm 1}$, $\phi_{\mathrm 2}$ were derived, which are presented with the other results in Section \ref{presen}.

Alias frequencies arise through gaps in the data and the total time span of the data set and are an issue in most ground-based surveys. Aliasing leads to an increased number of high-amplitude peaks in Fourier spectra, thereby rendering the identification of the true period difficult, which must not necessarily show the highest amplitude. This holds especially true for the low-cadence ASAS-3 data, which suffer from strong daily aliasing (cf. Section \ref{surveys_ASAS3}). To reduce the caveats of aliasing as much as possible, we carefully investigated the spectral windows of the corresponding data sets and double checked the period solution of all doubtful cases. We are confident that we have come up with the correct rotational period -- an assumption that is corroborated by the good agreement of our published ASAS-3 period solutions with the literature values (cf. \citealt{bernhard15a,huemmerich16}). Nevertheless, alias periods cannot be excluded from our results based on ASAS-3 data. Because of more favourable observing cadences, this problem is much reduced in KELT and MASCARA data and we do not expect aliasing to be a significant issue here -- at least not in the investigated frequency space.

In orientations where two chemical spots come into view during a single rotation cycle, the light curve becomes a double wave \citep[e.g.][]{maitzen80,jagelka19}. In the case of spots of similar extent and photometric properties, a symmetric double-wave light curve is to be expected, with 'maxima' of approximately the same height. Given data of sufficient precision, it is generally straighforward to identify the true rotational period by a visual inspection of the data folded on the single ($P$) and double ($2P$) periods. However, many of the investigated stars show amplitudes near the detection threshold of the data and a twice longer (or, in some cases, shorter) rotation period cannot be excluded -- in particular for objects with significant scatter in their light curves. This applies in particular to objects analyzed with ASAS-3 data. However, symmetric double-wave light curves are comparably rare (they make up 15\,\% of the sample of 650 stars analyzed by \citealt{jagelka19}), so we do not expect this to significantly influence our results.

\subsubsection{Classification}  \label{classi}

All stars satisfying the following criteria were classified as ACV variables: (i) a spectral type in agreement with a CP2 or CP4 star classification; (ii) a colour index in rough agreement with the given spectral type [consistency check]; (iii) a period longer than $P >$ 0.5 days; (iv) a variability pattern (shape, amplitude, light curve stability) in agreement with an ACV classification. In all objects, only one significant frequency was found (cf. Section \ref{datapr}), in agreement with the expectations for ACV variables; no pulsators like e.g. SPB and $\gamma$ Doradus stars, which generally show multiple periods and quite different frequency spectra from rotating variables, were identified. We will have missed the additional variability seen in roAp stars, which -- in addition to rotational light changes -- exhibit pulsational variability in the period range of about 5 to 20 min (high-overtone, low-degree, and nonradial pulsation modes; \citealt{kurtz82}). This variability, however, is of very small amplitude that is well below the detection limit of the here employed survey data.

Stars of other CP subclasses (mainly CP1 and CP3 stars) and stars from the RM09 catalogue with non-peculiar classifications that show light variability patterns strongly suggestive of rotational modulation have been classified here as ACV candidates. Further (spectroscopic) studies are necessary to decide whether these objects are in fact (i) non-mCP stars showing rotational variability, (ii) spectroscopically misclassified mCP stars or (iii) what other variability mechanisms besides rotational modulation might be at work. These stars are listed in a separate table (Table \ref{table_candidates}).

We have identified four eclipsing binary stars among our targets (HD\,244391, HD\,247441, HD\,248784 and HD\,252519), which, following GCVS standards, have been classified according to light curve shape. These objects are discussed in more detail in Section \ref{noteso}.

\begin{figure}
\centering
\includegraphics[width=0.47\textwidth]{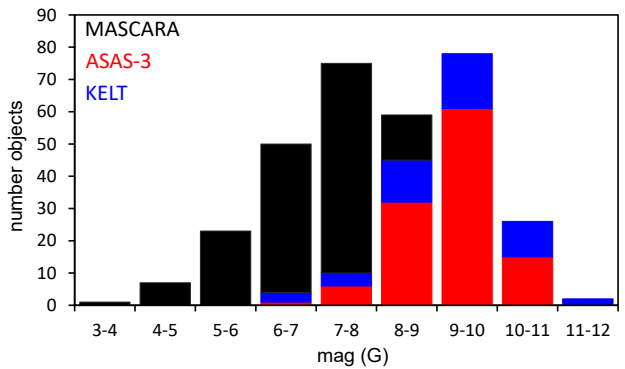}
\caption{$G$ magnitude histogram of our sample stars. Data sources are represented by different colours.}
 \label{histogram}
\end{figure}

\begin{figure}
\begin{center}
\includegraphics[width=80mm, clip]{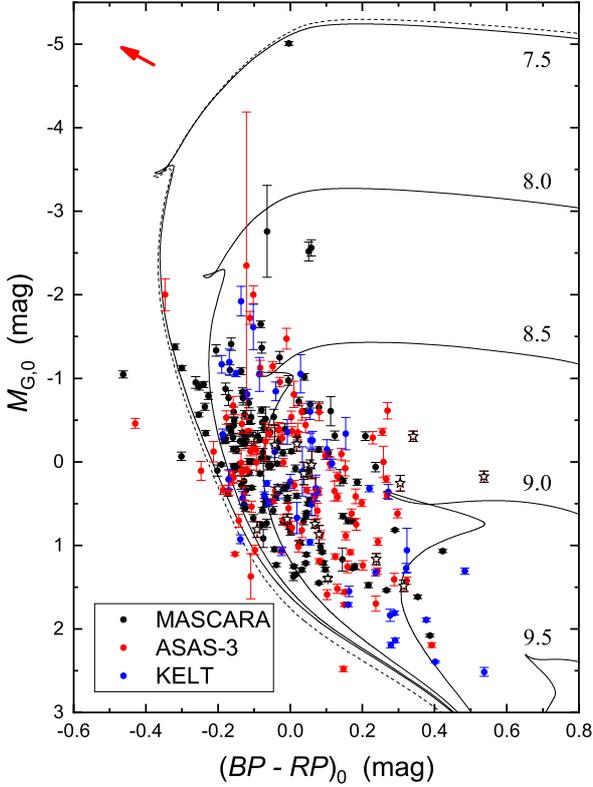}
\caption{The location of our sample stars in the $(BP - RP)_0$ versus $M_{\mathrm{G,0}}$ diagram. Also shown are PARSEC isochrones for a solar metallicity of [Z]\,=\,0.02\,dex (solid lines) and a smaller value of [Z]\,=\,0.0152\,dex (dotted line). Logarithmic ages are given. Open asterisks denote stars for which we were not able to derive reddening estimates and adopted zero values. The reddening vector is shown in the upper left corner. Five objects (HD\,40146, HD\,45530, HD\,68419, HD\,199122 and HD\,273763) are located (within 3\,$\sigma$) below the zero age main sequence. Data sources are represented by different colours.}
\label{HRD} 
\end{center} 
\end{figure}

\section{Results} \label{result}

\subsection{Presentation of results}  \label{presen}

The presentation of results is structured as follows. Table \ref{table_ASASKELT} lists essential data for the confirmed mCP stars analyzed with ASAS-3 ($N$\,=\,112) and KELT ($N$\,=\,38) data. With only a very few exceptions that are commented on in the table, these stars are here identified as photometric (and hence ACV) variables for the first time. With more than 150 new rotation periods from newly identified photometric variables or hitherto unidentified/misclassified ACV stars, Table \ref{table_ASASKELT} significantly adds to our knowledge of the rotational properties of mCP stars. It is organized as follows:
\begin{itemize}
\item Column 1: photometric survey source (ASAS-3 or KELT).
\item Column 2: HD number or other conventional identifier.
\item Column 3: identification number from RM09.
\item Column 4: right ascension (J2000). Positional information was taken from GAIA DR2 \citep{gaia1,gaia2,gaia3}.
\item Column 5: declination (J2000).
\item Column 6: Spectral type from RM09; as in this catalogue, the 'p' denoting peculiarity has been omitted from the spectral classifications.
\item Column 7: $V$ magnitude range.
\item Column 8: period (d).
\item Column 9: epoch (HJD-2450000); as is standard procedure for ACV variables, time of photometric maximum is indicated.
\item Column 10: Semi-amplitude of the fundamental variation ($A_{\mathrm 1}$). \footnote{Columns 10--13 have only been calculated for ACV variables or candidates.}
\item Column 11: Semi-amplitude of the first harmonic variation ($A_{\mathrm 2}$).
\item Column 12: Phase of the fundamental variation ($\phi_{\mathrm 1}$).\footnote{The calculation of the phase values has been based on the times of observations as provided by the ASAS-3 and KELT databases, i.e. HJD-2450000.}
\item Column 13: Phase of the first harmonic variation ($\phi_{\mathrm 2}$).
\item Column 14: $G$\,mag from GAIA DR2.
\item Column 15: $(G$\textsubscript{BP}$-G$\textsubscript{RP})\,index from GAIA DR2.
\end{itemize}
Where available, classifications and periods from the VSX are provided in footnotes. Only part of the table is printed here for guidance regarding its form and content. The complete table is given in the Appendix (Table \ref{table_masterASASKELT3}).

Known ACV variables whose rotational periods have been investigated with MASCARA data ($N$\,=\,144) are presented in Table \ref{table_MASCARA}. It is organized similar to Table \ref{table_ASASKELT} but also includes a comparison with the periods listed in the VSX at the time of this writing (June 2019). In the case of strongly discrepant values, the given periods are highlighted in bold font. The complete table is given in the Appendix (Table \ref{table_masterMASCARA}).

Potential rotational variables that were identified among stars of other CP subclasses (mainly CP1 and CP3 stars) or stars from RM09 with non-peculiar classifications are presented here as ACV candidates ($N$\,=\,24) and listed in Table \ref{table_candidates}. Spectroscopic confirmation of their suspected mCP star status is needed.

\begin{table*}
\caption{Essential data for the mostly new ACV variables analyzed with ASAS-3 and KELT data, sorted by data source and increasing right ascension. The columns denote: (1) Photometric survey source (A=ASAS-3; M=MASCARA; K=KELT). (2) HD number or other conventional identification. (3) Identification number from RM09. (4) Right ascension (J2000; GAIA DR2). (5) Declination (J2000; GAIA DR2). (6) Spectral type from RM09. (7) $V$ magnitude range. (8) Period, derived in the present investigation. (9) Epoch (HJD-2450000). (10) Semi-amplitude of the fundamental variation ($A_{\mathrm 1}$). (11) Semi-amplitude of the first harmonic variation ($A_{\mathrm 2}$). (12) Phase of the fundamental variation ($\phi_{\mathrm 1}$). (13) Phase of the first harmonic variation ($\phi_{\mathrm 2}$). (14) $G$\,mag (GAIA DR2). (15) $(BP-RP)$ index (GAIA DR2). Where available, classifications and periods from the VSX are provided in footnotes. Only part of the table is printed here for guidance regarding its form and content. The complete table is given in the Appendix (Table \ref{table_masterASASKELT3}).}
\label{table_ASASKELT}
\begin{center}
\begin{adjustbox}{max width=\textwidth}
\begin{tabular}{lllcclccccccccc}
\hline 
(1) & (2) & (3) & (4) & (5) & (6) & (7) & (8) & (9) & (10) & (11) & (12) & (13) & (14) & (15) \\
Src & Star & \#RM09 & $\alpha$(J2000) & $\delta$(J2000) & SpT & $V$\,range & Period & Epoch & $A_{\mathrm 1}$ & $A_{\mathrm 2}$ & $\phi_{\mathrm 1}$ & $\phi_{\mathrm 2}$ & $G$ & $(BP-RP)$ \\
     &        &           &                  &                  & [lit] & [mag]       & [d]    &   & [mag]           & [mag]           & [rad]              & [rad]              & [mag]   & [mag] \\
\hline
A	&	TYC 4701-644-1	&	3920	&	02 32 01.69	&	-03 49 26.9	&	A0 Sr Eu	&	10.13-10.15	&	7.296(4)	&	3736.63(7)	&	0.009	&	0.002	&	0.496	&	0.661	&	10.139	&	0.273	\\
A	&	HD 16504	&	4100	&	02 35 21.70	&	-68 09 40.0	&	B8 Si	&	9.02-9.03	&	3.3057(7)	&	4867.62(3)	&	0.005	&	0.002	&	0.287	&	0.838	&	9.021	&	-0.023	\\
A	&	HD 25258	&	6438	&	04 00 38.96	&	-04 05 49.6	&	A2 Sr Eu	&	10.31-10.33	&	2.4490(4)	&	3806.47(2)	&	0.009	&	0.001	&	0.477	&	0.922	&	10.304	&	0.300	\\
A	&	HD 28365$^{a}$	&	7280	&	04 27 56.40	&	-13 27 52.3	&	B9 Si	&	8.43-8.45	&	1.80606(9)	&	4685.91(2)	&	0.005	&	0.001	&	0.245	&	0.441	&	8.412	&	-0.107	\\
A	&	HD 29925	&	7690	&	04 43 00.38	&	+01 06 28.4	&	B9 Si	&	8.32-8.34	&	0.665558(8)	&	2558.775(7)	&	0.007	&	0.002	&	0.222	&	0.624	&	8.294	&	-0.107	\\
A	&	TYC 8510-712-1	&	8060	&	04 53 29.13	&	-53 39 26.4	&	A0 Si	&	10.54-10.55	&	10.263(8)	&	3652.8(1)	&	0.005	&	0.001	&	0.816	&	0.044	&	10.534	&	-0.092	\\
A	&	HD 273763	&	8460	&	05 05 29.78	&	-47 53 24.0	&	A0 Sr	&	10.96-10.98	&	0.77418(2)	&	3430.557(8)	&	0.008	&	0.000	&	0.507	&	0.814	&	10.970	&	0.147	\\
A	&	HD 36955	&	9740	&	05 35 04.54	&	-01 24 06.6	&	A2 Cr Eu	&	9.44-9.45	&	2.2848(2)	&	4590.46(2)	&	0.006	&	0.002	&	0.655	&	0.059	&	9.416	&	0.201	\\
A	&	HD 40146$^{b}$	&	10710	&	05 56 58.90	&	-03 45 20.4	&	A0 Si	&	9.35-9.37	&	1.78216(8)	&	3759.68(2)	&	0.009	&	0.001	&	0.168	&	0.098	&	9.306	&	0.250	\\
A	&	HD 42335$^{c}$	&	11290	&	06 10 05.33	&	-00 18 08.6	&	A0 Si	&	8.40-8.43	&	5.076(1)	&	3440.60(5)	&	0.013	&	0.004	&	0.876	&	0.451	&	8.358	&	0.139	\\
\hline
\multicolumn{15}{l}{$^{a}$NSV 16014 (ACV:); $^{b}$ASAS J055659-0345.3 (MISC; $P$\textsubscript{VSX}\,=\,1.781649\,d); $^{c}$HIP 29253 (VAR; $P$\textsubscript{VSX}\,=\,5.07872\,d)} \\
\hline     
\end{tabular}                                                                                                                                                                 
\end{adjustbox}                                                                                                                                         
\end{center}
\end{table*}  

\begin{table*}
\caption{Essential data for the known ACV variables whose rotational periods were investigated with MASCARA data, sorted by increasing right ascension. The columns denote: (1) HD number or other conventional identification. (2) Identification number from RM09. (3) Right ascension (J2000; GAIA DR2). (4) Declination (J2000; GAIA DR2). (5) Spectral type from RM09. (6) $V$ magnitude range. (7) Literature period (VSX). (8) Period (this work). (9) Epoch (HJD-2450000). (10) Semi-amplitude of the fundamental variation ($A_{\mathrm 1}$). (11) Semi-amplitude of the first harmonic variation ($A_{\mathrm 2}$). (12) Phase of the fundamental variation ($\phi_{\mathrm 1}$). (13) Phase of the first harmonic variation ($\phi_{\mathrm 2}$). (14) $G$\,mag (GAIA DR2). (15) $(BP-RP)$ index (GAIA DR2). Only part of the table is printed here for guidance regarding its form and content. The complete table is given in the Appendix (Table \ref{table_masterMASCARA}).}
\label{table_MASCARA}
\begin{center}
\begin{adjustbox}{max width=\textwidth}
\begin{tabular}{lllcclccccccccc}
\hline 
(1) & (2) & (3) & (4) & (5) & (6) & (7) & (8) & (9) & (10) & (11) & (12) & (13) & (14) & (15) \\
Star & \#RM09 & $\alpha$(J2000) & $\delta$(J2000) & SpT   & $V$\,range & P\textsubscript{VSX} & P\textsubscript{MASCARA} & Epoch          & $A_{\mathrm 1}$ & $A_{\mathrm 2}$ & $\phi_{\mathrm 1}$ & $\phi_{\mathrm 2}$ & $G$ & $(BP-RP)$ \\
     &           &                  &                  & [lit] & [mag]       & [d]   & [d]   & & [mag]           & [mag]           & [rad]              & [rad]              & [mag]   & [mag] \\
\hline
HD 3580	&	970	&	00 38 31.86	&	-20 17 47.6	&	B8 Si	&	6.71-6.73	&	1.4788	&	1.47561(4)	&	7219.71(1)	&	0.006	&	0.001	&	0.986	&	0.484	&	6.687		&	-0.157	\\
HD 4778	&	1250	&	00 50 18.27	&	+45 00 08.2	&	A1 Cr Sr Eu	&	6.10-6.13	&	2.5481	&	2.5619(1)	&	7185.70(2)	&	0.006	&	0.008	&	0.899	&	0.974	&	6.110		&	0.028	\\
HD 4796	&	1263	&	00 52 28.98	&	+79 50 25.6	&	A0 Sr Si	&	7.71-7.73	&	8	&	7.987(2)	&	7060.62(7)	&	0.009	&	0.000	&	0.765	&	0.323	&	7.695		&	0.152	\\
HD 7676	&	1880	&	01 16 06.81	&	-34 08 55.8	&	A5 Sr Cr Eu	&	8.36-8.41	&	5.0976	&	5.088(2)	&	7252.84(5)	&	0.022	&	0.007	&	0.010	&	0.744	&	8.388		&	0.217	\\
HD 7546	&	1870	&	01 16 24.50	&	+48 04 56.1	&	B8 Si	&	6.60-6.62	&	5.229	&	5.4060(9)	&	7057.38(5)	&	0.009	&	0.004	&	0.250	&	0.826	&	6.563		&	-0.002	\\
HD 8441	&	2050	&	01 24 18.69	&	+43 08 31.6	&	A2 Sr	&	6.66-6.68	&	69.92	&	69.9(1)	&	7230.2(7)	&	0.009	&	0.004	&	0.228	&	0.942	&	6.651		&	0.079	\\
HD 10221	&	2550	&	01 42 20.51	&	+68 02 34.9	&	A0 Si Sr Cr	&	5.54-5.58	&	3.1848	&	3.1528(2)	&	7216.49(3)	&	0.017	&	0.005	&	0.769	&	0.899	&	5.529		&	-0.064	\\
HD 10783	&	2670	&	01 45 42.52	&	+08 33 33.2	&	A2 Si Cr Sr	&	6.53-6.56	&	4.1321	&	4.1335(3)	&	7057.30(4)	&	0.009	&	0.005	&	0.374	&	0.063	&	6.514		&	-0.009	\\
HD 11415	&	2870	&	01 54 23.74	&	+63 40 12.4	&	B3 He wk.	&	3.34-3.35	&	\textbf{0.08946}	&	\textbf{14.53(1)}	&	7192.6(1)	&	0.006	&	0.002	&	0.528	&	0.503	&	3.261		&	0.070	\\
HD 12288	&	3130	&	02 03 30.49	&	+69 34 56.4	&	A2 Cr Si	&	7.72-7.74	&	34.9	&	35.73(3)	&	7218.6(3)	&	0.007	&	0.002	&	0.772	&	0.633	&	7.706		&	0.177	\\
\hline          
\end{tabular}                                                                                                                                                                 
\end{adjustbox}                                                                                                                                         
\end{center}
\end{table*} 

\begin{table*}
\caption{Essential data for the ACV variable candidates identified among stars of other CP subclasses or stars from RM09 with non-peculiar classifications, sorted by increasing right ascension. The columns denote: (1) Data source. (2) HD number or other conventional identification. (3) Identification number from RM09. (4) Right ascension (J2000; GAIA DR2). (5) Declination (J2000; GAIA DR2). (6) Spectral type from RM09. (7) $V$ magnitude range. (8) Period, derived in the present investigation. (9) Epoch (HJD-2450000). (10) Semi-amplitude of the fundamental variation ($A_{\mathrm 1}$). (11) Semi-amplitude of the first harmonic variation ($A_{\mathrm 2}$). (12) Phase of the fundamental variation ($\phi_{\mathrm 1}$). (13) Phase of the first harmonic variation ($\phi_{\mathrm 2}$). (14) $G$\,mag (GAIA DR2). (15) $(G$\textsubscript{BP}$-G$\textsubscript{RP})\,index (GAIA DR2).}
\label{table_candidates}
\begin{center}
\begin{adjustbox}{max width=\textwidth}
\begin{tabular}{lllcclcccccccccc}
\hline 
(1) & (2) & (3) & (4) & (5) & (6) & (7) & (8) & (9) & (10) & (11) & (12) & (13) & (14) & (15) \\
Source & Star & ID (RM09) & $\alpha$ (J2000) & $\delta$ (J2000) & SpT   & Range ($V$) & Period & Epoch          & $A_{\mathrm 1}$ & $A_{\mathrm 2}$ & $\phi_{\mathrm 1}$ & $\phi_{\mathrm 2}$ & $G$ & $(G$\textsubscript{BP}$-G$\textsubscript{RP}) \\
     &        &           &                  &                  & [lit] & [mag]       & [d]    & [HJD-2450000]  & [mag]           & [mag]           & [rad]              & [rad]              & [mag]   & [mag] \\
\hline
MASCARA	&	HD 7157	&	1795	&	01 13 09.78	&	+61 42 22.3	&	B9	&	6.43-6.46	&	1.00674(2)	&	7151.70(1)	&	0.015	&	0.001	&	0.944	&	0.389	&	6.477	&	-0.004	\\
KELT	&	TYC 2903-775-1	&	8178	&	05 02 37.13	&	+42 33 16.4	&	A2	&	10.86-10.87	&	3.2402(2)	&	4057.84(3)	&	0.007	&	0.001	&	0.396	&	0.411	&	10.828	&	0.346	\\
KELT	&	HD 36589$^{a}$	&	9520	&	05 33 38.84	&	+20 28 27.2	&	B7 HgMn?	&	6.18-6.19	&	7.656(2)	&	4057.77(7)	&	0.003	&	0.001	&	0.836	&	0.471	&	6.156	&	-0.006	\\
KELT	&	TYC 2412-1315-1	&	10035	&	05 39 53.12	&	+34 31 27.2	&	A0-	&	8.74-8.76	&	0.75402(2)	&	4038.903(7)	&	0.007	&	0.000	&	0.258	&	0.912	&	8.663	&	0.426	\\
KELT	&	HD 247209	&	10315	&	05 46 12.12	&	+31 23 16.0		&	B9-	&	9.33-9.35	&	5.710(1)	&	4379.99(5)	&	0.007	&	0.002	&	0.674	&	0.837	&	9.314	&	0.106	\\
KELT	&	HD 250237	&	10859	&	06 01 21.03	&	+32 57 50.0	&		A0-	&	9.98-9.99	&	2.6873(2)	&	4035.81(2)	&	0.004	&	0.001	&	0.064	&	0.954	&	9.960	&	0.340	\\
MASCARA	&	HD 55667	&	15134	&	07 21 50.63	&	+75 05 22.0	&	A2	&	6.93-6.96	&	1.79690(6)	&	7176.49(1)	&	0.014	&	0.003	&	0.936	&	0.077	&	6.938	&	0.024	\\
MASCARA	&	HD 101753	&	29335	&	11 42 44.02	&	+18 14 31.3	&	B9	&	7.35-7.41	&	7.828(1)	&	7056.80(7)	&	0.029	&	0.006	&	0.304	&	0.679	&	7.371	&	-0.184	\\
MASCARA	&	HD 180582	&	50098	&	19 15 17.24	&	+40 06 49.2	&	B9	&	8.07-8.15	&	2.2908(1)	&	7085.66(2)	&	0.035	&	0.017	&	0.452	&	0.590	&	8.092	&	-0.129	\\
MASCARA	&	HD 182865	&	50555	&	19 25 39.02	&	+26 06 14.3	&	B8	&	7.36-7.38	&	6.424(1)	&	7079.50(6)	&	0.012	&	0.003	&	0.693	&	0.300	&	7.304	&	0.219	\\
MASCARA	&	HD 183142	&	50623	&	19 26 01.21	&	+45 00 47.1	&	B8	&	7.04-7.09	&	1.08805(2)	&	7064.78(1)	&	0.022	&	0.001	&	0.698	&	0.067	&	7.034	&	-0.221	\\
MASCARA	&	HD 189775	&	52618	&	19 59 15.36	&	+52 03 20.6	&	B5	&	6.09-6.15	&	2.6074(1)	&	7135.63(2)	&	0.028	&	0.003	&	0.049	&	0.511	&	6.097	&	-0.239	\\
MASCARA	&	HD 193553	&	54027	&	20 19 49.81	&	+29 43 36.8	&	B8	&	6.75-6.79	&	6.8197(9)	&	7076.68(6)	&	0.019	&	0.003	&	0.023	&	0.520	&	6.733	&	-0.190	\\
MASCARA	&	HD 195447	&	54447	&	20 29 20.82	&	+50 27 37.2	&	B9	&	7.56-7.59	&	5.3970(6)	&	7060.57(5)	&	0.014	&	0.005	&	0.501	&	0.282	&	7.557	&	0.072	\\
KELT	&	HD 196655	&	54880	&	20 37 50.18	&	+29 43 17.9	&	A2-	&	8.24-8.25	&	2.3409(2)	&	4274.80(2)	&	0.005	&	0.000	&	0.606	&	0.133	&	8.214	&	0.145	\\
MASCARA	&	HD 199122	&	55433	&	20 53 39.89	&	+41 02 58.4	&	A2	&	7.56-7.60	&	1.24673(3)	&	7085.78(1)	&	0.024	&	0.001	&	0.261	&	0.091	&	7.575	&	-0.130	\\
KELT	&	TYC 3171-971-1	&	55450	&	20 54 03.30	&	+40 42 23.8	&	A4-	&	8.75-8.76	&	1.46591(5)	&	4276.93(1)	&	0.002	&	0.000	&	0.208	&	0.306	&	8.714	&	0.366	\\
KELT	&	HD 200407$^{b}$	&	55840	&	21 01 47.46	&	+44 11 12.9	&	A4-F1	&	6.75-6.76	&	8.325(3)	&	4386.60(8)	&	0.004	&	0.001	&	0.833	&	0.369	&	6.686	&	0.418	\\
KELT	&	TYC 2709-59-1	&	56010	&	21 05 11.27	&	+34 40 49.1	&	A5-	&	9.31-9.32	&	2.4770(2)	&	4274.89(2)	&	0.006	&	0.001	&	0.962	&	0.062	&	9.264	&	0.322	\\
MASCARA	&	HD 204374	&	56913	&	21 26 28.36	&	+49 16 49.0	&	A0	&	7.92-7.99	&	1.27985(2)	&	7151.68(1)	&	0.030	&	0.005	&	0.886	&	0.787	&	7.917	&	-0.116	\\
KELT	&	HD 204936	&	57090	&	21 31 20.66	&	+21 12 16.2	&	A2-	&	8.54-8.55	&	13.56(1)	&	4601.9(1)	&	0.004	&	0.003	&	0.401	&	0.263	&	8.437	&	0.512	\\
KELT	&	TYC 2712-2178-1	&	57270	&	21 35 16.02	&	+33 58 31.6	&	A2-	&	9.48-9.49	&	0.56069(1)	&	4260.906(5)	&	0.003	&	0.000	&	0.422	&	0.740	&	9.419	&	0.493	\\
MASCARA	&	HD 217062	&	59915	&	22 56 56.13	&	+59 57 42.2	&	B9	&	7.19-7.23	&	1.26737(3)	&	7176.69(1)	&	0.021	&	0.000	&	0.078	&	0.596	&	7.192	&	-0.073	\\
MASCARA	&	HD 220885	&	60550	&	23 27 07.39	&	+42 54 43.2	&	B9 Mn	&	5.72-5.76	&	1.47905(4)	&	7056.34(1)	&	0.016	&	0.005	&	0.856	&	0.135	&	5.728	&	-0.002	\\
\hline
\multicolumn{15}{l}{$^{a}$ NSV 2132; suspected of variability by \citet{crawford63}; non-variable according to the VSX.} \\
\multicolumn{15}{l}{$^{b}$ NSV 25432; possible microvariable according to \citet{rufener81}; range is 6.74-6.76\,mag ($V$) according to the VSX.} \\
\hline
\end{tabular}                                                                                                                                                                 
\end{adjustbox}                                                                                                                                         
\end{center}
\end{table*} 

The light curves of all objects, folded with the periods listed in Tables \ref{table_masterASASKELT3}, \ref{table_masterMASCARA} and \ref{table_candidates} are presented in the Appendix (Section \ref{appendixLCs}). The $G$ magnitude histogram of our sample stars is presented in Fig. \ref{histogram}. It shows a bimodal distribution that illustrates the different data sources used in the analysis, with the 'bright' peak corresponding to objects with MASCARA data and the other peak to sources with ASAS-3 and KELT data.

\subsection{Colour-magnitude diagram}
To investigate the astrophysical properties of our sample stars in a colour-magnitude diagram (CMD), we employed the homogeneous Gaia DR2 photometry from \citet{Arenou2018}. Only HD\,170000 ($\phi$ Dra) is not included in this data set because it is too bright. For two stars (HD\,122264 and HD\,245601), no parallax measurements exist, and two objects (HD\,164068 and GSC\,2413-00426) have significant negative parallaxes (-2.6 and -1.3\,mas, respectively). Because no Hipparcos parallaxes are available \citep{Leeuwen2007}, we excluded these stars from further analysis. Another object, HD\,46462, has a very small parallax (0.08\,mas) with a large error (1.09\,mas). We strongly suspect that, for some reason we cannot identify, the Gaia DR2 measurement is erroneous and have adopted the Hipparcos satellite parallax of 2.93(52)\,mas.
 
For the estimation of the interstellar reddening (absorption), we preferably used the dereddening calibrations in the Str{\"o}mgren-Crawford system \citep{Paunzen2005,Paunzen2006} valid for mCP stars \citep{Netopil2008}. For 177 stars, all four Str{\"o}mgren-Crawford indices ($b-y$, $m_1$, $c_1$ and $\beta$) are available in the catalogue of \citet{Paunzen2015}. The transformation of the reddening values was performed using the relations
\begin{eqnarray}
A_V &=& 1.1A_G = 3.1E(B - V) = 4.3E(b - y) = \nonumber \\
    &=& 2.1E(BP - RP).
\end{eqnarray}
As next step, the published reddening map by \citet{Green2018} and the individual reddening estimates by \citet{Chen2019} were employed using the distances listed in \citet{Bailer2018}. We note that these references do not cover the entire sky. Therefore, for sixteen objects (HD\,71006, HD\,88701, HD\,98956, HD\,111672, HD\,129189, HD\,147174, HD\,162306, HD\,162639, HD\,163583, HD\,167024, HD\,168108, HD\,172626, HD\,193382, TYC\,7415-2499-1, TYC\,8143-3244-1 and TYC\,8510-712-1), we were not able to derive reddening estimates and have thus adopted zero values.

In Fig. \ref{HRD}, we present the CMD of our sample stars together with PARSEC isochrones \citep{Bressan2012} for a solar metallicity of [Z]\,=\,0.02\,dex. We favour this value because it has been shown to be consistent with recent results of helioseismology \citep{Vagnozzi2019}. Adopting a smaller value, as e.g. suggested by \citet{Bressan2012}, shifts the main sequence slightly to the blue, as indicated in Figure \ref{HRD}. In general, our reddening estimates seem to be very reliable. However, five objects (HD\,40146, HD\,45530, HD\,68419, HD\,199122 and HD\,273763) are located within 3\,$\sigma$ below the zero-age main sequence (ZAMS). On the basis of Str{\"o}mgren-Crawford photometry, which generally is very reliable, we derive absorption values ($A_G$) between 0.45 and 0.64\,mag for HD\,45530, HD\,68419 and HD\,199122. In the case of HD\,41046, we derived a value of 1.31\,mag based on reddening maps. This star is a member of the Orion OB1 association \citep{Romanyuk2013} and hence situated in a part of the sky where strong differential reddening is evident which might cause an overestimation. HD\,273763, on the other hand, seems to exhibit no reddening at all, which is to be expected considering its location at a Galactic latitude of $-$37$\degr$. Interestingly, it is a high proper-motion star \citep{Tetzlaff2011}. Otherwise, no indications of measurement errors or false identifications were found. Two further stars (HD\,91590 and TYC\,6545-2278-1) are situated below the ZAMS in Fig. \ref{HRD} but do not satisfy the 3\,$\sigma$ criterion.

From the distribution of stars in Fig. \ref{HRD}, we find that almost all objects are between 100\,Myr and 1\,Gyr old, with an apparent lack of very young stars (pertaining, in particular, to objects with absolute magnitudes below 1.5\,mag). This is in agreement with the currently favoured model predicitions that almost all mCP stars are hydrogen burning main-sequence objects. \citet{landstreet06} and \citet{landstreet07,landstreet08} have shown that the magnetic field at the surface of mCP stars decreases with time because of magnetic flux conversation with increased stellar radius and additional decay (cf. also \citealt{fossati16,shultz19}). The time scales involved vary strongly with mass ($\sim$250\,Myr for stars of 2-3\,$M_{\odot}$ to $\sim$15\,Myr for stars of 4-5\,$M_{\odot}$ according to \citealt{landstreet08}). The more evolved stars of our sample (see below) may provide an interesting testbed to further elaborate on these findings.

\citet{landstreet07} find magnetic fields in mCP stars from the ZAMS to the terminal-age main sequence (TAMS). The assumption that the magnetic fields are already there at late stages of the pre-main sequence phase has been substantiated by the discovery of global fields in Herbig Ae/Be stars \citep[e.g.][]{wade05,alecian13b}. In contrast to this, as regards our sample stars, we find a conspicuous lack of mCP stars directly on the ZAMS, which seems to imply that a certain amount of time is necessary either for the magnetic field to build up in strength or for the peculiarities to develop. This recalls the observations of \citet{Hubrig2000}, who find that mCP stars with $M_{\odot}$\,$\le$\,3 are typically more than 30\,\% of the way through their life on the main sequence. The assumption that the magnetic fields appear only after a certain time, however, is in contrast with theoretical considerations that have narrowed down the time a fossil field requires to relax onto a stable configuration to just a few Alfv{\'e}n times \citep{braithwaite06} and has been disputed by several studies (cf. the discussion in \citealt{landstreet07}). Our study does not follow the methodological approach of \citet{Hubrig2000} and therefore cannot be affected by the same systematic issues. We further cannot identify systematic errors in our treatment of the data that have led to the construction of Fig. \ref{HRD}, such as a possible overestimation of the luminosity of our sample stars. In summary, with the currently available data, we are unable to satisfactorily explain and assess our observation.

Several stars of our sample seem to have already left the main sequence and are situated on the subgiant branch -- albeit well before the first dredge-up \citep{Vassiliadis1993}. In these evolutionary stages, diffusion seems to only play a minor role in the stellar atmospheres of intermediate mass stars. \citet{bailey14} have shown that several elements, such as the iron-peak elements Cr, Fe and Ti, decrease in abundance with age, approaching near-solar values close to the terminal-age main sequence, while the abundance of He increases. In this respect, it would be highly interesting to investigate the abundance patterns of the more evolved mCP stars in our sample.

\subsection{Rotational periods} \label{rotperiods}

\begin{figure}
\centering
\includegraphics[width=0.47\textwidth]{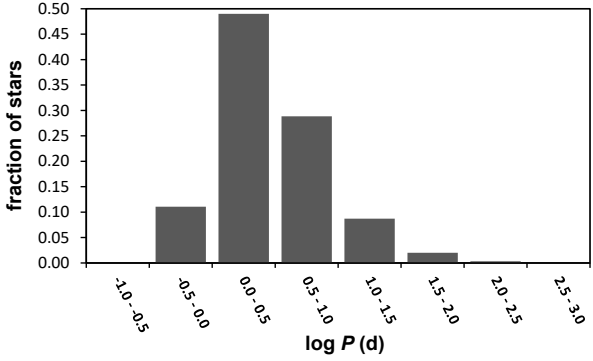}
\caption{Rotational period histogram of the 294 confirmed mCP stars of our sample.}
 \label{period_histogram}
\end{figure}

Fig. \ref{period_histogram} shows a histogram of the period distribution of the 294 confirmed mCP stars of our sample. The rotational period distribution is in very good agreement with the literature (e.g. \citealt{RM09,netopil17}) and shows the typical peak distribution in the $0.0$\,<\,$log(P[d])$\,<\,$0.5$ bin. We note that because of the time span and characteristics of the here employed photometric time-series data, we will have missed very slowly rotating ACV variables with periods of several years or more \citep{Mathys17}.

\begin{figure}
\centering
\includegraphics[width=0.47\textwidth]{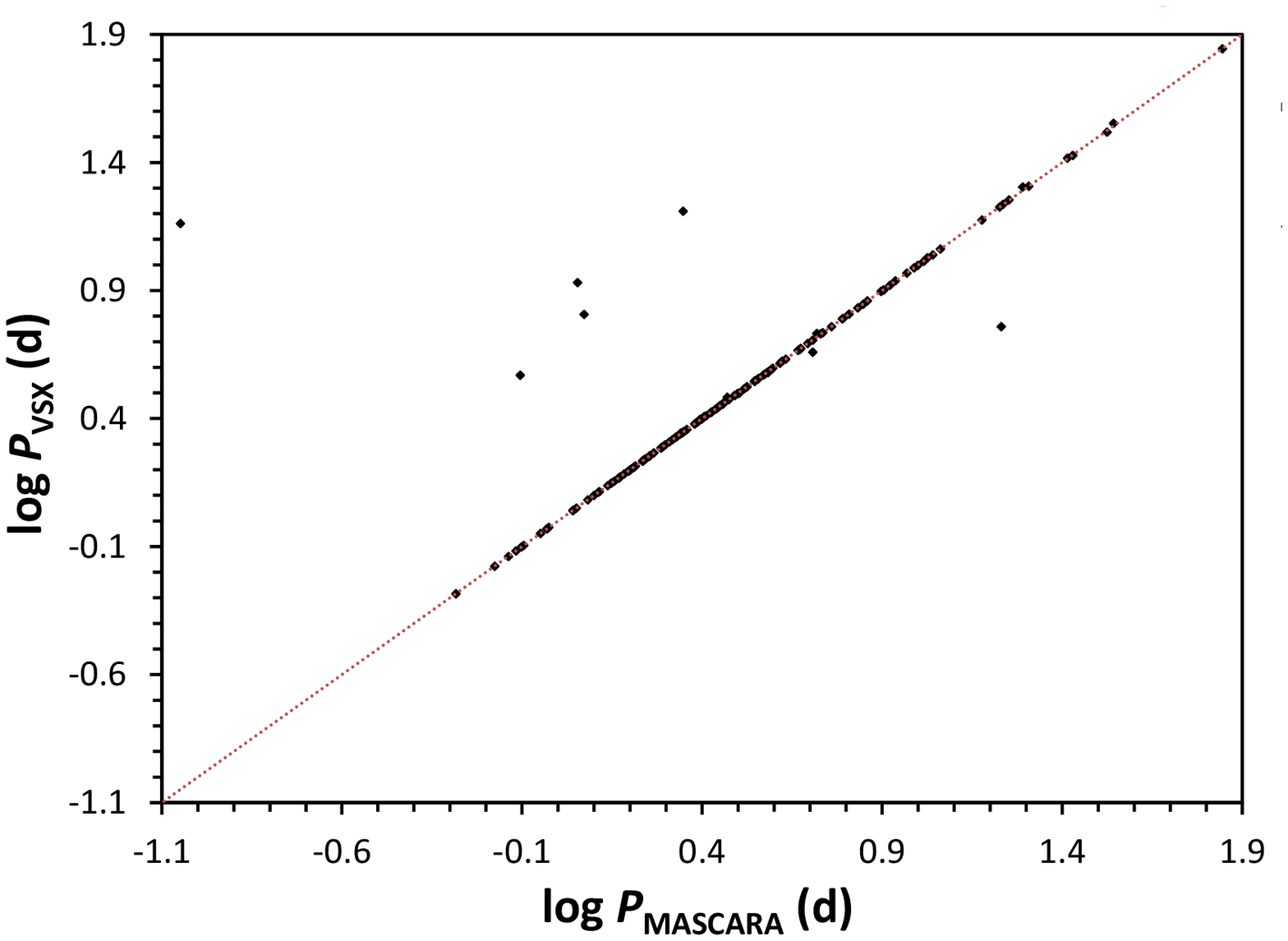}
\includegraphics[width=0.47\textwidth]{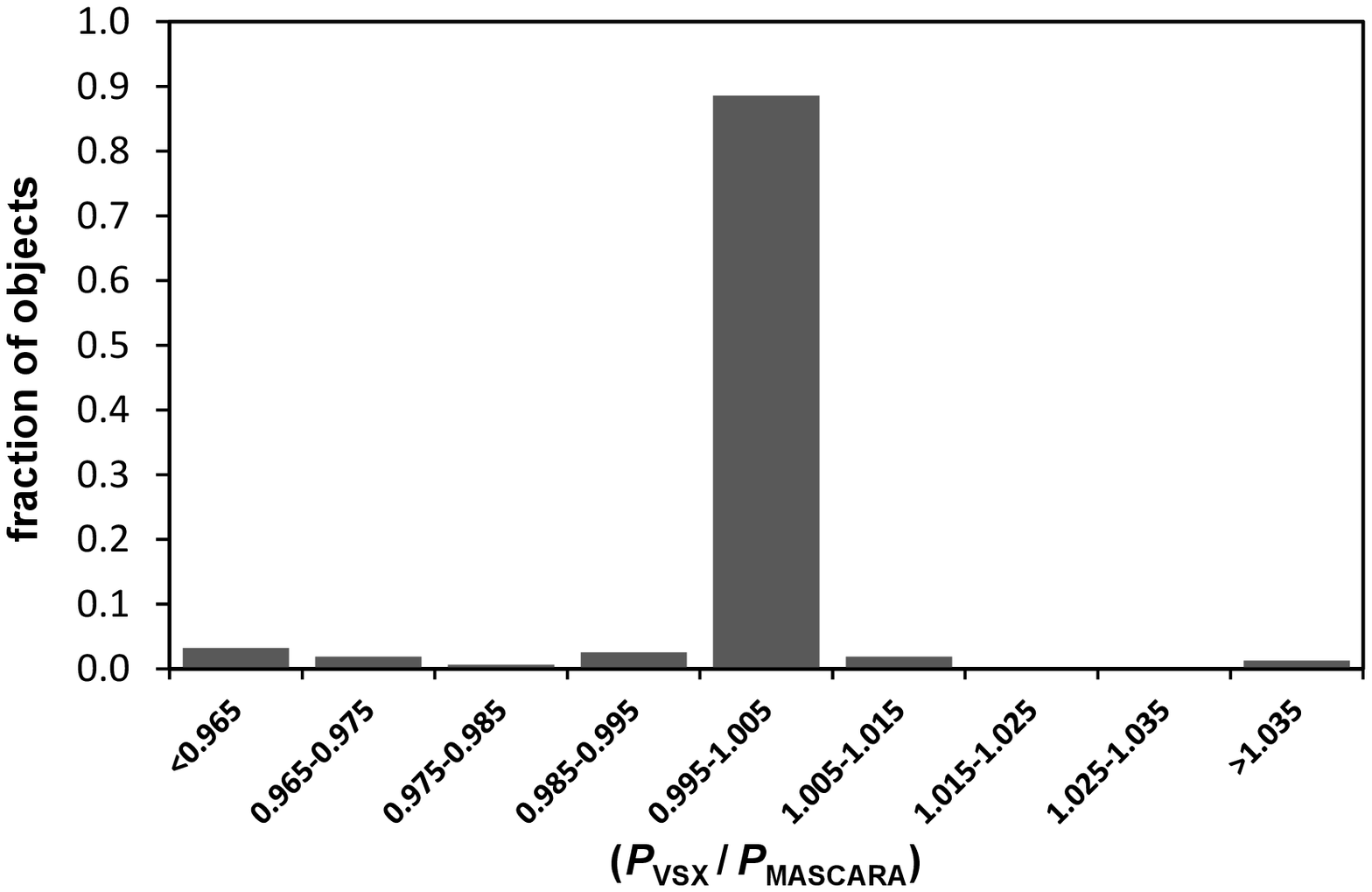}
\caption{The upper panel shows a comparison of the VSX periods and the periods derived from the analysis of MASCARA data in the present work. The lower panel gives the corresponding $\delta$$P$ value histogram.}
 \label{period_histogram_VSX_MASCARA}
\end{figure}

As has been pointed out (Section \ref{target}), at the time of this writing, MASCARA data were only publicly available for known variables with period information of differing quality. These data were therefore employed to verify or improve the published periods. A comparison of the VSX periods with the periods derived from the analysis of MASCARA data in the present work is presented in the upper panel of Fig. \ref{period_histogram_VSX_MASCARA}; the lower panel illustrates the corresponding $\delta$$P$ value histogram. The agreement is generally very good, highlighting the quality of the VSX period data which -- although based on very different sources -- are constantly being updated with new results from the literature. Only for some objects, the discrepancy between the VSX and MASCARA periods is striking. The observed differences are most likely due to alias periods, as e.g. in the case of HD\,193722 ($P$\textsubscript{VSX}\,=\,$1.132854$\,d vs. $P$\textsubscript{MASCARA}\,=\,$8.530(1)$\,d), or overintepretation of short or sparsely sampled datasets.

It is well known that some ACV variables show variations of the rotational (photometric) period over time \citep{Mikulasek17}. However, no detailed analysis of this phenomenon has been carried out for a statistically sound sample of stars. The four ACVs variables with the currently largest known period variations -- HD\,37479 (He strong star; \citealt{townsend10,oksala12}), HD\,37776 (He strong star; \citealt{mikulasek11}), HD\,124224 (classical Si star; \citealt{smith97,pyper98}) and HD\,125630 (cool SrCrEu star; \citealt{mikulasek15}) -- form a very heterogeneous sample of mCP stars. However, the observed period changes range from 0.08 to almost 3\,s yr$^{-1}$, which equals to about 9x10$^{-8}$\,d$^{-1}$. This is well below the accuracy of the derived periods (Tables \ref{table_ASASKELT} and \ref{table_MASCARA}). Therefore, the possible occurrence of period changes in our sample stars are not expected to account for even the slight discrepancies between the (mostly older) VSX periods and the here derived periods.

\subsection{Eclipsing binary systems} \label{noteso}
This sections contains a short discussion of the eclipsing binary systems identified among our sample stars.

\subsubsection{HD\,244391}  \label{HD244391}
HD\,244391 is listed with a spectral type of B8pSiSr in the RM09 catalogue. Only five references to this star are contained in the SIMBAD database; the only photometric study available is that of \citet{wraight12}, who investigated mCP stars with the STEREO satellite and found HD\,244391 to be constant.

\begin{figure*}
 \centering
	\mbox{
			\subfigure{\includegraphics[width=0.47\textwidth]{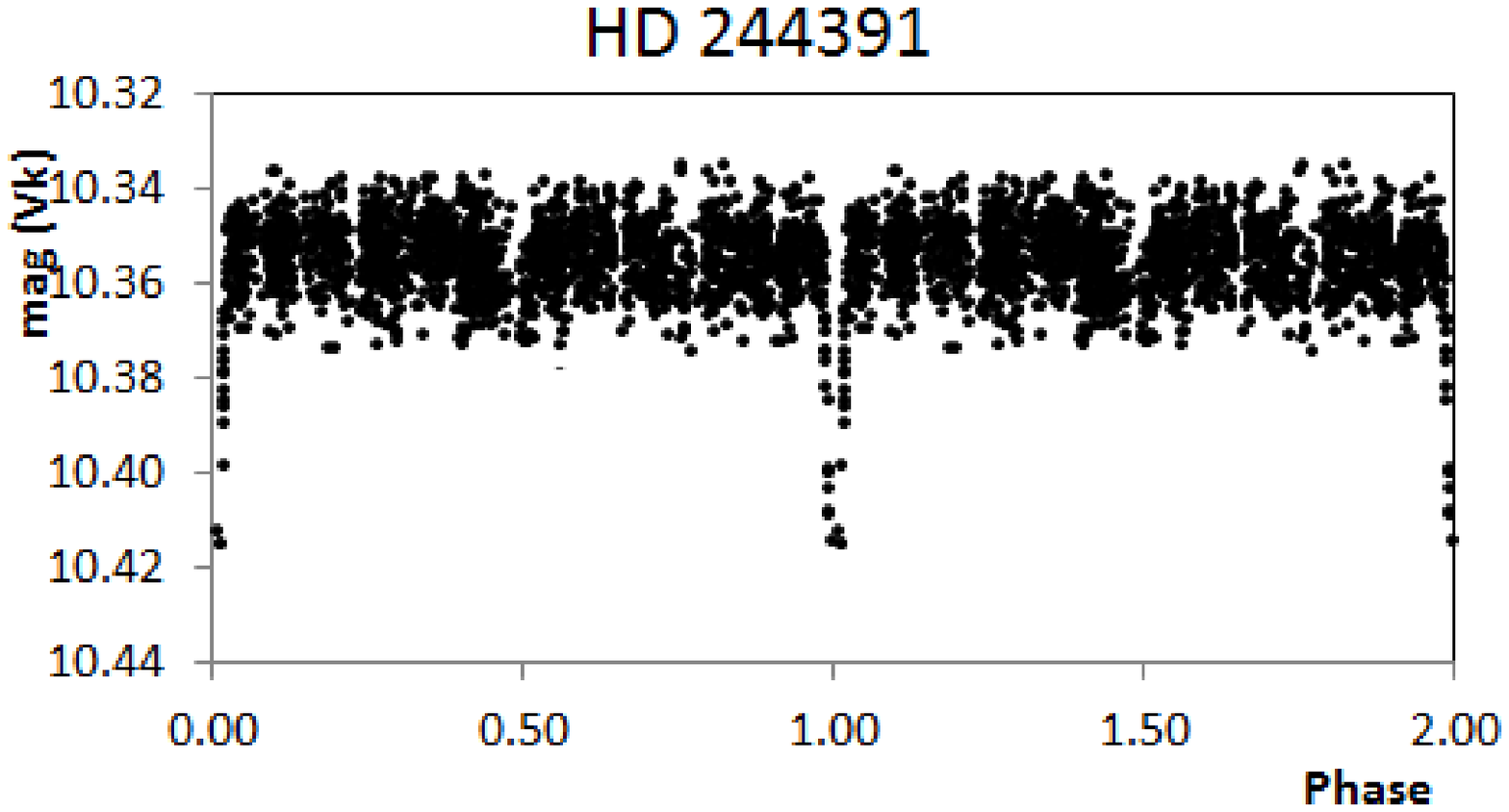}}
			\subfigure{\includegraphics[width=0.47\textwidth]{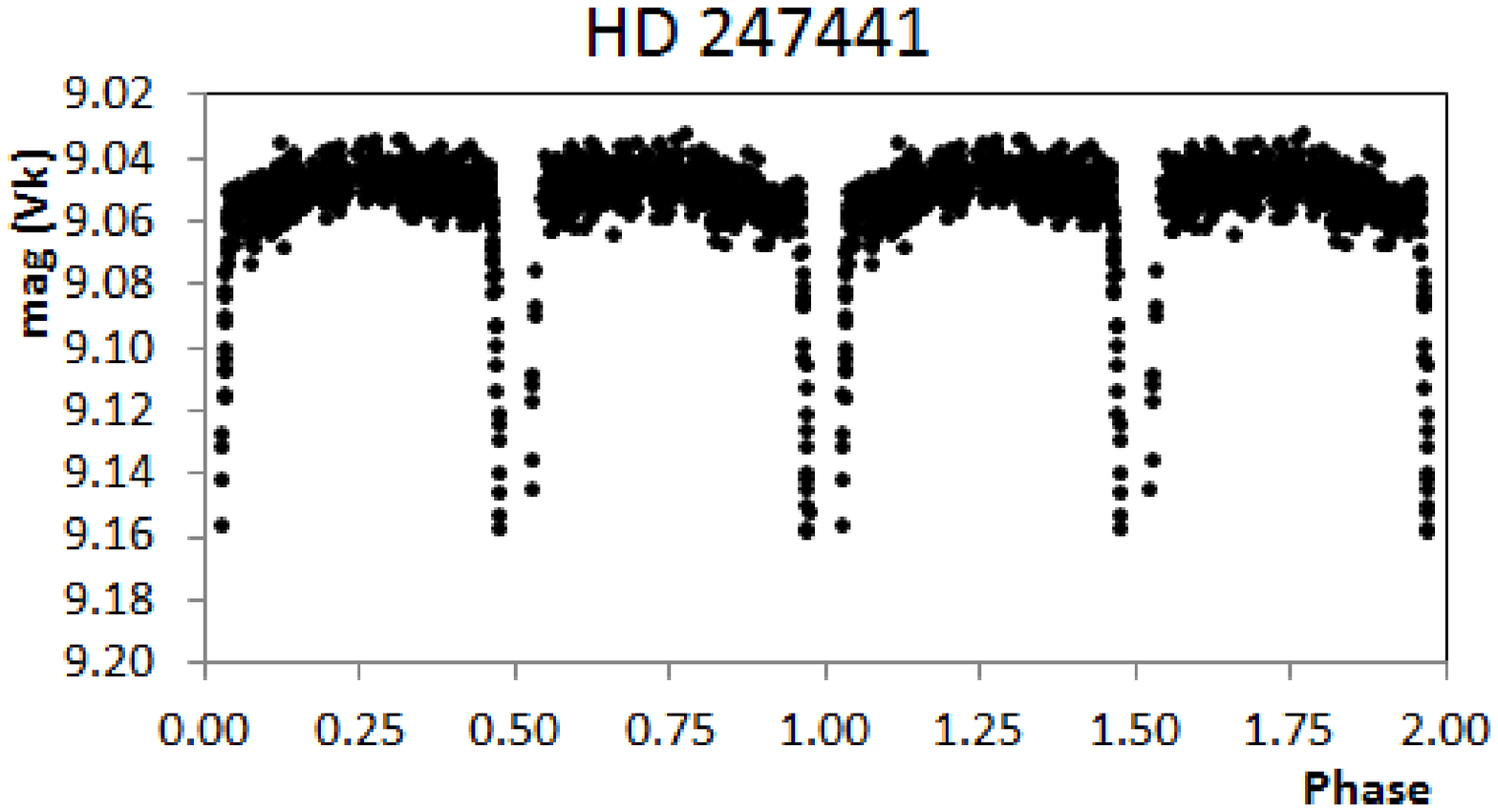}}
	} 
	\mbox{
			\subfigure{\includegraphics[width=0.47\textwidth]{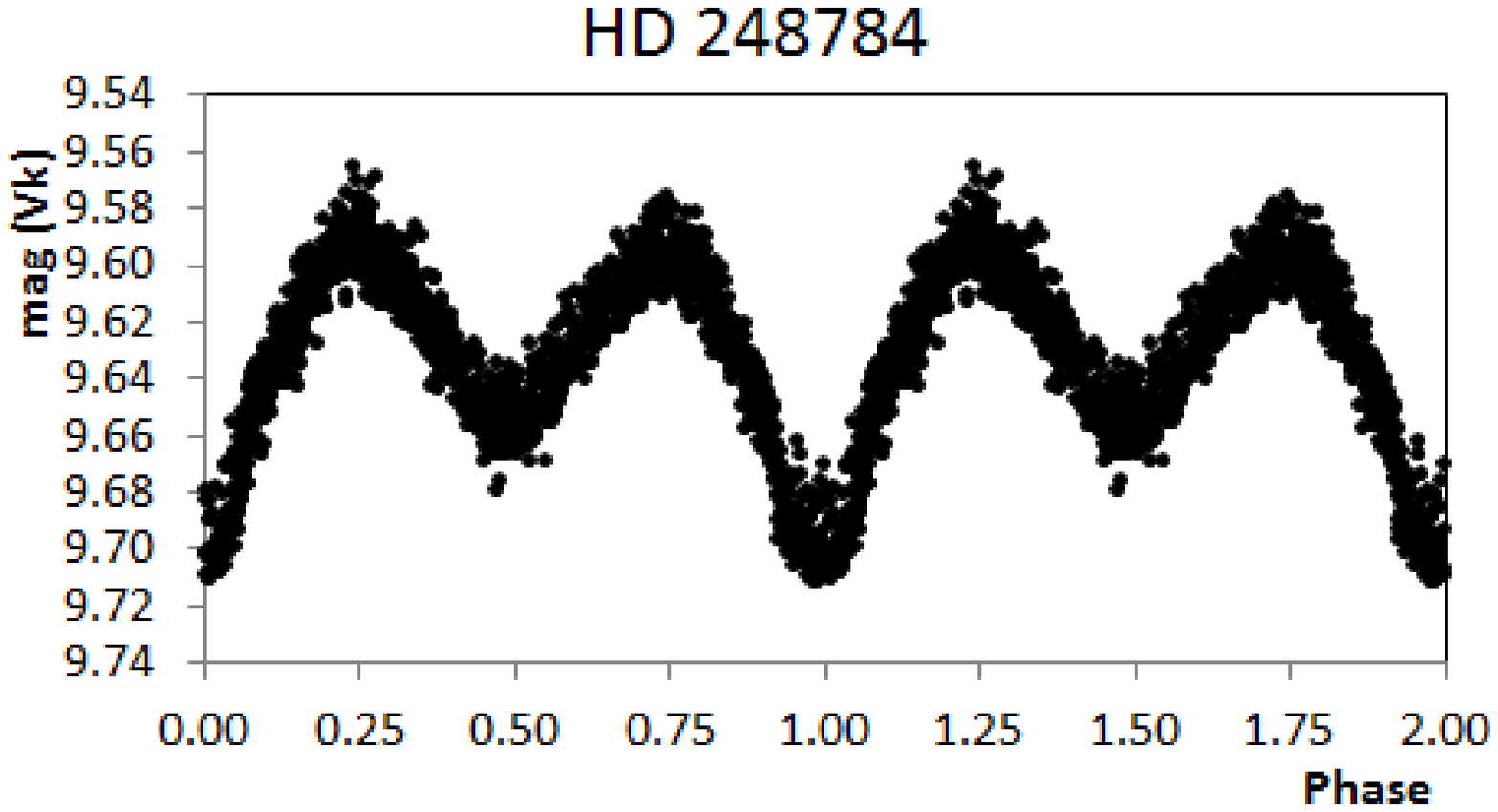}}
			\subfigure{\includegraphics[width=0.47\textwidth]{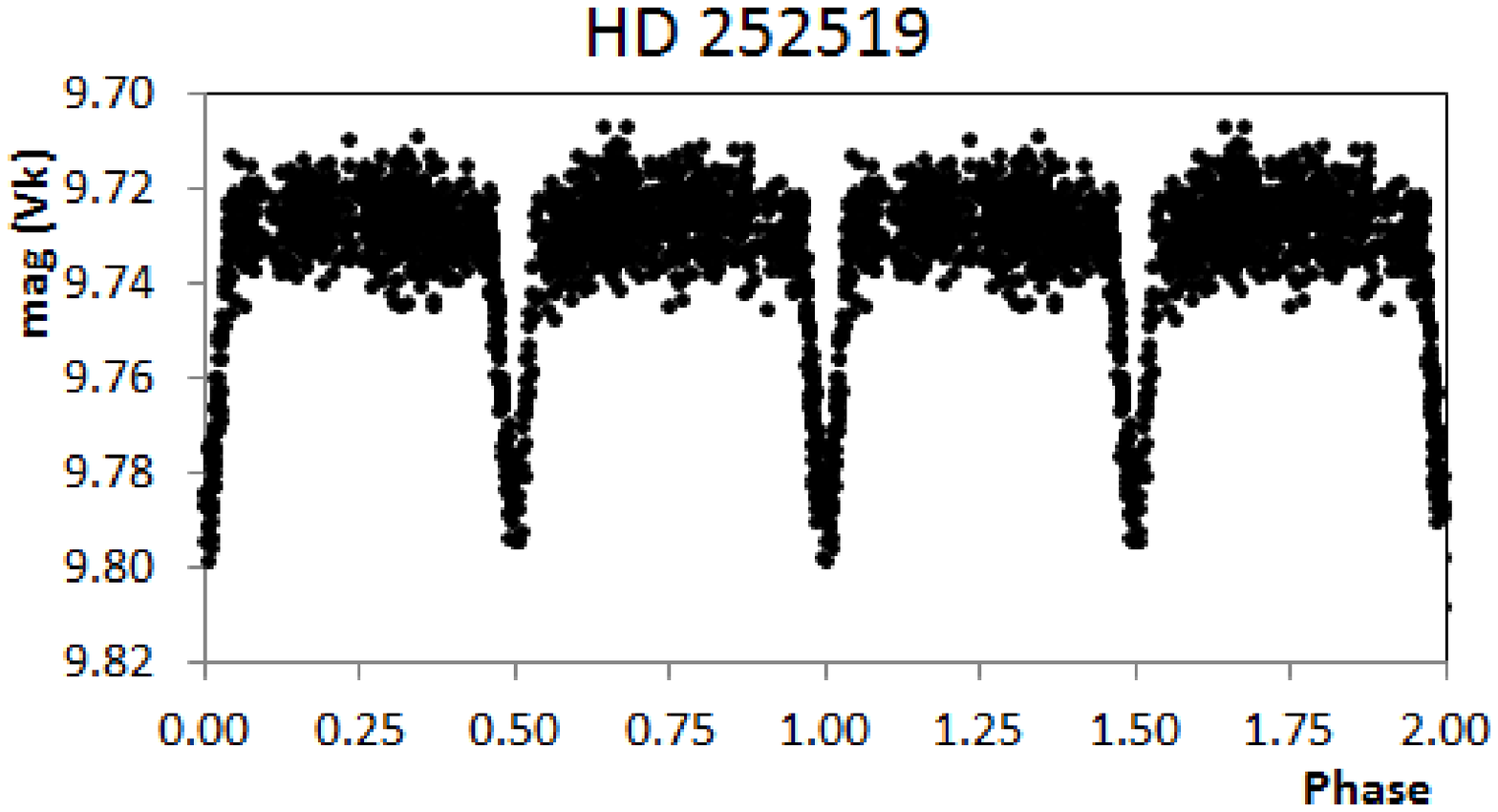}}
	} 
 \caption{The KELT light curves of the eclipsing binaries HD\,244391, HD\,247441, HD\,248784 and HD\,252519, folded on the orbital periods of, respectively, $P$\textsubscript{orb}\,=\,6.0783(4)\,d, $P$\textsubscript{orb}\,=\,4.26756(5)\,d, $P$\textsubscript{orb}\,=\,0.81821(1)\,d and $P$\textsubscript{orb}\,=\,3.07948(2)\,d.}
 \label{eclipsing_binaries}
\end{figure*}

KELT data clearly indicate that the star is an eclipsing binary with a period of $P$\,=\,6.0783(4)\,d. The primary minimum is sharp and suggests a detached or semi-detached system (cf. Figure \ref{eclipsing_binaries}); a secondary eclipse is visible at phase $\varphi$\,=\,0.5. No secondary variability due to ACV-type rotational variability of one (or both) components could be identified.

mCP stars are very rarely found in close binary systems, in particular eclipsing ones \citep{niemczura17,kochukhov18}; if proven that at least one component of this system is indeed a CP2 star, as suggested by the spectral type, HD\,244391 would be an object of high interest. We therefore strongly encourage spectroscopic and spectropolarimetric studies to shed more light on the nature of the HD\,244391 system.

\subsubsection{HD\,247441}  \label{HD247441}
HD\,247441 was identified as an eclipsing binary system by \citet{wraight11}, who derived a period of $P$\,=\,4.26760\,d from STEREO data and listed a spectral type of A0, which was gleaned from the SIMBAD database and obviously goes back to \citet{cannon31}. Newer spectroscopic studies, however, have suggested the presence of chemical peculiarities, and the star is listed with a spectral type of B9pSiSr in the RM09 catalogue. Other than that, not much is known about HD\,247441, to which the SIMBAD database lists only four references.

Employing KELT data, we have derived an orbital period of $P$\,=\,4.26756(5)\,d, in agreement with the results of \citet{wraight11}. The light curve of the star is shown in Fig. \ref{eclipsing_binaries}. No additional variability due to rotational variability of one (or both) components could be identified. We nevertheless consider HD\,247441 another good candidate for an eclipsing system containing an mCP star and encourage further studies.

\subsubsection{HD\,248784}  \label{HD248784}
HD\,248784 is identified here as a photometric variable for the first time. No detailed studies of this object exist, to which only two references are listed in the SIMBAD database. The star's KELT light curve resembles a close binary system showing proximity effects (Fig. \ref{eclipsing_binaries}), which is in agreement with the rather short photometric period of $P$\,=\,0.81821(1)\,d that we identify with the orbital period. Although certain spot configurations can lead to similar light curves in ACV variables (cf. the light curves of our sample stars shown in Section \ref{appendixLCs}), the large amplitude ($\sim$0.11\,mag) and the rather sharp primary minimum are strongly suggestive of an eclipsing binary system. No secondary variability is present in the light curve.

The star is listed with a spectral type of B8pSiSr in the RM09 catalogue and we encourage spectroscopic studies to investigate the presence of an mCP star component in the HD\,248784 system.

\begin{figure*}
 \centering
\includegraphics[width=0.98\textwidth]{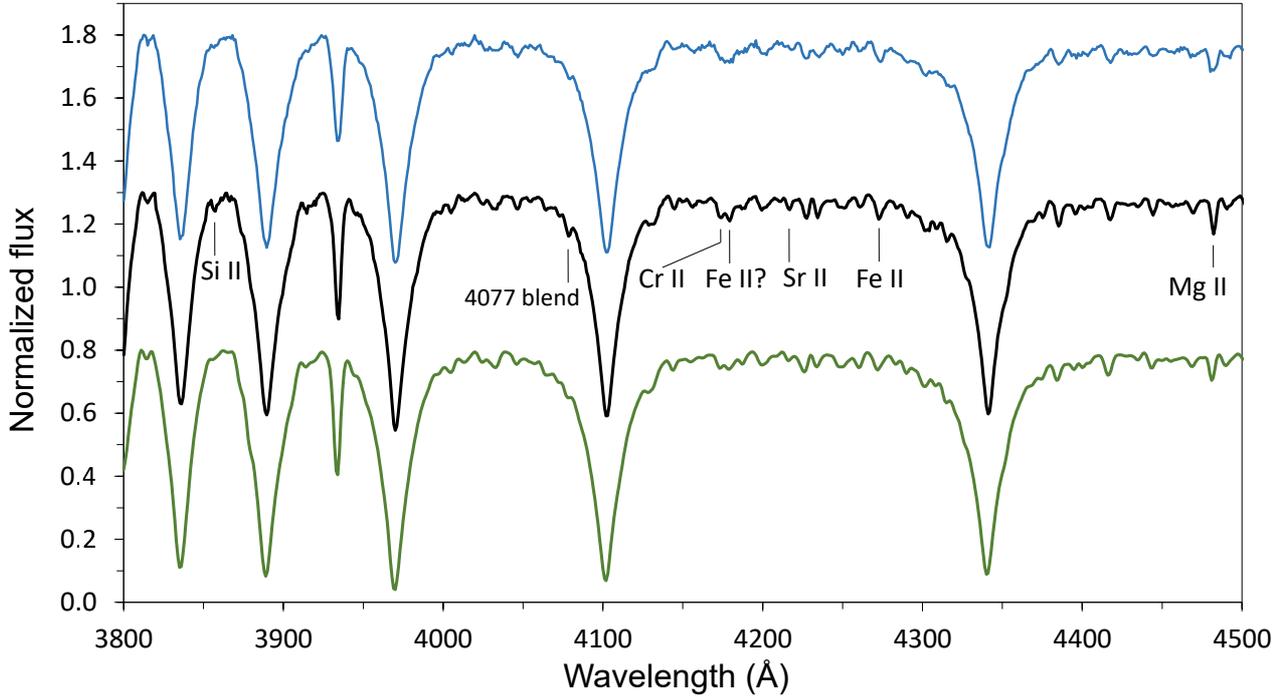}
\caption{A comparison of the LAMOST DR4 spectra of HD\,252519 taken on 5 January 2012 ('spectrum1': upper spectrum, blue) and 9 January 2014 ('spectrum2': middle spectrum, black) to a synthetic spectrum with ${T}_{\rm eff}$\,=\,8750\,K, $\log g$\,=\,4.0, [M/H]\,=\,0.0 and a microturbulent velocity of 2 km/s (lower spectrum, green). Only the blue-violet spectral region is shown. Some lines of interest are indicated.} 
 \label{spectra_HD252519}
\end{figure*}

\subsubsection{HD\,252519}  \label{HD252519}
In their variability study of TESS input catalogue sources with KELT data, \citet{oelkers18} classified HD\,252519 as non-variable. No other studies of this star exists, to which the SIMBAD database contains only two references. However, on the basis of the same data that \citet{oelkers18} used, we clearly find that the star is an eclipsing binary system and identify an orbital period of $P$\,=\,3.07948(2)\,d. Thus, the star is presented here as a photometric variable for the first time. It is listed with a spectral type of A1pCrSi in the RM09 catalogue.

The light curve of HD\,252519 is typical of a detached or semi-detached system (cf. Figure \ref{eclipsing_binaries}) and shows a secondary eclipse at phase $\varphi$\,=\,0.5. We could identify no secondary variability due to the possible presence of a rotationally variable mCP star in the system.

The star has been observed two times with the Large Sky Area Multi-Object Fiber Spectroscopic Telescope (LAMOST; \citealt{lamost1,lamost2}). We have downloaded the spectra in their DR4 reduction via the VizieR service. They are shown in Figure \ref{spectra_HD252519}, together with a synthetic spectrum which was computed using the program SPECTRUM\footnote{http://www.appstate.edu/\string~grayro/spectrum/spectrum.html} \citep{SPECTRUM} and an ATLAS9 model atmosphere \citep{ATLAS9} for ${T}_{\rm eff}$\,=\,8750\,K, $\log g$\,=\,4.0, [M/H]\,=\,0.0 and a microturbulent velocity of 2 km/s. The output spectrum was smoothed to a resolution of 4\,\AA\ and an output spacing of 1\,\AA\ to match the LAMOST resolution.

The first LAMOST spectrum (upper spectrum, blue; spectrum1 hereafter) was taken on 5 January 2012 at a JD of 2455932.1895 (S/N of g filter\,=\,249), which corresponds to an orbital phase of $\varphi$\,=\,0.818. The second LAMOST spectrum (middle spectrum, black; spectrum2 hereafter) was taken on 9 January 2014 at a JD of 2456667.1430 (S/N of g filter\,=\,712), which corresponds to an orbital phase of $\varphi$\,=\,0.479; hence, it was acquired during the secondary minimum.

\begin{figure}
 \centering
\includegraphics[width=0.47\textwidth]{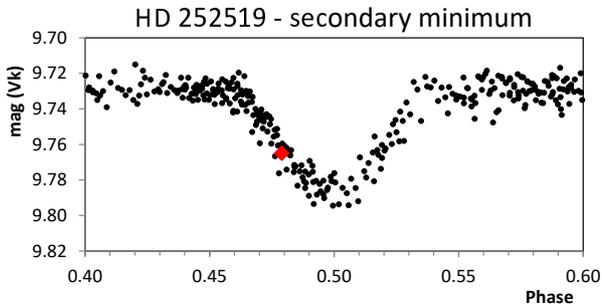}
\caption{A zoom-in on the secondary minimum of the eclipsing binary HD\,252519, based on KELT data. The orbital period is $P$\,=\,3.07948(2)\,d. The red dot marks the orbital phase during which the second LAMOST spectrum ('spectrum2') was acquired (JD 2455932.1895).}
 \label{HD252519_minII}
\end{figure}

Spectrum1 is well fitted with the shown synthetic spectrum (lower spectrum, green), whose parameters correspond to a main sequence object of spectral type A2. The hydrogen line profiles match perfectly; however, the observed \ion{Ca}{II} K line in HD\,252519 appears slightly weak. We nevertheless derive a spectral type of A2V. From the derived orbital phase, we conclude that spectrum1 belongs to the primary star of the system, although there is a contribution of unknown strength from the secondary component.

Spectrum2, which was taken during the secondary minimum (cf. Fig. \ref{HD252519_minII}) and is therefore dominated by light from the secondary star, looks superficially similar but exhibits some significant differences. The \ion{Ca}{II} K line perfectly matches that of an A2V star; the hydrogen line profile points to a slightly later type (A3V). We hence derive a spectral type of A2.5V from the available spectrum. Furthermore, we identify a minor strengthening of the absorption lines at 3856\,\AA\ (\ion{Si}{II}), 4077\,\AA\ (\ion{Si}{II}? \ion{Cr}{II}? \ion{Sr}{II}?), 4172\,\AA\ (\ion{Cr}{II}), 4179\,\AA\ (\ion{Fe}{II}?), 4216\,\AA\ (\ion{Sr}{II}), 4233\,\AA\ (\ion{Fe}{II}) and 4281\,\AA\ (\ion{Mg}{II}). This is well in line with a CP2 star classification and the RM09 spectral type of A1pCrSi and strong evidence that the system of HD 252519 indeed harbours an mCP star component, thus rendering it a potentially most interesting object (cf. \citealt{kochukhov18}). Further, detailed studies of this interesting object are strongly encouraged.

\section{Conclusion} \label{conclu}

Archival photometric time-series data from the ASAS-3, KELT and MASCARA surveys were employed to improve existing rotational period information and derive new rotational periods for mCP stars hitherto lacking this information. Our final sample consists of 294 mCP stars, a considerable amount of which (more than 40\%) are here presented as variable stars for the first time. In addition, 24 mCP star candidates have been identified that show light variability in agreement with rotational modulation but lack spectroscopic confirmation.

Summary data and folded light curves are presented for all objects. From an investigation of our sample stars in the colour-magnitude diagram, we conclude that the vast majority of our sample stars are between 100\,Myr and 1\,Gyr old. We find a conspicuous lack of mCP stars directly on the zero age main sequence; several stars of our sample seem to have left the main sequence and are situated on the subgiant branch, albeit well before the first dredge-up. Our results are in agreement with the currently-favoured model predicitions that, except for rare cases, mCP stars are hydrogen burning main-sequence objects.

The rotational period distribution of our sample stars is typical of ACV variables and agrees well with results from the literature. Our sample significantly enhances the sample size of mCP stars with accurate rotational period information and facilitates further research into the rotational properties of mCP stars and their evolution in time.

Four eclipsing binary systems have been identified among our sample stars (HD\,244391, HD\,247441, HD\,248784 and HD\,252519). As mCP stars are very rarely found in close binary systems -- in particular eclipsing ones --, these objects are potentially of great interest. Using archival LAMOST spectra, we find strong evidence that the system of HD\,252519 indeed harbours an mCP star component. All systems should be investigated with time-resolved spectroscopy for confirmation.

\section*{Acknowledgements}
We thank the anonymous referee for the thoughtful report that helped to improve the paper. Some of the data presented in this paper were obtained from the Mikulski Archive for Space Telescopes (MAST). STScI is operated by the Association of Universities for Research in Astronomy, Inc., under NASA contract NAS5-26555. Support for MAST for non-HST data is provided by the NASA Office of Space Science via grant NNX09AF08G and by other grants and contracts. Guoshoujing Telescope (the Large Sky Area Multi-Object Fiber Spectroscopic Telescope LAMOST) is a National Major Scientific Project built by the Chinese Academy of Sciences. Funding for the project has been provided by the National Development and Reform Commission. LAMOST is operated and managed by the National Astronomical Observatories, Chinese Academy of Sciences. This research has made use of the VizieR catalogue access tool (DOI : 10.26093/cds/vizier) and the SIMBAD database, both operated at CDS, Strasbourg, France.


\bibliographystyle{mnras}
\bibliography{MASCARA}


\appendix

\section{Essential data for our sample stars}

\setcounter{table}{0}  
\begin{table*}
\caption{Essential data for the mostly new ACV variables analyzed with ASAS-3 and KELT data, sorted by data source and increasing right ascension. The columns denote: (1) Photometric survey source (A=ASAS-3; M=MASCARA; K=KELT). (2) HD number or other conventional identification. (3) Identification number from RM09. (4) Right ascension (J2000; GAIA DR2). (5) Declination (J2000; GAIA DR2). (6) Spectral type from RM09. (7) $V$ magnitude range. (8) Period, derived in the present investigation. (9) Epoch (HJD-2450000). (10) Semi-amplitude of the fundamental variation ($A_{\mathrm 1}$). (11) Semi-amplitude of the first harmonic variation ($A_{\mathrm 2}$). (12) Phase of the fundamental variation ($\phi_{\mathrm 1}$). (13) Phase of the first harmonic variation ($\phi_{\mathrm 2}$). (14) $G$\,mag (GAIA DR2). (15) $(BP-RP)$ index (GAIA DR2). Where available, classifications and periods from the VSX are provided in footnotes. The HD numbers of the four eclipsing binary systems in our sample are highlighted in bold font.}
\label{table_masterASASKELT}
\begin{center}
\begin{adjustbox}{max width=\textwidth}
\begin{tabular}{lllcclccccccccc}
\hline 
(1) & (2) & (3) & (4) & (5) & (6) & (7) & (8) & (9) & (10) & (11) & (12) & (13) & (14) & (15) \\
Src & Star & \#RM09 & $\alpha$(J2000) & $\delta$(J2000) & SpT & $V$\,range & Period & Epoch & $A_{\mathrm 1}$ & $A_{\mathrm 2}$ & $\phi_{\mathrm 1}$ & $\phi_{\mathrm 2}$ & $G$ & $(BP-RP)$ \\
     &        &           &                  &                  & [lit] & [mag]       & [d]    &   & [mag]           & [mag]           & [rad]              & [rad]              & [mag]   & [mag] \\
\hline
A	&	TYC 4701-644-1	&	3920	&	02 32 01.69	&	-03 49 26.9	&	A0 Sr Eu	&	10.13-10.15	&	7.296(4)	&	3736.63(7)	&	0.009	&	0.002	&	0.496	&	0.661	&	10.139	&	0.273	\\
A	&	HD 16504	&	4100	&	02 35 21.70	&	-68 09 40.0	&	B8 Si	&	9.02-9.03	&	3.3057(7)	&	4867.62(3)	&	0.005	&	0.002	&	0.287	&	0.838	&	9.021	&	-0.023	\\
A	&	HD 25258	&	6438	&	04 00 38.96	&	-04 05 49.6	&	A2 Sr Eu	&	10.31-10.33	&	2.4490(4)	&	3806.47(2)	&	0.009	&	0.001	&	0.477	&	0.922	&	10.304	&	0.300	\\
A	&	HD 28365$^{a}$	&	7280	&	04 27 56.40	&	-13 27 52.3	&	B9 Si	&	8.43-8.45	&	1.80606(9)	&	4685.91(2)	&	0.005	&	0.001	&	0.245	&	0.441	&	8.412	&	-0.107	\\
A	&	HD 29925	&	7690	&	04 43 00.38	&	+01 06 28.4	&	B9 Si	&	8.32-8.34	&	0.665558(8)	&	2558.775(7)	&	0.007	&	0.002	&	0.222	&	0.624	&	8.294	&	-0.107	\\
A	&	TYC 8510-712-1	&	8060	&	04 53 29.13	&	-53 39 26.4	&	A0 Si	&	10.54-10.55	&	10.263(8)	&	3652.8(1)	&	0.005	&	0.001	&	0.816	&	0.044	&	10.534	&	-0.092	\\
A	&	HD 273763	&	8460	&	05 05 29.78	&	-47 53 24.0	&	A0 Sr	&	10.96-10.98	&	0.77418(2)	&	3430.557(8)	&	0.008	&	0.000	&	0.507	&	0.814	&	10.970	&	0.147	\\
A	&	HD 36955	&	9740	&	05 35 04.54	&	-01 24 06.6	&	A2 Cr Eu	&	9.44-9.45	&	2.2848(2)	&	4590.46(2)	&	0.006	&	0.002	&	0.655	&	0.059	&	9.416	&	0.201	\\
A	&	HD 40146$^{b}$	&	10710	&	05 56 58.90	&	-03 45 20.4	&	A0 Si	&	9.35-9.37	&	1.78216(8)	&	3759.68(2)	&	0.009	&	0.001	&	0.168	&	0.098	&	9.306	&	0.250	\\
A	&	HD 42335$^{c}$	&	11290	&	06 10 05.33	&	-00 18 08.6	&	A0 Si	&	8.40-8.43	&	5.076(1)	&	3440.60(5)	&	0.013	&	0.004	&	0.876	&	0.451	&	8.358	&	0.139	\\
A	&	HD 42510$^{d}$	&	11322	&	06 11 13.84	&	+03 51 46.5	&	A0 Si	&	9.01-9.10	&	1.90142(9)	&	3056.70(2)	&	0.012	&	0.035	&	0.255	&	0.551	&	9.052	&	0.149	\\
A	&	HD 258686$^{e}$	&	12290	&	06 30 47.06	&	+10 03 46.4	&	B8 Si	&	9.12-9.13	&	1.47938(7)	&	2526.89(1)	&	0.006	&	0.002	&	0.749	&	0.552	&	9.398	&	-0.078	\\
A	&	HD 46305	&	12400	&	06 32 06.36	&	-12 26 36.2	&	A0 Si	&	10.12-10.13	&	16.19(1)	&	2578.8(2)	&	0.007	&	0.001	&	0.526	&	0.292	&	10.082	&	-0.009	\\
A	&	HD 46771	&	12540	&	06 34 34.15	&	-17 21 52.1	&	B9 Si	&	8.75-8.76	&	0.548205(6)	&	3734.691(5)	&	0.004	&	0.002	&	0.242	&	0.547	&	8.736	&	-0.011	\\
A	&	HD 295354	&	12970	&	06 43 32.24	&	-03 24 09.6	&	A3-A9 Sr Eu	&	9.44-9.47	&	1.30194(3)	&	3015.71(1)	&	0.008	&	0.003	&	0.399	&	0.121	&	9.405	&	0.427	\\
A	&	HD 49223	&	13280	&	06 47 47.77	&	+07 19 34.1	&	A0 Sr Eu	&	9.02-9.04	&	0.561189(7)	&	4531.584(6)	&	0.010	&	0.001	&	0.772	&	0.892	&	8.982	&	0.015	\\
A	&	HD 49310	&	13320	&	06 48 10.73	&	+10 11 22.2	&	A0 Si	&	9.12-9.14	&	1.91912(9)	&	4762.85(2)	&	0.008	&	0.003	&	0.866	&	0.329	&	9.101	&	-0.094	\\
A	&	HD 50090	&	13630	&	06 51 13.16	&	-11 25 47.5	&	A0 Si	&	9.32-9.34	&	0.86490(1)	&	3055.690(9)	&	0.005	&	0.002	&	0.837	&	0.649	&	9.977	&	-0.032	\\
A	&	HD 50620	&	13880	&	06 51 38.38	&	-47 19 40.3	&	A3 Sr Eu Cr	&	9.74-9.75	&	12.73(1)	&	2171.8(1)	&	0.005	&	0.002	&	0.331	&	0.585	&	9.738	&	0.196	\\
A	&	HD 50773	&	13950	&	06 54 37.04	&	-00 27 09.5	&	A2 Sr Cr Eu	&	9.37-9.39	&	1.04551(2)	&	3411.62(1)	&	0.009	&	0.002	&	0.663	&	0.594	&	9.362	&	0.168	\\
A	&	HD 52156	&	14320	&	07 00 01.75	&	-02 06 07.3	&	A0 Si	&	9.30-9.32	&	1.65880(8)	&	3823.55(2)	&	0.009	&	0.002	&	0.782	&	0.602	&	9.275	&	-0.041	\\
A	&	HD 55067$^{f}$	&	15016	&	07 10 51.44	&	-14 28 08.7	&	A4 Sr	&	9.22-9.28	&	1.09012(2)	&	5154.85(1)	&	0.030	&	0.002	&	0.025	&	0.563	&	9.257	&	0.441	\\
A	&	HD 55094	&	15020	&	07 11 05.50	&	-11 32 34.7	&	B8 Si	&	10.26-10.30	&	4.1833(9)	&	3490.49(4)	&	0.018	&	0.005	&	0.294	&	0.053	&	10.344	&	0.146	\\
A	&	TYC 6532-2200-1	&	15090	&	07 12 02.29	&	-27 43 04.9	&	B5 He	&	9.25-9.27	&	2.6412(5)	&	2877.91(3)	&	0.004	&	0.002	&	0.156	&	0.409	&	9.220	&	-0.251	\\
A	&	HD 56882	&	15520	&	07 16 41.14	&	-46 21 17.4	&	F0 Sr Cr Eu	&	8.37-8.38	&	1.65475(7)	&	3089.62(2)	&	0.004	&	0.001	&	0.551	&	0.756	&	8.352	&	0.249	\\
A	&	TYC 6545-2278-1	&	15830	&	07 22 55.93	&	-27 47 30.8	&	Si	&	10.10-10.15	&	0.695415(9)	&	2566.818(7)	&	0.021	&	0.001	&	0.710	&	0.738	&	10.301	&	-0.138	\\
A	&	TYC 7104-2089-1	&	15860	&	07 23 21.48	&	-31 50 06.8	&	A Si	&	10.68-10.70	&	0.81842(2)	&	4487.617(8)	&	0.006	&	0.003	&	0.558	&	0.184	&	10.685	&	0.160	\\
A	&	HD 57981	&	15800	&	07 23 24.47	&	-06 38 10.2	&	A0 Si	&	8.86-8.89	&	0.639702(8)	&	3636.883(6)	&	0.007	&	0.005	&	0.376	&	0.236	&	8.848	&	-0.069	\\
A	&	HD 58097	&	15820	&	07 24 17.59	&	+05 47 30.8	&	A0 Si	&	9.41-9.43	&	0.567468(7)	&	4876.694(6)	&	0.007	&	0.002	&	0.902	&	0.352	&	9.425	&	-0.079	\\
A	&	HD 58675	&	16030	&	07 25 14.91	&	-33 33 10.0	&	B9 Si	&	9.62-9.63	&	1.34887(3)	&	3278.84(1)	&	0.005	&	0.001	&	0.953	&	0.121	&	9.590	&	-0.063	\\
A	&	HD 62562	&	17190	&	07 42 32.02	&	-45 02 00.0	&	A0 Sr Eu Cr	&	8.90-8.92	&	1.43748(6)	&	3731.81(1)	&	0.005	&	0.002	&	0.632	&	0.695	&	8.896	&	0.214	\\
A	&	TYC 8143-3244-1$^{g}$	&	17740	&	07 52 48.67	&	-49 37 31.9	&	Si	&	9.14-9.20	&	2.1587(2)	&	1925.67(2)	&	0.025	&	0.008	&	0.770	&	0.323	&	9.111	&	-0.034	\\
A	&	HD 68419	&	18920	&	08 10 00.54	&	-48 20 05.4	&	A2 Sr Cr Eu	&	8.18-8.19	&	1.50518(7)	&	4599.57(2)	&	0.005	&	0.001	&	0.867	&	0.092	&	8.174	&	0.079	\\
A	&	HD 68998	&	19120	&	08 14 21.67	&	-16 11 08.7	&	A5 Eu Cr Sr	&	8.67-8.68	&	2.3768(3)	&	4539.59(2)	&	0.004	&	0.001	&	0.748	&	0.825	&	8.620	&	0.309	\\
A	&	HD 71006	&	19610	&	08 20 28.08	&	-68 09 38.6	&	A0 Si	&	9.29-9.30	&	0.74340(1)	&	1884.758(7)	&	0.006	&	0.002	&	0.487	&	0.068	&	9.263	&	-0.107	\\
A	&	TYC 8149-1763-1	&	19560	&	08 22 31.73	&	-45 18 12.8	&	Si	&	9.96-9.98	&	3.1844(7)	&	2961.78(3)	&	0.007	&	0.004	&	0.493	&	0.618	&	9.912	&	-0.050	\\
A	&	HD 73455	&	20390	&	08 38 03.18	&	-16 02 15.3	&	A0 Sr Cr	&	10.26-10.29	&	5.977(1)	&	3127.58(6)	&	0.006	&	0.005	&	0.432	&	0.218	&	10.265	&	0.163	\\
A	&	TYC 7144-2934-1	&	20500	&	08 38 24.32	&	-35 31 24.5	&	Si	&	9.75-9.76	&	2.04627(9)	&	1948.68(2)	&	0.005	&	0.001	&	0.496	&	0.028	&	9.709	&	0.107	\\
A	&	HD 74611	&	20850	&	08 44 32.73	&	-13 11 36.9	&	B9 Si	&	9.53-9.54	&	7.551(4)	&	4468.75(8)	&	0.007	&	0.001	&	0.927	&	0.171	&	9.620	&	-0.172	\\
A	&	HD 77546	&	21940	&	09 02 18.27	&	-28 35 48.4	&	A0 Si	&	9.42-9.45	&	3.9725(8)	&	3055.61(4)	&	0.007	&	0.004	&	0.661	&	0.334	&	9.436	&	0.024	\\
A	&	HD 81141	&	23030	&	09 20 30.18	&	-66 29 20.9	&	B9 Si	&	8.02-8.04	&	1.27486(2)	&	3474.60(1)	&	0.007	&	0.001	&	0.244	&	0.150	&	7.974	&	0.077	\\
A	&	HD 82038	&	23330	&	09 27 05.79	&	-59 22 10.2	&	A2 Sr Eu Cr	&	9.62-9.64	&	4.1808(9)	&	3184.50(4)	&	0.005	&	0.003	&	0.889	&	0.409	&	9.573	&	0.365	\\
A	&	HD 83957	&	23960	&	09 39 51.91	&	-56 31 07.5	&	A0 Si	&	9.65-9.67	&	3.8513(8)	&	2947.84(4)	&	0.005	&	0.004	&	0.476	&	0.861	&	9.644	&	0.045	\\
A	&	HD 86978	&	24860	&	10 00 44.94	&	-49 53 54.9	&	A0 Si	&	9.62-9.64	&	0.76767(1)	&	1954.678(8)	&	0.005	&	0.001	&	0.577	&	0.145	&	9.558	&	0.197	\\
A	&	HD 87653$^{h}$	&	25130	&	10 04 45.97	&	-54 57 03.0	&	B9 Si	&	8.05-8.07	&	3.3065(7)	&	2676.56(3)	&	0.009	&	0.002	&	0.175	&	0.913	&	7.989	&	0.037	\\
A	&	HD 88208	&	25260	&	10 08 29.43	&	-57 43 19.9	&	A0 Si	&	8.84-8.85	&	1.17626(3)	&	3495.57(1)	&	0.005	&	0.001	&	0.094	&	0.093	&	8.816	&	0.097	\\
A	&	HD 88242	&	25280	&	10 09 05.39	&	-52 07 16.7	&	A0 Si Cr	&	9.41-9.43	&	0.581006(7)	&	4473.795(6)	&	0.007	&	0.001	&	0.686	&	0.035	&	9.386	&	0.037	\\
A	&	HD 88701	&	25390	&	10 13 00.23	&	-37 30 12.5	&	B9 Cr Eu	&	9.27-9.33	&	25.77(2)	&	4756.9(3)	&	0.015	&	0.016	&	0.182	&	0.573	&	9.287	&	0.069	\\
A	&	HD 89680$^{i}$	&	25750	&	10 19 53.49	&	-47 05 56.4	&	A0 Cr	&	8.42-8.43	&	2.1226(1)	&	1954.70(2)	&	0.007	&	0.001	&	0.863	&	0.936	&	8.409	&	0.110	\\
A	&	HD 91337	&	26280	&	10 32 09.85	&	-29 16 39.2	&	A0 Cr Eu Sr	&	9.44-9.46	&	1.48162(6)	&	3775.81(1)	&	0.005	&	0.004	&	0.304	&	0.915	&	9.427	&	0.063	\\
A	&	TYC 6074-361-1	&	26320	&	10 33 19.50	&	-21 05 11.5	&	Sr Cr Eu	&	9.73-9.75	&	5.656(1)	&	2027.55(6)	&	0.010	&	0.001	&	0.257	&	0.297	&	9.739	&	0.327	\\
A	&	HD 91590	&	26350	&	10 33 32.41	&	-46 58 33.2	&	B9 Si	&	7.04-7.05	&	3.3169(7)	&	3410.80(3)	&	0.005	&	0.001	&	0.547	&	0.064	&	7.064	&	-0.103	\\
A	&	HD 95198$^{j}$	&	27420	&	10 59 04.25	&	-34 52 34.0	&	B9 Si	&	7.82-7.84	&	0.97441(2)	&	3183.54(1)	&	0.011	&	0.001	&	0.585	&	0.562	&	7.829	&	-0.019	\\
A	&	HD 95569	&	27530	&	11 00 35.89	&	-65 45 24.6	&	B9 Si	&	8.57-8.59	&	6.812(3)	&	3731.83(7)	&	0.005	&	0.001	&	0.925	&	0.841	&	8.514	&	0.256	\\
A	&	HD 95508	&	27520	&	11 00 44.91	&	-50 25 46.7	&	A0 Cr	&	9.52-9.54	&	9.385(7)	&	1918.71(9)	&	0.006	&	0.001	&	0.315	&	0.852	&	9.500	&	0.078	\\
A	&	HD 98956	&	28520	&	11 22 29.17	&	-62 27 19.6	&	B9 Si	&	8.63-8.66	&	46.62(6)	&	3052.7(5)	&	0.012	&	0.002	&	0.264	&	0.692	&	9.308	&	0.341	\\
A	&	HD 100357$^{k}$	&	28840	&	11 32 06.06	&	-67 02 18.0	&	A0 Eu Cr Sr	&	8.99-9.02	&	1.62794(7)	&	2062.56(2)	&	0.007	&	0.012	&	0.982	&	0.766	&	8.970	&	0.125	\\
A	&	HD 101089	&	29130	&	11 36 59.84	&	-74 02 16.6	&	B9 Si	&	9.25-9.26	&	0.540369(6)	&	3728.838(5)	&	0.005	&	0.001	&	0.199	&	0.717	&	9.128	&	0.534	\\
A	&	HD 101600$^{l}$	&	29270	&	11 41 01.34	&	-60 36 48.2	&	A0 Si	&	8.55-8.58	&	4.4259(9)	&	4296.55(4)	&	0.011	&	0.003	&	0.009	&	0.208	&	8.525	&	0.067	\\
A	&	HD 103302$^{m}$	&	29780	&	11 53 38.19	&	-49 19 13.9	&	A0 Sr Cr Eu	&	8.32-8.34	&	1.48098(7)	&	2676.78(1)	&	0.008	&	0.006	&	0.363	&	0.839	&	8.293	&	0.080	\\
A	&	HD 110293$^{n}$	&	32050	&	12 41 45.29	&	-62 41 22.5	&	B9 Si	&	9.57-9.60	&	10.624(9)	&	3160.7(1)	&	0.012	&	0.006	&	0.399	&	0.634	&	9.478	&	0.362	\\
A	&	HD 110446	&	32100	&	12 42 40.42	&	-56 52 25.2	&	B9 Cr Eu Si	&	9.40-9.42	&	4.2992(9)	&	3067.79(4)	&	0.005	&	0.004	&	0.152	&	0.641	&	9.394	&	0.323	\\
A	&	HD 111672	&	32390	&	12 51 35.67	&	-52 35 31.8	&	B9 Si	&	9.84-9.86	&	1.00358(2)	&	4305.56(1)	&	0.010	&	0.003	&	0.625	&	0.085	&	9.810	&	-0.043	\\
A	&	HD 112555	&	32740	&	12 58 23.36	&	-58 19 47.5	&	B8 Si	&	9.23-9.28	&	6.664(2)	&	3406.79(7)	&	0.020	&	0.005	&	0.588	&	0.280	&	9.162	&	0.225	\\
A	&	HD 116419$^{o}$	&	33570	&	13 25 04.30	&	-63 10 33.0	&	B9 Si	&	8.57-8.58	&	4.638(1)	&	4507.78(5)	&	0.006	&	0.000	&	0.863	&	0.010	&	8.469	&	0.165	\\
\hline
\multicolumn{15}{l}{$^{a}$NSV 16014 (ACV:); $^{b}$ASAS J055659-0345.3 (MISC; $P$\textsubscript{VSX}\,=\,1.781649\,d); $^{c}$HIP 29253 (VAR; $P$\textsubscript{VSX}\,=\,5.07872\,d); $^{d}$ASAS J061114+0351.8 (EC|ESD; $P$\textsubscript{VSX}\,=\,1.9014\,d)} \\
\multicolumn{15}{l}{$^{e}$NSV 16881 (VAR:); $^{f}$ASAS J071051-1428.1 (DCEP-FO|ESD; $P$\textsubscript{VSX}\,=\,1.09015\,d); $^{g}$ASAS J075249-4937.5 (ESD|DCEP-FU|ED; $P$\textsubscript{VSX}\,=\,4.3169\,d); $^{h}$HIP 49373 (VAR; $P$\textsubscript{VSX}\,=\,3.30513\,d)} \\
\multicolumn{15}{l}{$^{i}$HIP 50578 (VAR; $P$\textsubscript{VSX}\,=\,2.12287\,d); $^{j}$	HIP 53684 (VAR; $P$\textsubscript{VSX}\,=\,0.97435\,d); $^{k}$HIP 56269 (VAR; $P$\textsubscript{VSX}\,=\,0.49205\,d); $^{l}$ASAS J114101-6036.8 (MISC; $P$\textsubscript{VSX}\,=\,4.4110\,d)} \\
\multicolumn{15}{l}{$^{m}$HIP 57987 (VAR; $P$\textsubscript{VSX}\,=\,1.67687\,d); $^{n}$ASAS J124145-6241.4 (VAR; $P$\textsubscript{VSX}\,=\,21.318367\,d); $^{o}$NSV 19772 (VAR:)} \\
\hline
\end{tabular}                                                                                                                                                                   
\end{adjustbox}
\end{center}                                                                                                                                             
\end{table*}
\setcounter{table}{0}  
\begin{table*}
\caption{continued.}
\label{table_masterASASKELT2}
\begin{center}
\begin{adjustbox}{max width=\textwidth}
\begin{tabular}{lllcclccccccccc}
\hline 
(1) & (2) & (3) & (4) & (5) & (6) & (7) & (8) & (9) & (10) & (11) & (12) & (13) & (14) & (15) \\
Src & Star & \#RM09 & $\alpha$(J2000) & $\delta$(J2000) & SpT & $V$\,range & Period & Epoch & $A_{\mathrm 1}$ & $A_{\mathrm 2}$ & $\phi_{\mathrm 1}$ & $\phi_{\mathrm 2}$ & $G$ & $(BP-RP)$ \\
     &        &           &                  &                  & [lit] & [mag]       & [d]    &   & [mag]           & [mag]           & [rad]              & [rad]              & [mag]   & [mag] \\
\hline
A	&	HD 116423	&	33580	&	13 25 26.47	&	-68 10 39.2	&	A0 Eu Sr Si	&	8.53-8.55	&	17.34(1)	&	5057.5(2)	&	0.008	&	0.002	&	0.021	&	0.174	&	8.425	&	0.269	\\
A	&	HD 120059$^{p}$	&	34630	&	13 48 43.12	&	-58 47 16.3	&	B8 Si	&	8.83-8.85	&	4.1042(9)	&	2658.82(4)	&	0.011	&	0.001	&	0.936	&	0.054	&	8.790	&	0.098	\\
A	&	HD 122264$^{q}$	&	35060	&	14 02 24.91	&	-55 22 56.3	&	B9 Si	&	9.58-9.60	&	4.0490(8)	&	3815.70(4)	&	0.005	&	0.003	&	0.276	&	0.970	&	9.508	&	0.173	\\
A	&	HD 123206	&	35320	&	14 08 15.62	&	-61 03 32.7	&	B9 Si	&	9.84-9.86	&	2.4328(4)	&	4879.84(2)	&	0.008	&	0.002	&	0.014	&	0.859	&	9.786	&	0.348	\\
A	&	HD 123960$^{r}$	&	35460	&	14 12 47.73	&	-61 18 18.4	&	B9 Si	&	9.71-9.73	&	0.80499(1)	&	1964.715(8)	&	0.007	&	0.001	&	0.044	&	0.556	&	9.683	&	0.153	\\
A	&	HD 129189$^{s}$	&	36840	&	14 45 32.98	&	-72 05 14.9	&	B9 Cr Eu	&	8.59-8.62	&	1.35563(4)	&	4109.87(1)	&	0.012	&	0.003	&	0.077	&	0.311	&	8.536	&	0.104	\\
A	&	HD 133061	&	37780	&	15 04 23.30	&	-51 30 17.4	&	B9 Si	&	9.08-9.09	&	0.73512(1)	&	3901.669(7)	&	0.005	&	0.002	&	0.322	&	0.585	&	9.041	&	0.058	\\
A	&	HD 134305$^{t}$	&	38130	&	15 08 44.86	&	+12 29 20.0	&	A6 Sr Eu Cr	&	7.25-7.27	&	1.03218(2)	&	2819.58(1)	&	0.006	&	0.001	&	0.184	&	0.235	&	7.207	&	0.275	\\
A	&	HD 134121	&	38080	&	15 10 09.32	&	-52 10 03.3	&	B9 Si	&	10.00-10.02	&	1.27933(3)	&	2783.89(1)	&	0.005	&	0.001	&	0.652	&	0.737	&	9.974	&	0.167	\\
A	&	HD 135459	&	38470	&	15 17 22.32	&	-53 43 39.0	&	A2 Eu Cr	&	9.94-9.97	&	0.92542(2)	&	2055.619(9)	&	0.011	&	0.005	&	0.297	&	0.349	&	9.950	&	0.526	\\
A	&	HD 135714	&	38540	&	15 19 00.95	&	-56 34 41.2	&	A0 Si	&	10.14-10.16	&	4.0112(8)	&	3127.70(4)	&	0.007	&	0.001	&	0.008	&	0.116	&	10.085	&	0.102	\\
A &	HD 137436	&	39080	&	15 28 25.02	&	-53 25 12.1	&	B8 Si	&	8.80-8.83	&	2.2985(2)	&	4862.87(2)	&	0.009	&	0.004	&	0.970	&	0.558	&	8.826	&	0.004	\\
A	&	HD 146971	&	41510	&	16 19 29.59	&	-09 37 30.1	&	A0 Sr Cr Eu	&	8.64-8.66	&	3.0512(6)	&	3128.76(3)	&	0.007	&	0.002	&	0.384	&	0.854	&	8.568	&	0.368	\\
A	&	HD 147039	&	41540	&	16 22 18.20	&	-55 07 17.9	&	B8 Si	&	9.98-10.00	&	3.6142(7)	&	3657.51(4)	&	0.009	&	0.004	&	0.931	&	0.679	&	9.886	&	0.102	\\
A	&	HD 147174	&	41590	&	16 25 11.94	&	-68 07 35.8	&	A0 Si Cr Sr	&	8.85-8.86	&	21.99(1)	&	3588.6(2)	&	0.004	&	0.001	&	0.651	&	0.220	&	9.248	&	0.080	\\
A	&	HD 149046	&	42120	&	16 32 26.97	&	-07 11 04.4	&	A0 Sr Cr Eu	&	9.57-9.59	&	2.2103(2)	&	1949.88(2)	&	0.011	&	0.003	&	0.407	&	0.550	&	9.504	&	0.409	\\
A	&	HD 149136	&	42170	&	16 34 42.53	&	-43 28 04.4	&	A0 Si	&	10.07-10.09	&	1.53894(8)	&	4298.75(2)	&	0.008	&	0.001	&	0.439	&	0.163	&	9.988	&	0.330	\\
A	&	HD 150347$^{u}$	&	42510	&	16 41 35.48	&	-28 35 10.9	&	B9 Si	&	9.00-9.02	&	1.27630(3)	&	2031.73(1)	&	0.009	&	0.003	&	0.896	&	0.984	&	8.938	&	0.204	\\
A	&	HD 150957	&	42670	&	16 46 31.78	&	-45 45 55.5	&	B8 Si	&	9.27-9.32	&	4.823(1)	&	4705.65(5)	&	0.020	&	0.002	&	0.029	&	0.561	&	9.251	&	0.113	\\
A	&	HD 153149	&	43317	&	16 58 20.86	&	-17 28 05.1	&	A5 Sr Eu Cr	&	9.45-9.47	&	0.673791(9)	&	3189.591(7)	&	0.008	&	0.002	&	0.034	&	0.997	&	9.350	&	0.460	\\
A	&	HD 154187	&	43585	&	17 04 27.12	&	-14 03 36.5	&	A0 Si	&	9.26-9.29	&	8.096(5)	&	3467.83(8)	&	0.011	&	0.003	&	0.409	&	0.072	&	9.086	&	0.813	\\
A	&	HD 154253	&	43610	&	17 08 09.00	&	-60 45 03.2	&	A0 Sr Cr Eu	&	9.05-9.06	&	3.0919(6)	&	3487.79(3)	&	0.006	&	0.002	&	0.670	&	0.846	&	8.990	&	0.281	\\
A	&	HD 154458	&	43660	&	17 09 12.81	&	-59 48 49.3	&	B9 Si	&	8.27-8.30	&	4.4573(9)	&	3564.62(4)	&	0.009	&	0.002	&	0.096	&	0.043	&	8.253	&	-0.045	\\
A	&	TYC 8336-2364-1	&	43760	&	17 11 15.23	&	-50 09 56.1	&	Si	&	9.94-9.95	&	2.0840(1)	&	3571.59(2)	&	0.006	&	0.002	&	0.840	&	0.309	&	9.869	&	0.252	\\
A	&	HD 156366	&	44050	&	17 18 15.66	&	-28 07 21.4	&	A2 Sr Eu	&	9.68-9.70	&	3.8057(8)	&	4227.79(4)	&	0.008	&	0.003	&	0.676	&	0.049	&	9.548	&	0.606	\\
A &	HD 158234$^{v}$	&	44510	&	17 29 38.16	&	-35 46 26.3	&	A0 Si	&	9.50-9.58	&	2.6825(5)	&	1953.84(3)	&	0.042	&	0.004	&	0.421	&	0.706	&	9.446	&	0.214	\\
A	&	HD 161445	&	45496	&	17 45 23.11	&	+05 33 45.6	&	A0 Si Eu Sr	&	10.09-10.11	&	2.1568(1)	&	2711.86(2)	&	0.006	&	0.002	&	0.335	&	0.128	&	10.057	&	0.198	\\
A	&	HD 162306$^{w}$	&	45790	&	17 51 55.23	&	-35 04 57.6	&	B9 Si	&	8.67-8.70	&	1.70249(8)	&	3582.74(2)	&	0.014	&	0.002	&	0.309	&	0.321	&	8.859	&	0.019	\\
A	&	HD 162639	&	45930	&	17 54 41.33	&	-50 26 45.9	&	A5 Sr Eu Cr	&	9.86-9.90	&	39.82(4)	&	3093.9(4)	&	0.013	&	0.001	&	0.018	&	0.634	&	9.821	&	0.537	\\
A	&	HD 163583	&	46280	&	18 00 05.31	&	-55 37 58.4	&	A3 Sr Cr Eu	&	10.49-10.51	&	1.38313(6)	&	3584.81(1)	&	0.006	&	0.001	&	0.919	&	0.137	&	10.434	&	0.315	\\
A	&	HD 164068	&	46388	&	18 00 24.15	&	-22 57 52.2	&	A0 Si	&	9.70-9.73	&	2.4133(3)	&	3573.64(2)	&	0.013	&	0.003	&	0.984	&	0.867	&	9.783	&	0.271	\\
A	&	HD 167024	&	47010	&	18 15 07.57	&	-38 53 39.7	&	A2 Sr Eu	&	9.19-9.22	&	3.0934(7)	&	3178.66(3)	&	0.006	&	0.006	&	0.088	&	0.708	&	9.142	&	0.239	\\
A	&	HD 168071	&	47115	&	18 17 58.30	&	+03 28 47.7	&	A0 Sr Eu Cr	&	8.61-8.62	&	2.3122(3)	&	2529.52(2)	&	0.005	&	0.001	&	0.795	&	0.733	&	8.509	&	0.505	\\
A	&	HD 168108	&	47120	&	18 20 33.49	&	-47 58 39.0	&	B8 Si	&	8.67-8.69	&	1.42776(5)	&	2441.69(1)	&	0.006	&	0.002	&	0.710	&	0.277	&	8.621	&	-0.008	\\
A	&	HD 169021$^{x}$	&	47410	&	18 24 03.93	&	-34 20 02.7	&	B9 Si	&	7.00-7.05	&	3.0955(7)	&	4578.86(3)	&	0.019	&	0.008	&	0.601	&	0.273	&	6.954	&	-0.009	\\
A	&	TYC 7415-2499-1	&	48010	&	18 36 08.66	&	-35 08 34.5	&	Si	&	9.76-9.79	&	149.4(1)	&	2904.5(9)	&	0.015	&	0.002	&	0.260	&	0.991	&	9.759	&	0.038	\\
A	&	HD 171914	&	48170	&	18 37 12.19	&	+02 58 34.4	&	A0 Si Sr Eu	&	7.87-7.89	&	1.65793(7)	&	2882.61(2)	&	0.007	&	0.001	&	0.053	&	0.609	&	7.838	&	0.133	\\
A	&	HD 172251	&	48263	&	18 39 17.24	&	-03 09 04.2	&	B9 Si	&	9.21-9.23	&	4.5498(9)	&	2862.73(5)	&	0.009	&	0.001	&	0.587	&	0.204	&	9.060	&	0.649	\\
A	&	HD 172626	&	48360	&	18 42 51.08	&	-37 04 37.4	&	A2 Sr Cr Eu	&	9.69-9.70	&	12.32(1)	&	5024.7(1)	&	0.005	&	0.001	&	0.905	&	0.373	&	9.630	&	0.305	\\
A	&	HD 177016	&	49374	&	19 03 40.57	&	-20 34 36.9	&	F0 Eu Sr Cr	&	9.29-9.31	&	1.64041(7)	&	1994.85(2)	&	0.006	&	0.002	&	0.804	&	0.500	&	9.213	&	0.498	\\
A	&	HD 230952	&	49970	&	19 14 13.57	&	+14 23 30.4	&	A0 Si	&	9.47-9.49	&	1.42820(5)	&	3133.84(1)	&	0.010	&	0.001	&	0.475	&	0.614	&	9.399	&	0.421	\\
A	&	HD 181028	&	50184	&	19 19 06.74	&	-06 59 22.9	&	A2 Sr Eu Cr	&	10.11-10.12	&	7.457(3)	&	2134.59(7)	&	0.006	&	0.001	&	0.582	&	0.039	&	10.006	&	0.675	\\
A	&	HD 184020$^{y}$	&	50800	&	19 34 42.78	&	-49 41 05.4	&	A0 Sr Cr Eu	&	8.15-8.17	&	2.5515(5)	&	3552.85(3)	&	0.007	&	0.003	&	0.292	&	0.842	&	8.126	&	0.017	\\
A	&	HD 185280	&	51180	&	19 40 11.30	&	-40 52 32.0	&	A2 Cr Eu	&	8.42-8.44	&	1.35464(3)	&	3190.75(1)	&	0.006	&	0.001	&	0.260	&	0.132	&	9.428	&	0.150	\\
A	&	HD 193382	&	54000	&	20 21 56.75	&	-47 43 24.6	&	A0 Si Cr Eu	&	9.86-9.89	&	8.374(6)	&	1996.86(4)	&	0.009	&	0.003	&	0.450	&	0.683	&	9.850	&	0.060	\\
A	&	HD 209051	&	58130	&	22 00 39.98	&	-06 25 57.0	&	A0 Si Cr Eu	&	9.76-9.78	&	0.69640(1)	&	2216.569(7)	&	0.006	&	0.002	&	0.714	&	0.100	&	8.738	&	-0.048	\\
A	&	HD 222638	&	61030	&	23 42 33.97	&	-57 28 40.9	&	A0 Sr Eu Cr	&	8.65-8.66	&	1.17352(3)	&	2145.71(1)	&	0.004	&	0.002	&	0.324	&	0.864	&	8.650	&	0.057	\\
K	&	HD 7410	&	1860	&	01 14 40.41	&	+33 00 04.8	&	A5 Sr Cr Eu	&	9.07-9.09	&	37.08(2)	&	4054.7(3)	&	0.010	&	0.002	&	0.435	&	0.047	&	9.066	&	0.325	\\
K	&	HD 9393	&	2270	&	01 33 15.52	&	+43 53 45.1	&	B9 Si Cr Sr	&	8.60-8.61	&	2.4998(3)	&	4035.52(2)	&	0.003	&	0.002	&	0.528	&	0.075	&	8.600	&	-0.082	\\
K	&	HD 9492	&	2300	&	01 34 08.06	&	+44 05 38.3	&	A0 Si Cr	&	8.46-8.48	&	10.898(6)	&	4421.7(1)	&	0.006	&	0.002	&	0.918	&	0.288	&	8.432	&	0.096	\\
K	&	TYC 2816-447-1	&	3063	&	01 58 12.70	&	+37 34 40.5	&	F2 Sr	&	10.95-10.96	&	0.73356(2)	&	4057.684(7)	&	0.002	&	0.001	&	0.296	&	0.459	&	10.853	&	0.538	\\
K	&	HD 13404	&	3470	&	02 11 54.03	&	+36 57 58.3	&	A2 Sr Eu	&	8.74-8.75	&	4.358(2)	&	4037.71(4)	&	0.002	&	0.000	&	0.212	&	0.686	&	8.733	&	0.371	\\
K	&	HD 14522	&	3660	&	02 21 20.03	&	+28 04 15.8	&	A2 Sr Eu	&	8.80-8.81	&	26.36(1)	&	4059.3(2)	&	0.004	&	0.002	&	0.681	&	0.900	&	8.749	&	0.327	\\
K	&	HD 16605	&	4140	&	02 40 58.94	&	+42 52 16.6	&	A1 Si Cr Sr	&	9.56-9.57	&	8.820(4)	&	4034.79(8)	&	0.003	&	0.003	&	0.169	&	0.940	&	9.618	&	0.072	\\
K	&	TYC 2850-263-1	&	4480	&	02 53 02.19	&	+39 55 12.5	&	A Sr Cr Eu	&	9.79-9.80	&	12.440(8)	&	4052.5(1)	&	0.003	&	0.001	&	1.000	&	0.277	&	9.646	&	0.510	\\
K	&	TYC 2854-1633-1	&	4520	&	02 57 10.63	&	+43 04 43.8	&	Sr Eu	&	9.73-9.74	&	13.448(8)	&	4387.7(1)	&	0.006	&	0.003	&	0.617	&	0.128	&	9.695	&	0.452	\\
K	&	TYC 2394-739-1	&	8844	&	05 20 31.61	&	+32 36 14.1	&	B9 Si	&	11.00-11.01	&	4.464(2)	&	4061.80(4)	&	0.007	&	0.001	&	0.993	&	0.969	&	10.972	&	0.371	\\
K	&	HD 243308	&	9006	&	05 24 58.17	&	+30 04 46.5	&	B9 Si	&	10.78-10.79	&	1.7209(2)	&	4385.92(1)	&	0.003	&	0.001	&	0.136	&	0.525	&	10.771	&	0.228	\\
K	&	HD 243494	&	9080	&	05 26 12.31	&	+32 02 03.5	&	B9 Si	&	9.65-9.67	&	3.0124(2)	&	4417.84(2)	&	0.008	&	0.004	&	0.076	&	0.825	&	9.520	&	0.241	\\
K	&	HD 243492	&	9082	&	05 26 16.49	&	+33 15 44.3	&	A0 Si Sr	&	10.63-10.65	&	2.9317(2)	&	4386.98(2)	&	0.007	&	0.000	&	0.366	&	0.036	&	10.602	&	0.350	\\
K	&	\textbf{HD 244391}	&	9357	&	05 31 41.47	&	+31 37 25.0	&	B8 Si Sr	&	10.35-10.41	&	6.0783(4)	&	4422.27(2)	&	-	&	-	&	-	&	-	&	10.260	&	0.507	\\
K	&	TYC 2412-1235-1	&	9655	&	05 36 02.39	&	+34 06 50.9	&	B9 Si	&	9.87-9.88	&	0.55386(1)	&	4040.938(5)	&	0.003	&	0.001	&	0.750	&	0.845	&	9.959	&	0.089	\\
K	&	HD 245601	&	9953	&	05 38 08.91	&	+29 00 36.1	&	B9 Si Sr	&	9.84-9.85	&	1.57095(7)	&	4056.79(1)	&	0.003	&	0.000	&	0.358	&	0.008	&	9.170	&	1.901	\\
K	&	HD 246820	&	10239	&	05 44 16.39	&	+27 57 18.7	&	A0 Si Sr	&	10.54-10.57	&	1.29620(3)	&	4117.67(1)	&	0.014	&	0.001	&	0.022	&	0.408	&	10.511	&	0.315	\\
K	&	TYC 2413-822-1	&	10316	&	05 46 29.38	&	+34 13 34.6	&	B8 Si	&	10.18-10.19	&	1.41839(7)	&	4059.77(1)	&	0.003	&	0.001	&	0.457	&	0.337	&	10.156	&	0.156	\\
K	&	\textbf{HD 247441}$^{z}$	&	10334	&	05 47 04.35	&	+28 13 44.7	&	B9 Si Sr	&	9.05-<9.16	&	4.26756(5)	&	4061.70(1)	&	-	&	-	&	-	&	-	&	9.050	&	0.119	\\
K	&	GSC 2413-00426	&	10391	&	05 48 36.91	&	+33 52 56.1	&	A3 Sr	&	11.92-11.93	&	1.5810(2)	&	4037.81(1)	&	0.006	&	0.001	&	0.896	&	0.562	&	11.870	&	0.475	\\
K	&	HD 38979	&	10476	&	05 50 53.58	&	+30 43 24.5	&	B7 Si	&	9.17-9.21	&	2.56099(4)	&	4746.89(2)	&	0.021	&	0.003	&	0.269	&	0.409	&	9.197	&	0.077	\\
K	&	\textbf{HD 248784}	&	10593	&	05 54 04.81	&	+34 01 50.2	&	B8 Si Sr	&	9.59-9.71	&	0.81821(1)	&	4057.835(4)	&	-	&	-	&	-	&	-	&	9.569	&	0.134	\\
K	&	HD 39865	&	10643	&	05 56 30.82	&	+29 45 07.2	&	B8 Si	&	8.60-8.64	&	1.8372(1)	&	4051.83(1)	&	0.012	&	0.012	&	0.333	&	0.825	&	8.587	&	0.114	\\
K	&	HD 249478	&	10668	&	05 57 26.08	&	+29 53 16.7	&	A1 Si Sr	&	10.38-10.39	&	2.4246(4)	&	4061.85(2)	&	0.002	&	0.001	&	0.536	&	0.285	&	10.398	&	0.264	\\
K	&	HD 40562	&	10840	&	06 01 47.08	&	+44 18 31.3	&	B9 Si	&	8.77-8.78	&	2.7522(3)	&	4057.77(2)	&	0.004	&	0.000	&	0.374	&	0.104	&	8.767	&	0.003	\\
K	&	\textbf{HD 252519}	&	11255	&	06 10 33.31	&	+34 00 29.2	&	A1 Cr Si	&	9.73-9.80	&	3.07948(2)	&	4038.87(1)	&	-	&	-	&	-	&	-	&	9.651	&	0.286	\\
K	&	HD 44903	&	11880	&	06 25 20.80	&	+23 03 24.5	&	A5 Sr Eu	&	8.35-8.38	&	1.41143(3)	&	4066.80(1)	&	0.012	&	0.007	&	0.412	&	0.101	&	8.356	&	0.161	\\
K	&	HD 49040	&	13200	&	06 48 22.12	&	+40 57 45.1	&	B9 Sr Cr Eu	&	8.86-8.87	&	9.008(5)	&	4496.59(9)	&	0.003	&	0.003	&	0.611	&	0.436	&	8.822	&	0.208	\\
\hline
\multicolumn{15}{l}{$^{p}$NSV 19966 (VAR:); $^{q}$NSV 20024 (VAR:); $^{r}$NSV 20059 (VAR:); $^{s}$HIP 72159 (VAR; $P$\textsubscript{VSX}\,=\,1.35579\,d); $^{t}$NSV 20254 (VAR:); $^{u}$EPIC 202727506 (VAR; $P$\textsubscript{VSX}\,=\,1.275979\,d)} \\
\multicolumn{15}{l}{$^{v}$ASAS J172938-3546.4 (ACV|ESD; $P$\textsubscript{VSX}\,=\,2.682024\,d); $^{w}$ASAS J175155-3505.0 (VAR); $^{x}$NSV 24422 (no type and period given); $^{y}$HIP 96290 (VAR; $P$\textsubscript{VSX}\,=\,0.08582\,d)} \\
\multicolumn{15}{l}{$^{z}$HD 247441 (EA; $P$\textsubscript{VSX}\,=\,4.2676\,d)} \\
\hline
\end{tabular}                                                                                                                                                                   
\end{adjustbox}
\end{center}                                                                                                                                             
\end{table*}  
\setcounter{table}{0}  
\begin{table*}
\caption{continued.}
\label{table_masterASASKELT3}
\begin{center}
\begin{adjustbox}{max width=\textwidth}
\begin{tabular}{lllcclccccccccc}
\hline 
(1) & (2) & (3) & (4) & (5) & (6) & (7) & (8) & (9) & (10) & (11) & (12) & (13) & (14) & (15) \\
Src & Star & \#RM09 & $\alpha$(J2000) & $\delta$(J2000) & SpT & $V$\,range & Period & Epoch & $A_{\mathrm 1}$ & $A_{\mathrm 2}$ & $\phi_{\mathrm 1}$ & $\phi_{\mathrm 2}$ & $G$ & $(BP-RP)$ \\
     &        &           &                  &                  & [lit] & [mag]       & [d]    &   & [mag]           & [mag]           & [rad]              & [rad]              & [mag]   & [mag] \\
\hline
K	&	TYC 3161-135-1$^{aa}$	&	54600	&	20 32 12.87	&	+43 03 57.5	&	B9 Si	&	9.11-9.14	&	1.09626(3)	&	4273.88(1)	&	0.018	&	0.001	&	0.136	&	0.626	&	9.111	&	-0.025	\\
K	&	HD 196542	&	54800	&	20 36 43.34	&	+39 06 14.8	&	A4 Sr Cr Eu	&	9.03-9.05	&	1.7929(1)	&	4265.85(1)	&	0.010	&	0.002	&	0.528	&	0.256	&	9.038	&	0.112	\\
K	&	HD 197560	&	55080	&	20 43 17.78	&	+35 44 30.1	&	A0 Si	&	7.99-8.03	&	2.8812(2)	&	4270.81(2)	&	0.012	&	0.007	&	0.420	&	0.261	&	7.981	&	-0.052	\\
K	&	HD 198195	&	55170	&	20 47 09.92	&	+42 24 35.4	&	B8 Si	&	7.43-7.45	&	1.49665(7)	&	4588.87(1)	&	0.008	&	0.001	&	0.673	&	0.454	&	7.417	&	-0.061	\\
K	&	TYC 2696-483-1	&	55400	&	20 52 44.75	&	+34 13 00.8	&	Si	&	10.18-10.20	&	3.5031(2)	&	4380.61(3)	&	0.011	&	0.003	&	0.235	&	0.841	&	10.226	&	0.074	\\
K	&	TYC 2713-1881-1	&	56156	&	21 07 40.45	&	+35 54 02.8	&	B8 Si	&	10.56-10.58	&	0.75430(2)	&	4268.851(7)	&	0.007	&	0.003	&	0.528	&	0.003	&	10.444	&	-0.089	\\
K	&	HD 203786	&	56730	&	21 23 54.71	&	+21 55 50.8	&	B9 Si	&	8.00-8.01	&	5.398(2)	&	4277.80(5)	&	0.004	&	0.003	&	0.133	&	0.853	&	7.988	&	-0.091	\\
K	&	HD 205087$^{ab}$	&	57120	&	21 32 27.01	&	+23 23 40.5	&	A0 Sr Si Cr	&	6.67-6.68	&	0.59774(1)	&	4275.933(5)	&	0.002	&	0.000	&	0.222	&	0.988	&	6.657	&	-0.065	\\
K	&	HD 206898	&	57560	&	21 44 20.80	&	+35 15 18.3	&	A0 Si	&	8.57-8.60	&	1.6485(1)	&	4261.87(1)	&	0.009	&	0.006	&	0.553	&	0.138	&	8.565	&	0.047	\\
K	&	HD 209059	&	58140	&	22 00 06.67	&	+26 47 01.8	&	B9 Si	&	7.72-7.73	&	3.3551(2)	&	4276.81(3)	&	0.002	&	0.001	&	0.029	&	0.371	&	7.714	&	-0.029	\\
\hline
\multicolumn{15}{l}{$^{aa}$HIP 101323 (SPB; $P$\textsubscript{VSX}\,=\,1.09613\,d); $^{ab}$NSV 13774 (no type and period given)} \\
\hline
\end{tabular}                                                                                                                                                                   
\end{adjustbox}
\end{center}                                                                                                                                             
\end{table*}

\setcounter{table}{1}  
\begin{table*}
\caption{Essential data for the known ACV variables whose rotational periods were investigated with MASCARA data, sorted by increasing right ascension. The columns denote: (1) HD number or other conventional identification. (2) Identification number from RM09. (3) Right ascension (J2000; GAIA DR2). (4) Declination (J2000; GAIA DR2). (5) Spectral type from RM09. (6) $V$ magnitude range. (7) Literature period (VSX). (8) Period (this work). (9) Epoch (HJD-2450000). (10) Semi-amplitude of the fundamental variation ($A_{\mathrm 1}$). (11) Semi-amplitude of the first harmonic variation ($A_{\mathrm 2}$). (12) Phase of the fundamental variation ($\phi_{\mathrm 1}$). (13) Phase of the first harmonic variation ($\phi_{\mathrm 2}$). (14) $G$\,mag (GAIA DR2). (15) $(BP-RP)$ index (GAIA DR2).}
\label{table_masterMASCARA}
\begin{center}
\begin{adjustbox}{max width=\textwidth}
\begin{tabular}{lllcclccccccccc}
\hline 
(1) & (2) & (3) & (4) & (5) & (6) & (7) & (8) & (9) & (10) & (11) & (12) & (13) & (14) & (15) \\
Star & \#RM09 & $\alpha$(J2000) & $\delta$(J2000) & SpT   & $V$\,range & P\textsubscript{VSX} & P\textsubscript{MASCARA} & Epoch          & $A_{\mathrm 1}$ & $A_{\mathrm 2}$ & $\phi_{\mathrm 1}$ & $\phi_{\mathrm 2}$ & $G$ & $(BP-RP)$ \\
     &           &                  &                  & [lit] & [mag]       & [d]   & [d]   & & [mag]           & [mag]           & [rad]              & [rad]              & [mag]   & [mag] \\
\hline
HD 3580	&	970	&	00 38 31.86	&	-20 17 47.6	&	B8 Si	&	6.71-6.73	&	1.4788	&	1.47561(4)	&	7219.71(1)	&	0.006	&	0.001	&	0.986	&	0.484	&	6.687		&	-0.157	\\
HD 4778	&	1250	&	00 50 18.27	&	+45 00 08.2	&	A1 Cr Sr Eu	&	6.10-6.13	&	2.5481	&	2.5619(1)	&	7185.70(2)	&	0.006	&	0.008	&	0.899	&	0.974	&	6.110		&	0.028	\\
HD 4796	&	1263	&	00 52 28.98	&	+79 50 25.6	&	A0 Sr Si	&	7.71-7.73	&	8	&	7.987(2)	&	7060.62(7)	&	0.009	&	0.000	&	0.765	&	0.323	&	7.695		&	0.152	\\
HD 7676	&	1880	&	01 16 06.81	&	-34 08 55.8	&	A5 Sr Cr Eu	&	8.36-8.41	&	5.0976	&	5.088(2)	&	7252.84(5)	&	0.022	&	0.007	&	0.010	&	0.744	&	8.388		&	0.217	\\
HD 7546	&	1870	&	01 16 24.50	&	+48 04 56.1	&	B8 Si	&	6.60-6.62	&	5.229	&	5.4060(9)	&	7057.38(5)	&	0.009	&	0.004	&	0.250	&	0.826	&	6.563		&	-0.002	\\
HD 8441	&	2050	&	01 24 18.69	&	+43 08 31.6	&	A2 Sr	&	6.66-6.68	&	69.92	&	69.9(1)	&	7230.2(7)	&	0.009	&	0.004	&	0.228	&	0.942	&	6.651		&	0.079	\\
HD 10221	&	2550	&	01 42 20.51	&	+68 02 34.9	&	A0 Si Sr Cr	&	5.54-5.58	&	3.1848	&	3.1528(2)	&	7216.49(3)	&	0.017	&	0.005	&	0.769	&	0.899	&	5.529		&	-0.064	\\
HD 10783	&	2670	&	01 45 42.52	&	+08 33 33.2	&	A2 Si Cr Sr	&	6.53-6.56	&	4.1321	&	4.1335(3)	&	7057.30(4)	&	0.009	&	0.005	&	0.374	&	0.063	&	6.514		&	-0.009	\\
HD 11415	&	2870	&	01 54 23.74	&	+63 40 12.4	&	B3 He wk.	&	3.34-3.35	&	\textbf{0.08946}	&	\textbf{14.53(1)}	&	7192.6(1)	&	0.006	&	0.002	&	0.528	&	0.503	&	3.261		&	0.070	\\
HD 12288	&	3130	&	02 03 30.49	&	+69 34 56.4	&	A2 Cr Si	&	7.72-7.74	&	34.9	&	35.73(3)	&	7218.6(3)	&	0.007	&	0.002	&	0.772	&	0.633	&	7.706		&	0.177	\\
HD 14392	&	3620	&	02 20 58.20	&	+50 09 05.4	&	B9 Si	&	5.55-5.57	&	4.189	&	4.2143(5)	&	7060.59(4)	&	0.006	&	0.002	&	0.298	&	0.078	&	5.522		&	-0.098	\\
HD 14437	&	3640	&	02 21 02.67	&	+42 56 38.2	&	B9 Cr Eu Si	&	7.24-7.27	&	26.87	&	26.78(1)	&	7077.7(2)	&	0.013	&	0.002	&	0.427	&	0.530	&	7.221		&	0.005	\\
HD 16545	&	4120	&	02 40 36.94	&	+44 05 28.4	&	A0 Si	&	7.33-7.37	&	1.61942	&	1.61905(5)	&	7068.33(1)	&	0.022	&	0.002	&	0.061	&	0.080	&	7.333		&	-0.107	\\
HD 18296	&	4560	&	02 57 17.28	&	+31 56 03.3	&	A0 Si Sr	&	5.08-5.10	&	2.88422	&	2.8836(2)	&	7065.37(2)	&	0.007	&	0.006	&	0.585	&	0.386	&	5.074		&	-0.039	\\
HD 18473	&	4610	&	03 00 53.79	&	+59 39 57.6	&	B9 Si	&	7.33-7.35	&	0.666832	&	0.66690(1)	&	7057.334(6)	&	0.009	&	0.001	&	0.480	&	0.039	&	7.306		&	0.026	\\
HD 19712	&	4880	&	03 10 18.09	&	-01 41 41.0	&	A0 Cr Eu	&	7.31-7.33	&	2.1945	&	2.2045(1)	&	7064.37(2)	&	0.007	&	0.007	&	0.230	&	0.574	&	7.306		&	-0.039	\\
HD 19832	&	4910	&	03 12 14.25	&	+27 15 25.1	&	B8 Si	&	5.76-5.81	&	0.727902	&	0.72795(1)	&	7059.386(7)	&	0.015	&	0.011	&	0.965	&	0.645	&	5.752		&	-0.133	\\
HD 20629	&	5160	&	03 19 47.76	&	+19 04 34.6	&	A0 Si Sr Cr	&	7.36-7.45	&	2.4994	&	2.49934(6)	&	7059.38(2)	&	0.045	&	0.004	&	0.264	&	0.680	&	7.377		&	-0.043	\\
HD 21590	&	5400	&	03 29 42.57	&	+16 45 44.4	&	B9 Si	&	7.03-7.05	&	2.7609	&	2.7586(2)	&	7057.27(2)	&	0.004	&	0.001	&	0.487	&	0.325	&	7.009		&	-0.018	\\
HD 21699	&	5410	&	03 32 08.60	&	+48 01 24.6	&	B8 He wk. Si	&	7.45-7.47	&	2.4761	&	2.4928(2)	&	7059.37(2)	&	0.011	&	0.001	&	0.848	&	0.013	&	5.410		&	-0.109	\\
HD 22136	&	5570	&	03 35 58.49	&	+47 05 27.8	&	B8 Si	&	6.86-6.88	&	0.9312	&	0.93111(3)	&	7077.412(9)	&	0.011	&	0.001	&	0.737	&	0.230	&	6.852		&	-0.009	\\
HD 22470	&	5710	&	03 36 17.39	&	-17 28 01.6	&	B9 Si	&	5.21-5.27	&	1.93	&	1.92911(7)	&	7067.48(1)	&	0.021	&	0.013	&	0.013	&	0.669	&	5.145		&	-0.157	\\
HD 22374	&	5660	&	03 36 58.03	&	+23 12 39.9	&	A1 Cr Sr Si	&	6.71-6.73	&	10.61	&	10.656(2)	&	7083.4(1)	&	0.008	&	0.003	&	0.174	&	0.111	&	6.687		&	0.172	\\
HD 22920	&	5840	&	03 40 38.34	&	-05 12 38.6	&	B8 Si	&	5.51-5.55	&	3.9474	&	3.9489(3)	&	7060.57(3)	&	0.019	&	0.006	&	0.651	&	0.045	&	5.485		&	-0.173	\\
HD 24155	&	6190	&	03 51 15.86	&	+13 02 46.0	&	B9 Si	&	6.27-6.33	&	2.53465	&	2.5348(1)	&	7080.55(2)	&	0.029	&	0.003	&	0.422	&	0.726	&	6.247		&	-0.050	\\
HD 24769	&	6330	&	03 57 03.80	&	+23 10 32.2	&	B9 Si	&	6.03-6.06	&	1.48778	&	1.48759(4)	&	7066.50(1)	&	0.019	&	0.000	&	0.436	&	0.597	&	6.009		&	0.104	\\
HD 25267	&	6440	&	03 59 55.48	&	-24 00 58.8	&	A0 Si	&	4.62-4.63	&	1.2094	&	1.20971(4)	&	7262.65(1)	&	0.007	&	0.001	&	0.095	&	0.652	&	4.561		&	-0.126	\\
HD 25354	&	6460	&	04 03 10.85	&	+38 03 17.3	&	A2 Eu Cr	&	7.82-7.84	&	3.90072	&	3.9021(5)	&	7066.61(3)	&	0.014	&	0.001	&	0.758	&	0.111	&	7.825		&	0.044	\\
HD 25823	&	6560	&	04 06 36.41	&	+27 35 59.5	&	B9 Sr Si	&	7.72-7.73	&	7.227424	&	7.238(2)	&	7080.45(7)	&	0.009	&	0.001	&	0.513	&	0.708	&	5.133		&	-0.159	\\
HD 26792	&	6840	&	04 17 21.08	&	+57 10 41.9	&	B8 Sr	&	6.67-6.71	&	3.8031	&	3.8023(3)	&	7128.33(3)	&	0.019	&	0.008	&	0.995	&	0.225	&	6.683		&	0.068	\\
HD 26961	&	6880	&	04 18 14.61	&	+50 17 43.8	&	A2 Si	&	4.57-4.62	&	1.5273643	&	1.52730(7)	&	7056.37(1)	&	0.006	&	0.023	&	0.421	&	0.463	&	4.511		&	0.112	\\
HD 27309	&	7010	&	04 19 36.70	&	+21 46 24.7	&	A0 Si Cr	&	5.33-5.38	&	1.56896	&	1.56882(5)	&	7056.40(1)	&	0.025	&	0.007	&	0.758	&	0.178	&	5.319		&	-0.124	\\
HD 27404	&	7030	&	04 20 37.78	&	+28 53 31.3	&	A0 Si	&	7.90-7.95	&	2.77929	&	2.7789(1)	&	7087.42(2)	&	0.024	&	0.003	&	0.366	&	0.851	&	7.833		&	0.413	\\
HD 28843	&	7380	&	04 32 37.56	&	-03 12 34.4	&	B9 He wk. Si	&	5.70-5.80	&	1.374	&	1.37393(2)	&	7059.50(1)	&	0.047	&	0.003	&	0.558	&	0.378	&	5.740		&	-0.161	\\
HD 29009	&	7420	&	04 33 54.73	&	-06 44 20.1	&	B9 Si	&	5.71-5.73	&	3.82	&	3.7993(4)	&	7057.40(3)	&	0.007	&	0.003	&	0.209	&	0.627	&	5.723		&	-0.164	\\
HD 30466	&	7870	&	04 49 16.00	&	+29 34 16.9	&	A0 Si	&	7.24-7.27	&	1.4069	&	1.40687(4)	&	7059.49(1)	&	0.011	&	0.003	&	0.898	&	0.048	&	7.194		&	0.190	\\
HD 32145	&	8190	&	05 01 06.03	&	+03 43 02.4	&	B7 Si	&	7.22-7.25	&	2.42082	&	2.4190(1)	&	7067.48(2)	&	0.009	&	0.004	&	0.151	&	0.454	&	7.199		&	-0.148	\\
HD 32549	&	8280	&	05 04 34.14	&	+15 24 15.0	&	B9 Si Cr	&	4.65-4.68	&	4.6397	&	4.6393(4)	&	7060.31(4)	&	0.015	&	0.002	&	0.942	&	0.707	&	4.610		&	-0.045	\\
HD 32633	&	8320	&	05 06 08.35	&	+33 55 07.3	&	B9 Si Cr	&	7.04-7.07	&	6.43	&	6.4303(8)	&	7059.85(6)	&	0.009	&	0.004	&	0.774	&	0.997	&	7.001		&	0.010	\\
HD 32966	&	8420	&	05 06 22.15	&	-14 41 47.6	&	B9 Si	&	7.06-7.17	&	3.0929	&	3.0928(1)	&	7057.20(3)	&	0.052	&	0.006	&	0.868	&	0.213	&	7.097		&	-0.098	\\
HD 32650	&	8350	&	05 12 22.43	&	+73 56 47.9	&	B9 Si	&	5.40-5.45	&	2.73475	&	2.7353(1)	&	7143.31(2)	&	0.021	&	0.002	&	0.202	&	0.650	&	5.392		&	-0.128	\\
HD 34719	&	8850	&	05 20 18.29	&	+19 34 41.7	&	A0 Hg Si Cr	&	6.63-6.66	&	1.63988	&	1.64000(5)	&	7059.43(1)	&	0.012	&	0.007	&	0.193	&	0.752	&	6.625		&	-0.036	\\
HD 36916	&	9700	&	05 34 53.96	&	-04 06 37.6	&	B8 He wk. Si	&	6.71-6.75	&	1.564	&	1.56520(5)	&	7066.32(1)	&	0.017	&	0.004	&	0.042	&	0.594	&	6.680		&	-0.165	\\
HD 37808	&	10210	&	05 40 46.19	&	-10 24 31.1	&	B9 Si	&	6.44-6.46	&	1.099	&	1.09852(2)	&	7096.33(1)	&	0.009	&	0.002	&	0.888	&	0.969	&	6.419		&	-0.157	\\
HD 38823	&	10440	&	05 48 25.50	&	-00 45 34.6	&	A5 Sr Eu Cr	&	7.31-7.33	&	8.635	&	8.677(2)	&	7087.98(8)	&	0.010	&	0.006	&	0.062	&	0.875	&	7.315		&	0.387	\\
HD 39082	&	10500	&	05 50 23.86	&	+04 57 24.1	&	B9 Sr Cr Eu	&	7.39-7.41	&	0.764776	&	0.76477(2)	&	7059.487(7)	&	0.010	&	0.001	&	0.900	&	0.484	&	7.389		&	0.032	\\
HD 39317	&	10560	&	05 52 22.30	&	+14 10 18.5	&	B9 Si Eu Cr	&	5.59-5.60	&	2.6569	&	2.6558(3)	&	7059.49(2)	&	0.005	&	0.001	&	0.974	&	0.216	&	5.561		&	-0.036	\\
HD 39575	&	10600	&	05 52 23.85	&	-26 17 28.1	&	A0 Si Cr Eu	&	7.82-7.85	&	3.1009	&	3.1014(2)	&	7301.74(3)	&	0.015	&	0.004	&	0.461	&	0.901	&	7.811		&	0.001	\\
HD 42616	&	11390	&	06 13 43.15	&	+41 41 50.2	&	A1 Sr Cr Eu	&	7.15-7.16	&	\textbf{17}	&	\textbf{5.739(1)}	&	7057.16(5)	&	0.008	&	0.000	&	0.137	&	0.511	&	7.113		&	0.186	\\
HD 43819	&	11620	&	06 19 01.86	&	+17 19 30.9	&	B9 Si	&	6.26-6.28	&	15.0285	&	14.981(6)	&	7065.8(1)	&	0.007	&	0.004	&	0.092	&	0.483	&	6.236		&	-0.079	\\
HD 45439	&	12040	&	06 25 42.46	&	-35 41 50.5	&	B9 Si	&	7.84-7.90	&	1.10064	&	1.10059(2)	&	7064.44(1)	&	0.028	&	0.003	&	0.979	&	0.320	&	7.839		&	-0.043	\\
HD 45530	&	12070	&	06 28 13.96	&	+05 16 20.2	&	A0 Si	&	7.47-7.49	&	0.7919	&	0.79187(2)	&	7065.622(7)	&	0.011	&	0.000	&	0.024	&	0.029	&	7.607		&	-0.043	\\
HD 46462	&	12430	&	06 31 45.20	&	-37 10 23.2	&	B9 Si	&	7.50-7.56	&	10.363	&	10.346(5)	&	7080.9(1)	&	0.022	&	0.015	&	0.388	&	0.989	&	7.519		&	-0.115	\\
HD 47144	&	12660	&	06 35 24.18	&	-36 46 47.5	&	B9 Si	&	5.76-5.77	&	2.21004	&	2.2111(2)	&	7081.32(2)	&	0.007	&	0.002	&	0.014	&	0.442	&	5.905		&	-0.128	\\
HD 47714	&	12810	&	06 39 32.66	&	-11 41 51.6	&	B8 Si	&	7.92-7.96	&	4.6953	&	4.6955(4)	&	7057.48(4)	&	0.017	&	0.004	&	0.725	&	0.209	&	7.949		&	-0.051	\\
HD 49333	&	13340	&	06 47 01.48	&	-21 00 55.4	&	B7 He wk. Si	&	6.05-6.08	&	2.181	&	2.1805(1)	&	7066.71(2)	&	0.014	&	0.005	&	0.716	&	0.142	&	6.027		&	-0.221	\\
HD 49484	&	13390	&	06 47 47.12	&	-22 10 10.3	&	B9 Si	&	8.32-8.35	&	7.039	&	7.033(1)	&	7068.55(7)	&	0.017	&	0.002	&	0.697	&	0.929	&	8.338		&	-0.134	\\
HD 49713	&	13480	&	06 49 44.29	&	-01 20 23.7	&	B9 Cr Eu Si	&	7.29-7.34	&	2.13503	&	2.13512(8)	&	7060.56(2)	&	0.023	&	0.004	&	0.924	&	0.844	&	7.281		&	-0.077	\\
HD 49976	&	13560	&	06 50 42.30	&	-08 02 27.6	&	A1 Sr Cr Eu	&	6.28-6.32	&	2.976	&	2.9768(2)	&	7067.30(2)	&	0.009	&	0.010	&	0.454	&	0.908	&	6.280		&	0.043	\\
HD 50304	&	13774	&	06 51 41.36	&	-23 48 10.4	&	A0 Eu Cr	&	7.54-7.58	&	7.884	&	7.886(1)	&	7287.92(7)	&	0.022	&	0.002	&	0.589	&	0.352	&	7.544		&	0.058	\\
HD 50461	&	13810	&	06 53 08.00	&	-07 45 56.1	&	B9 Si Cr	&	7.79-7.82	&	0.89403	&	0.89400(2)	&	7059.427(8)	&	0.011	&	0.005	&	0.377	&	0.695	&	7.778		&	-0.065	\\
HD 50341	&	13780	&	06 54 13.60	&	+33 00 09.1	&	B9 Sr Cr Eu	&	8.16-8.20	&	2.50919	&	2.5094(1)	&	7060.58(2)	&	0.019	&	0.008	&	0.036	&	0.514	&	8.166		&	0.036	\\
HD 51418	&	14180	&	06 59 20.14	&	+42 18 53.2	&	A0 Ho Dy	&	6.60-6.73	&	5.4379	&	5.4377(3)	&	7065.44(5)	&	0.069	&	0.011	&	0.413	&	0.048	&	6.664		&	0.099	\\
HD 52993	&	14550	&	07 01 47.21	&	-35 32 52.5	&	B9 Si	&	6.54-6.61	&	1.29644	&	1.29645(3)	&	7057.37(1)	&	0.025	&	0.018	&	0.295	&	0.460	&	6.540		&	-0.182	\\
HD 56273	&	15300	&	07 15 57.31	&	-12 32 35.2	&	B8 Si	&	7.88-7.92	&	1.78661	&	1.78678(6)	&	7064.46(1)	&	0.017	&	0.008	&	0.027	&	0.288	&	7.892		&	-0.060	\\
HD 59256	&	16180	&	07 27 59.16	&	-29 09 21.2	&	B9 Si	&	5.54-5.55	&	0.943	&	0.94343(3)	&	7067.558(9)	&	0.006	&	0.000	&	0.411	&	0.647	&	5.508		&	-0.029	\\
HD 62140	&	17050	&	07 46 27.40	&	+62 49 50.0	&	A8 Sr Eu	&	6.45-6.48	&	4.28679	&	4.2872(4)	&	7136.44(4)	&	0.011	&	0.010	&	0.257	&	0.554	&	6.458		&	0.278	\\
HD 63347	&	17473	&	07 54 20.48	&	+71 04 45.3	&	B8 Sr Cr Eu	&	7.34-7.36	&	1.74984	&	1.74939(9)	&	7147.54(1)	&	0.004	&	0.005	&	0.072	&	0.273	&	7.330		&	-0.014	\\
HD 65339	&	17910	&	08 01 42.43	&	+60 19 27.8	&	A3 Sr Eu Cr	&	6.01-6.02	&	8.0278	&	7.993(2)	&	7058.44(7)	&	0.003	&	0.001	&	0.548	&	0.679	&	5.994		&	0.179	\\
HD 68292	&	18860	&	08 11 03.36	&	-18 58 26.8	&	B9 Si	&	7.49-7.55	&	5.7426	&	5.7377(6)	&	7067.24(5)	&	0.028	&	0.004	&	0.060	&	0.992	&	7.489		&	-0.121	\\
HD 70325	&	19410	&	08 20 18.76	&	-29 32 28.7	&	A0 Si	&	7.30-7.34	&	17.878	&	17.962(6)	&	7082.4(1)	&	0.012	&	0.008	&	0.397	&	0.225	&	7.317		&	-0.122	\\
\hline
\end{tabular}                                                                                                                                                                   
\end{adjustbox}
\end{center}                                                                                                                                             
\end{table*} 
\setcounter{table}{1}  
\begin{table*}
\caption{continued.}
\label{table_masterMASCARA2}
\begin{center}
\begin{adjustbox}{max width=\textwidth}
\begin{tabular}{lllcclccccccccc}
\hline 
(1) & (2) & (3) & (4) & (5) & (6) & (7) & (8) & (9) & (10) & (11) & (12) & (13) & (14) & (15) \\
Star & \#RM09 & $\alpha$(J2000) & $\delta$(J2000) & SpT   & $V$\,range & P\textsubscript{VSX} & P\textsubscript{MASCARA} & Epoch          & $A_{\mathrm 1}$ & $A_{\mathrm 2}$ & $\phi_{\mathrm 1}$ & $\phi_{\mathrm 2}$ & $G$ & $(BP-RP)$ \\
     &           &                  &                  & [lit] & [mag]       & [d]   & [d]   &  & [mag]           & [mag]           & [rad]              & [rad]              & [mag]   & [mag] \\
\hline
HD 71866	&	19850	&	08 31 10.65	&	+40 13 29.7	&	A1 Eu Sr Si	&	6.70-6.72	&	6.80054	&	6.797(1)	&	7097.20(6)	&	0.004	&	0.005	&	0.478	&	0.301	&	6.703		&	0.089	\\
HD 77314	&	21834	&	09 01 58.77	&	+02 40 16.5	&	A2 Sr Cr Eu	&	7.20-7.28	&	2.8646	&	2.86445(8)	&	7056.66(2)	&	0.030	&	0.016	&	0.216	&	0.704	&	7.267		&	0.104	\\
HD 93226	&	26920	&	10 45 48.22	&	-10 42 48.8	&	A0 Si	&	7.43-7.48	&	1.72907	&	1.72913(6)	&	7057.77(1)	&	0.018	&	0.004	&	0.108	&	0.233	&	7.459		&	-0.033	\\
HD 95442	&	27500	&	11 00 44.05	&	-25 43 02.4	&	A0 Sr Cr Eu	&	7.82-7.84	&	3.3386	&	3.3384(3)	&	7122.58(3)	&	0.010	&	0.005	&	0.220	&	0.639	&	7.823		&	0.084	\\
HD 96707	&	27890	&	11 09 39.78	&	+67 12 36.9	&	A8 Sr	&	7.06-7.07	&	3.51556	&	3.5129(5)	&	7140.34(3)	&	0.006	&	0.002	&	0.189	&	0.424	&	6.023		&	0.302	\\
HD 98457	&	28360	&	11 19 25.95	&	-30 19 22.9	&	A0 Si	&	7.90-7.98	&	11.534	&	11.533(2)	&	7085.4(1)	&	0.043	&	0.009	&	0.387	&	0.958	&	7.919		&	-0.058	\\
HD 102454	&	29530	&	11 47 26.22	&	-31 15 17.0	&	B9 Si	&	7.30-7.32	&	2.07906	&	2.0792(1)	&	7071.51(2)	&	0.008	&	0.006	&	0.747	&	0.339	&	7.287		&	-0.029	\\
HD 111133	&	32310	&	12 47 02.30	&	+05 57 01.3	&	A1 Sr Cr Eu	&	6.28-6.31	&	\textbf{2.2246}	&	\textbf{16.230(5)}	&	7060.0(1)	&	0.014	&	0.002	&	0.577	&	0.591	&	6.276		&	0.003	\\
HD 118022	&	34020	&	13 34 07.95	&	+03 39 32.4	&	A2 Cr Eu Sr	&	4.90-4.92	&	3.722	&	3.7216(4)	&	7065.52(3)	&	0.008	&	0.002	&	0.212	&	0.838	&	4.854		&	0.058	\\
HD 119213	&	34410	&	13 40 21.37	&	+57 12 27.6	&	A3 Sr Cr	&	6.27-6.28	&	2.449967	&	2.4478(2)	&	7132.73(2)	&	0.007	&	0.001	&	0.815	&	0.681	&	6.266		&	0.088	\\
HD 124224	&	35560	&	14 12 15.81	&	+02 24 33.8	&	B9 Si	&	4.95-5.02	&	0.52071601	&	0.520700(3)	&	7064.698(5)	&	0.036	&	0.004	&	0.064	&	0.894	&	4.950		&	-0.103	\\
HD 125248	&	35760	&	14 18 38.26	&	-18 42 57.5	&	A1 Eu Cr	&	5.83-5.86	&	9.2954	&	9.302(1)	&	7067.40(9)	&	0.006	&	0.008	&	0.985	&	0.216	&	5.844		&	0.009	\\
HD 133029	&	37770	&	15 00 38.72	&	+47 16 38.8	&	B9 Si Cr Sr	&	6.34-6.36	&	2.8881	&	2.8885(2)	&	7095.68(2)	&	0.008	&	0.001	&	0.208	&	0.115	&	6.319		&	-0.106	\\
HD 138764	&	39510	&	15 34 26.51	&	-09 11 00.3	&	B6 Si	&	5.12-5.17	&	1.25881	&	1.25869(3)	&	7070.74(1)	&	0.020	&	0.001	&	0.228	&	0.643	&	5.103		&	-0.108	\\
HD 140160	&	39840	&	15 41 47.42	&	+12 50 51.2	&	A1 Sr	&	5.31-5.33	&	1.59584	&	1.59590(8)	&	7057.78(1)	&	0.010	&	0.004	&	0.167	&	0.057	&	5.292		&	0.045	\\
HD 142884	&	40510	&	15 57 48.81	&	-23 31 38.3	&	B9 Si	&	6.76-6.78	&	0.803	&	0.80296(2)	&	7058.762(8)	&	0.006	&	0.004	&	0.829	&	0.922	&	6.719		&	0.068	\\
HD 148112	&	41850	&	16 25 24.95	&	+14 01 59.8	&	A0 Cr Eu	&	4.56-4.58	&	2.951	&	3.0439(3)	&	7058.87(3)	&	0.011	&	0.004	&	0.813	&	0.577	&	4.497		&	0.037	\\
HD 151199	&	42760	&	16 42 58.46	&	+55 41 24.4	&	A3 Sr	&	6.16-6.17	&	2.226	&	2.2267(2)	&	7058.66(2)	&	0.005	&	0.003	&	0.795	&	0.820	&	6.158		&	0.097	\\
HD 150714	&	42640	&	16 43 39.68	&	-22 44 11.1	&	A0 Si	&	7.54-7.59	&	1.6288	&	1.62906(5)	&	7087.73(1)	&	0.018	&	0.013	&	0.930	&	0.109	&	7.523		&	0.245	\\
HD 152107	&	43050	&	16 49 14.23	&	+45 59 00.1	&	A3 Sr Cr Eu	&	4.81-4.82	&	3.8567	&	3.8586(6)	&	7056.80(3)	&	0.003	&	0.001	&	0.749	&	0.293	&	4.798		&	0.093	\\
HD 159376	&	44850	&	17 35 18.50	&	-22 02 37.7	&	B9 Si	&	6.46-6.50	&	9.75	&	9.738(2)	&	7134.70(9)	&	0.020	&	0.008	&	0.250	&	0.300	&	6.410		&	0.118	\\
HD 162613	&	45910	&	17 53 30.00	&	-36 30 30.8	&	A0 Si	&	7.97-8.01	&	3.5832	&	3.5821(3)	&	7116.75(3)	&	0.019	&	0.005	&	0.046	&	0.990	&	7.940		&	0.042	\\
HD 166053	&	46770	&	18 09 36.66	&	-19 21 27.0	&	B9 Si	&	8.32-8.43	&	3.6043	&	3.6054(5)	&	7151.54(3)	&	0.054	&	0.005	&	0.216	&	0.619	&	8.288		&	0.248	\\
HD 166469	&	46860	&	18 11 58.16	&	-28 54 05.5	&	A0 Si Cr Sr	&	8.51-8.52	&	2.8855	&	2.8890(3)	&	7121.58(2)	&	0.006	&	0.001	&	0.619	&	0.805	&	6.472		&	0.040	\\
HD 170000	&	47620	&	18 20 45.38	&	+71 20 16.0	&	A0 Si	&	4.25-4.27	&	1.71646	&	1.71665(9)	&	7135.57(1)	&	0.012	&	0.004	&	0.132	&	0.274	&	-	&	-	\\
HD 168856	&	47330	&	18 22 10.82	&	-07 29 55.8	&	B9 Si	&	7.04-7.07	&	2.427	&	2.4277(1)	&	7060.80(2)	&	0.013	&	0.007	&	0.139	&	0.024	&	6.959		&	0.270	\\
HD 169952	&	47600	&	18 25 23.11	&	+38 26 27.2	&	B9 Si	&	7.34-7.37	&	2.1223	&	2.12296(8)	&	7070.69(2)	&	0.011	&	0.001	&	0.183	&	0.515	&	7.319		&	-0.080	\\
HD 171184	&	47930	&	18 33 50.81	&	-14 25 32.9	&	A0 Si	&	7.88-7.93	&	2.79926	&	2.7989(2)	&	7120.66(2)	&	0.022	&	0.009	&	0.702	&	0.554	&	7.825		&	0.401	\\
HD 172977	&	48501	&	18 41 40.21	&	+34 05 55.5	&	B9 Si	&	7.40-7.44	&	2.2791	&	2.2799(1)	&	7067.84(2)	&	0.017	&	0.005	&	0.565	&	0.814	&	7.386		&	-0.189	\\
HD 173650	&	48660	&	18 45 35.62	&	+21 59 05.0	&	A0 Si Sr Cr	&	6.49-6.53	&	9.975	&	9.976(2)	&	7058.26(9)	&	0.020	&	0.006	&	0.202	&	0.628	&	6.476		&	0.075	\\
HD 173406	&	48570	&	18 45 53.23	&	-18 21 50.8	&	B9 Si	&	7.40-7.45	&	\textbf{5.0945}	&	\textbf{4.5627(4)}	&	7120.59(4)	&	0.014	&	0.010	&	0.200	&	0.507	&	7.366		&	0.197	\\
HD 173657	&	48690	&	18 47 37.16	&	-28 16 47.0	&	B9 Si Cr	&	7.39-7.41	&	1.93789	&	1.9378(1)	&	7124.77(1)	&	0.013	&	0.003	&	0.986	&	0.204	&	7.367		&	0.207	\\
HD 174646	&	48910	&	18 51 37.93	&	-01 02 39.9	&	B9 Si	&	8.19-8.22	&	\textbf{1.1807}	&	\textbf{6.4193(8)}	&	7124.69(6)	&	0.012	&	0.006	&	0.907	&	0.942	&	8.189		&	0.138	\\
HD 177410	&	49490	&	18 58 52.61	&	+69 31 52.5	&	B9 Si	&	6.49-6.52	&	1.1232524	&	1.12318(3)	&	7131.57(1)	&	0.010	&	0.008	&	0.279	&	0.819	&	6.458		&	-0.191	\\
HD 176582	&	49280	&	18 59 12.29	&	+39 13 02.3	&	B5 He wk.	&	6.39-6.41	&	1.58193	&	1.58221(5)	&	7057.81(1)	&	0.010	&	0.005	&	0.019	&	0.264	&	6.383		&	-0.232	\\
HD 178591	&	49677	&	19 07 19.19	&	+41 03 14.7	&	B5 Si	&	7.10-7.15	&	4.9396	&	4.9404(5)	&	7117.02(4)	&	0.011	&	0.022	&	0.084	&	0.647	&	7.114		&	-0.026	\\
HD 178929	&	49744	&	19 11 19.28	&	-20 20 50.2	&	B8 Si Sr	&	7.68-7.75	&	3.3453	&	3.3444(2)	&	7128.75(3)	&	0.032	&	0.003	&	0.241	&	0.525	&	7.677		&	0.010	\\
HD 180029	&	49955	&	19 14 45.83	&	+04 37 12.4	&	A2 Si	&	7.71-7.74	&	3.2858	&	3.2857(2)	&	7099.76(3)	&	0.015	&	0.004	&	0.867	&	0.271	&	7.946		&	0.250	\\
HD 184905	&	51000	&	19 34 43.92	&	+43 56 45.1	&	A0 Si Cr	&	6.59-6.63	&	1.84534	&	1.84548(7)	&	7068.79(1)	&	0.017	&	0.008	&	0.426	&	0.095	&	6.599		&	-0.020	\\
HD 185183	&	51130	&	19 38 56.22	&	-28 36 25.8	&	B9 Si	&	6.70-6.74	&	1.737	&	1.73776(6)	&	7167.58(1)	&	0.020	&	0.005	&	0.270	&	0.328	&	6.687		&	-0.091	\\
HD 187128	&	51570	&	19 47 39.75	&	+15 54 27.6	&	B9 Si Sr	&	7.58-7.59	&	6.1858	&	6.1880(8)	&	7085.42(6)	&	0.005	&	0.002	&	0.597	&	0.899	&	7.579		&	0.034	\\
HD 187473	&	51690	&	19 51 10.24	&	-27 28 19.3	&	B9 Eu Sr Si	&	7.26-7.36	&	4.718	&	4.7184(2)	&	7176.56(4)	&	0.040	&	0.021	&	0.743	&	0.844	&	7.283		&	-0.035	\\
HD 191287	&	53290	&	20 09 41.98	&	-18 20 48.0	&	B9 Eu	&	8.08-8.26	&	1.62345	&	1.62342(5)	&	7137.71(1)	&	0.087	&	0.030	&	0.040	&	0.327	&	8.188		&	0.082	\\
HD 191980	&	53520	&	20 12 03.85	&	+15 21 28.1	&	B5 He wk.	&	7.89-7.91	&	\textbf{19.5}	&	\textbf{20.15(1)}	&	7098.5(2)	&	0.014	&	0.005	&	0.616	&	0.010	&	7.876		&	-0.190	\\
HD 192913	&	53840	&	20 16 27.20	&	+27 46 34.0	&	A0 Si Cr	&	6.62-6.66	&	16.846	&	16.835(5)	&	7126.6(1)	&	0.018	&	0.006	&	0.403	&	0.125	&	6.620		&	-0.047	\\
HD 193325	&	53960	&	20 19 00.95	&	+20 27 50.9	&	B9 Si	&	7.50-7.52	&	3.1529	&	3.1520(3)	&	7129.73(3)	&	0.010	&	0.004	&	0.917	&	0.695	&	7.471		&	-0.183	\\
HD 193722	&	54060	&	20 19 56.05	&	+46 50 14.5	&	B9 Si	&	6.45-6.51	&	\textbf{1.132854}	&	\textbf{8.530(1)}	&	7091.65(8)	&	0.029	&	0.004	&	0.390	&	0.016	&	6.426		&	-0.069	\\
HD 196502	&	54780	&	20 31 30.42	&	+74 57 16.6	&	A2 Sr Cr Eu	&	5.17-5.19	&	20.2747	&	20.31(1)	&	7160.3(2)	&	0.003	&	0.007	&	0.179	&	0.699	&	5.143		&	0.123	\\
HD 196270	&	54710	&	20 35 06.20	&	+33 33 08.6	&	B9 Si	&	8.16-8.21	&	1.3021519	&	1.30209(3)	&	7095.73(1)	&	0.020	&	0.009	&	0.228	&	0.804	&	8.188		&	-0.017	\\
HD 197374	&	55020	&	20 41 37.07	&	+43 50 12.3	&	B9 Si	&	8.33-8.36	&	1.98413	&	1.98363(8)	&	7076.81(1)	&	0.018	&	0.002	&	0.152	&	0.092	&	8.338		&	-0.055	\\
HD 197451	&	55040	&	20 43 56.96	&	-05 35 24.8	&	F1 Sr Eu Cr	&	7.17-7.20	&	1.803	&	1.80268(6)	&	7139.71(1)	&	0.012	&	0.004	&	0.284	&	0.313	&	7.114		&	0.453	\\
HD 199180	&	55460	&	20 55 02.19	&	+17 15 34.4	&	A0 Si Cr	&	7.83-7.85	&	33.45	&	33.00(3)	&	7120.3(3)	&	0.012	&	0.003	&	0.983	&	0.177	&	7.822		&	0.081	\\
HD 199728	&	55630	&	20 59 36.14	&	-19 02 07.0	&	B9 Si	&	6.24-6.29	&	2.2411	&	2.2400(1)	&	7194.70(2)	&	0.018	&	0.018	&	0.671	&	0.090	&	6.232		&	-0.121	\\
HD 200177	&	55750	&	21 00 06.62	&	+48 40 46.0	&	A1 Cr Sr Eu	&	7.32-7.34	&	1.47	&	1.46937(4)	&	7120.74(1)	&	0.008	&	0.002	&	0.574	&	0.892	&	7.317		&	0.014	\\
HD 200311	&	55810	&	21 01 14.32	&	+43 43 18.5	&	B9 Si Cr Hg	&	7.68-7.70	&	25.96728	&	26.13(2)	&	7084.9(2)	&	0.012	&	0.002	&	0.605	&	0.665	&	7.658		&	-0.102	\\
HD 201834	&	56290	&	21 10 15.56	&	+53 33 47.2	&	B9 Si	&	5.74-5.75	&	3.53544	&	3.5388(5)	&	7137.45(3)	&	0.005	&	0.001	&	0.849	&	0.698	&	5.714		&	-0.139	\\
HD 203819	&	56750	&	21 22 43.03	&	+54 13 50.4	&	A0 Cr Si Sr	&	7.85-7.87	&	2.58	&	2.5765(2)	&	7147.67(2)	&	0.009	&	0.005	&	0.567	&	0.234	&	7.846		&	0.060	\\
HD 205938	&	57340	&	21 35 25.92	&	+68 13 09.3	&	B9 Si	&	6.43-6.48	&	8.34	&	8.335(1)	&	7137.54(8)	&	0.022	&	0.009	&	0.353	&	0.990	&	6.433		&	-0.088	\\
HD 206028	&	57370	&	21 38 45.78	&	+24 23 53.5	&	A0 Si	&	8.14-8.17	&	\textbf{0.785856}	&	\textbf{3.7092(3)}	&	7139.61(3)	&	0.015	&	0.004	&	0.846	&	0.421	&	8.118		&	-0.103	\\
HD 207188	&	57640	&	21 47 36.42	&	-17 17 41.1	&	A0 Si	&	7.63-7.69	&	2.6733	&	2.6735(1)	&	7194.67(2)	&	0.015	&	0.016	&	0.675	&	0.498	&	7.649		&	-0.120	\\
HD 208835	&	58070	&	21 57 51.37	&	+46 51 51.8	&	B9 Si	&	7.54-7.56	&	1.71677	&	1.71667(9)	&	7145.71(1)	&	0.010	&	0.000	&	0.215	&	0.788	&	7.525		&	-0.034	\\
HD 210071	&	58470	&	22 06 13.56	&	+56 20 36.3	&	B9 Si Cr Hg	&	6.36-6.41	&	1.43246	&	1.43264(4)	&	7147.65(1)	&	0.015	&	0.013	&	0.766	&	0.387	&	6.346		&	-0.105	\\
HD 212939	&	59060	&	22 26 55.18	&	+49 42 43.0	&	A0 Si	&	8.14-8.18	&	2.3934	&	2.3912(1)	&	7148.69(2)	&	0.018	&	0.001	&	0.131	&	0.583	&	8.149		&	0.053	\\
HD 213871	&	59320	&	22 33 37.58	&	+46 33 55.0	&	B9 Si	&	7.35-7.40	&	1.9505	&	1.95070(8)	&	7183.53(1)	&	0.023	&	0.007	&	0.203	&	0.702	&	7.339		&	0.007	\\
HD 215038	&	59450	&	22 39 22.90	&	+75 39 27.4	&	B9 Si	&	8.13-8.18	&	2.037638	&	2.03725(8)	&	7141.71(1)  	&	0.025	&	0.003	&	0.228	&	0.326	&	8.119		&	0.002	\\
HD 216533	&	59810	&	22 52 41.92	&	+58 48 23.4	&	A1 Sr Cr Si	&	7.86-7.89	&	17.22	&	17.256(6)	&	7150.4(1)	&	0.013	&	0.001	&	0.360	&	0.158	&	7.864		&	0.146	\\
HD 217833	&	60060	&	23 02 43.92	&	+55 14 11.0	&	B8 He wk.	&	6.50-6.55	&	5.36	&	5.3866(6)	&	7151.76(5)	&	0.024	&	0.008	&	0.957	&	0.530	&	6.484		&	-0.072	\\
HD 220147	&	60450	&	23 20 48.89	&	+62 24 45.4	&	B9 Cr Si Eu	&	8.08-8.12	&	10.99	&	10.965(2)	&	7134.7(1)	&	0.016	&	0.008	&	0.117	&	0.380	&	8.077		&	0.208	\\
HD 220668	&	60500	&	23 25 20.56	&	+36 09 52.0	&	A0 Si	&	7.66-7.69	&	6.16	&	6.1606(7)	&	7241.25(6)	&	0.016	&	0.005	&	0.307	&	0.952	&	7.660		&	-0.011	\\
HD 221394	&	60670	&	23 31 43.03	&	+28 24 12.7	&	A0 Sr Cr Si	&	6.37-6.41	&	2.8605	&	2.8600(2)	&	7152.80(2)	&	0.015	&	0.010	&	0.806	&	0.848	&	6.385		&	0.045	\\
HD 223640	&	61290	&	23 51 21.33	&	-18 54 32.8	&	B9 Si Sr Cr	&	5.16-5.20	&	3.735239	&	3.7384(8)	&	7219.72(3)	&	0.017	&	0.007	&	0.397	&	0.489	&	5.125		&	-0.153	\\
HD 223660	&	61310	&	23 51 26.93	&	+47 45 15.7	&	B8 Si	&	8.08-8.11	&	2.82	&	2.8258(2)	&	7060.25(2)	&	0.012	&	0.003	&	0.221	&	0.736	&	8.180		&	-0.034	\\
HD 224166	&	61440	&	23 55 37.93	&	+46 21 29.9	&	B9 Si	&	6.92-6.94	&	3.5112	&	3.5139(4)	&	7173.52(3)	&	0.008	&	0.006	&	0.250	&	0.791	&	6.904		&	-0.044	\\
\hline
\end{tabular}                                                                                                                                                                   
\end{adjustbox}
\end{center}                                                                                                                                             
\end{table*} 

\clearpage

\section{Light curves} \label{appendixLCs}

In this section, the light curves of all objects are presented, folded with the periods listed in 
Tables \ref{table_masterASASKELT}, \ref{table_masterMASCARA} and \ref{table_candidates}. Because of the 
high number of data points, a five-point-binning was applied to the MASCARA data for the construction of the phase plots to increase clarity.
The complete light curve figures are available from the authors and in the final publication.


\begin{figure*}
 \centering
\includegraphics[width=0.42\textwidth]{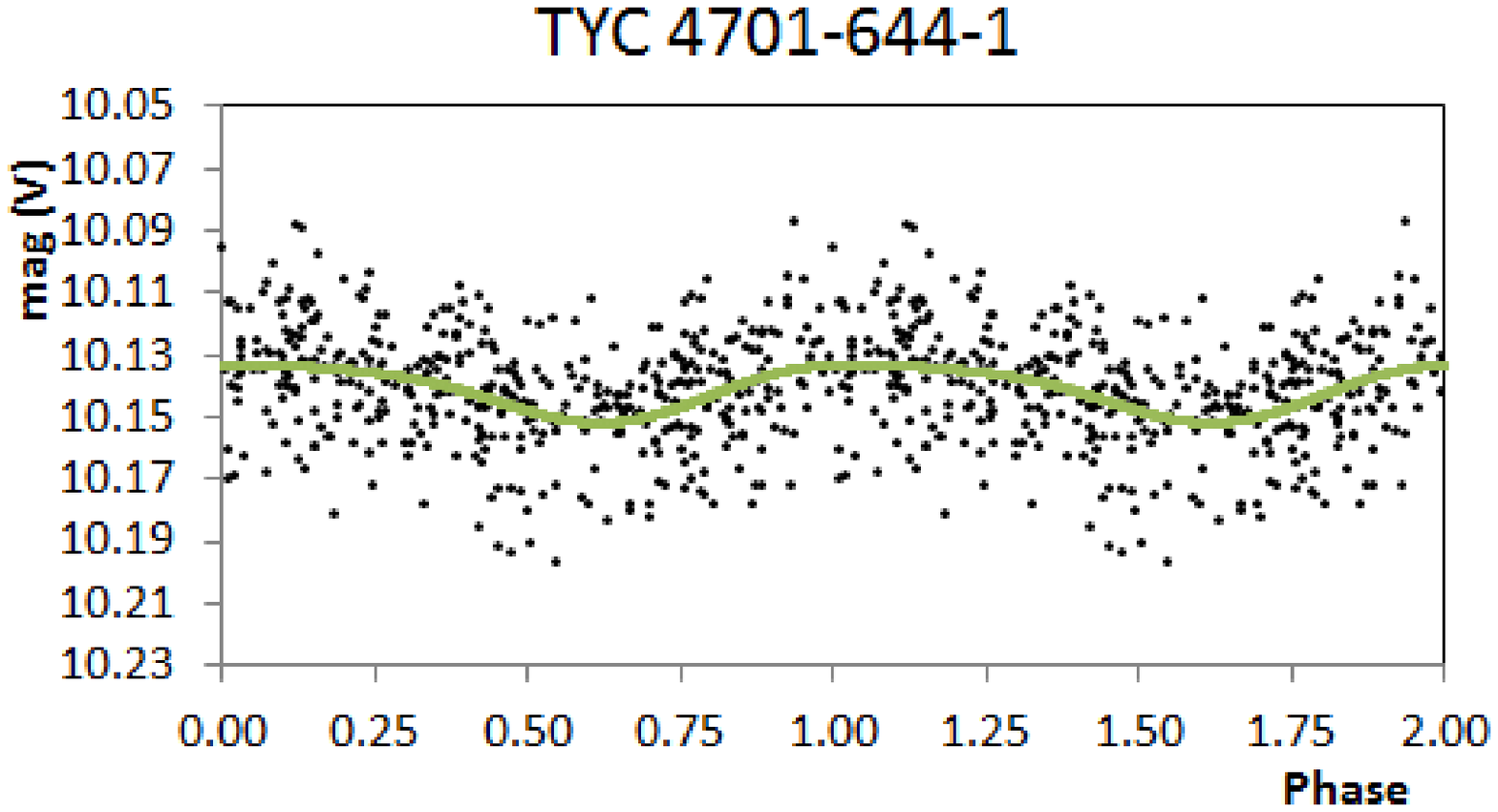}
\includegraphics[width=0.42\textwidth]{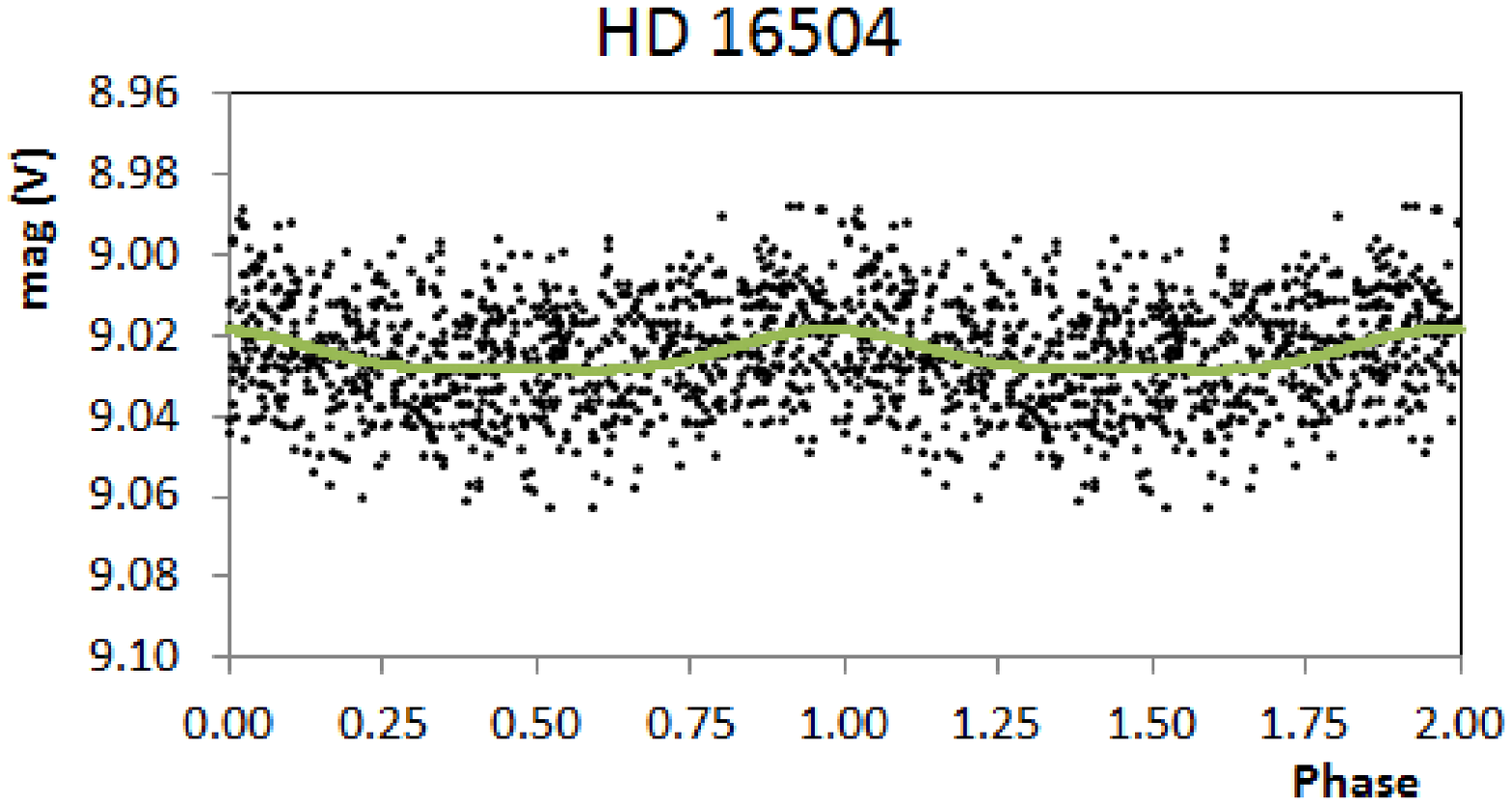}
\includegraphics[width=0.42\textwidth]{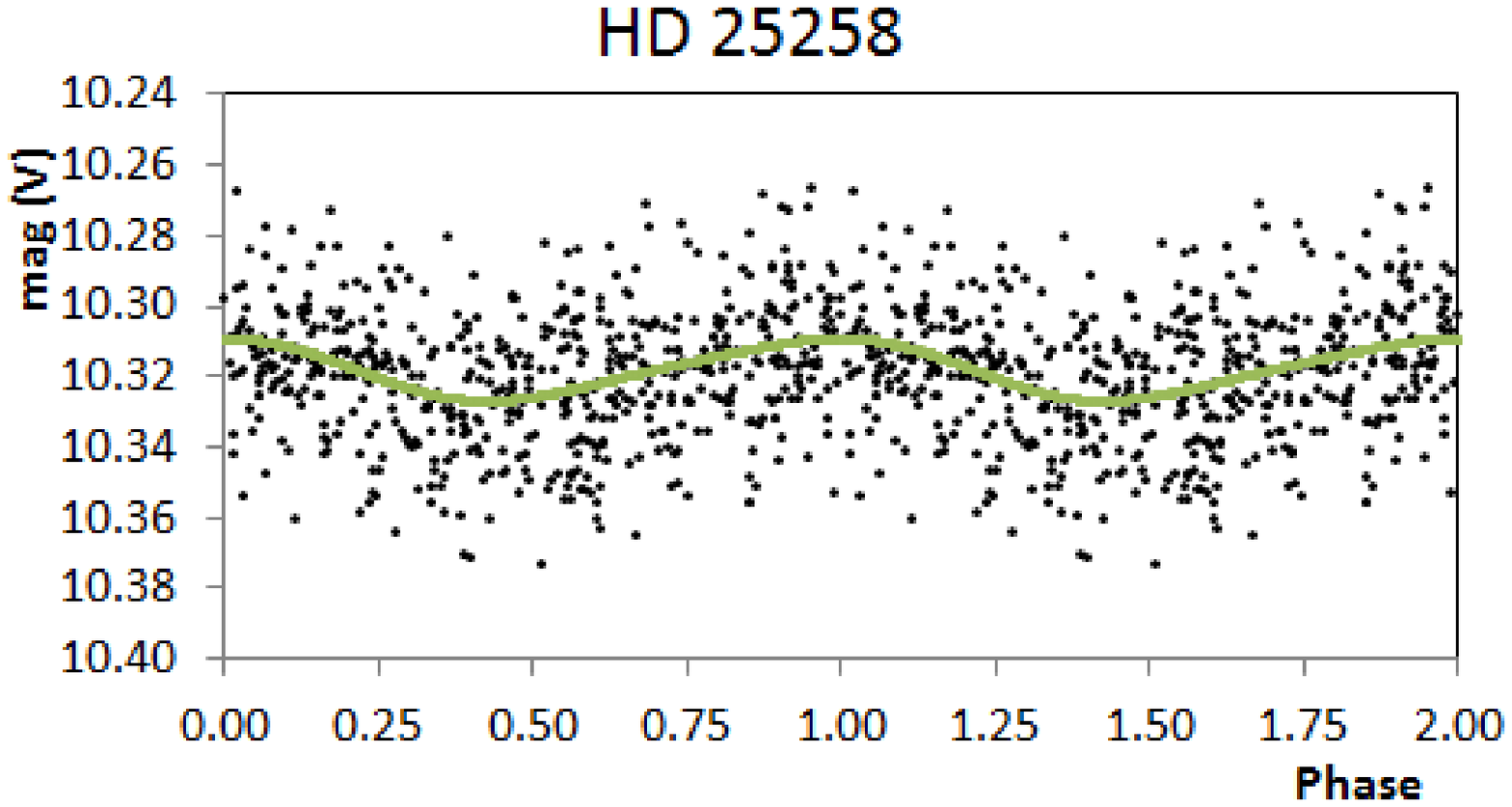}
\includegraphics[width=0.42\textwidth]{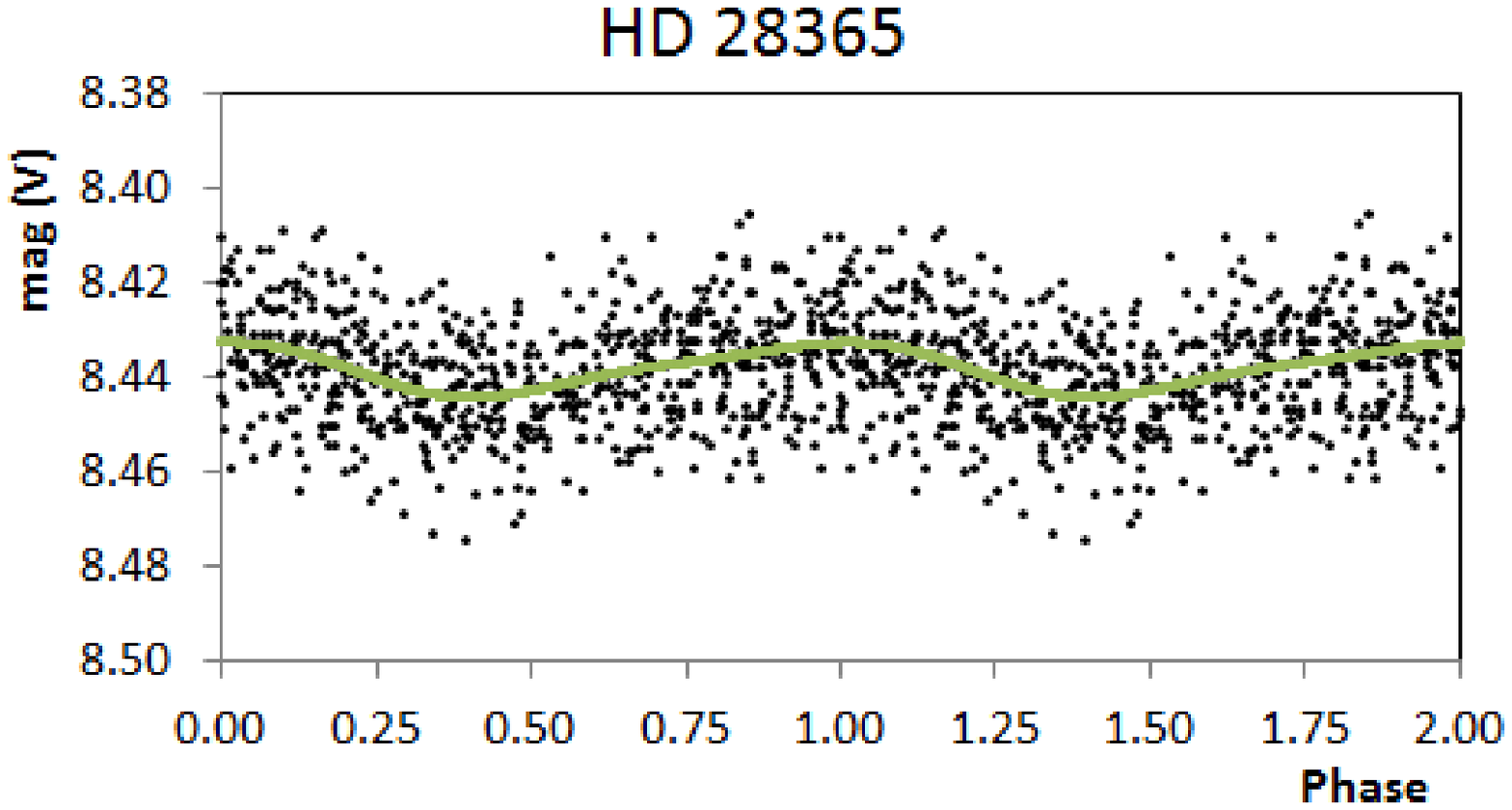}
\includegraphics[width=0.42\textwidth]{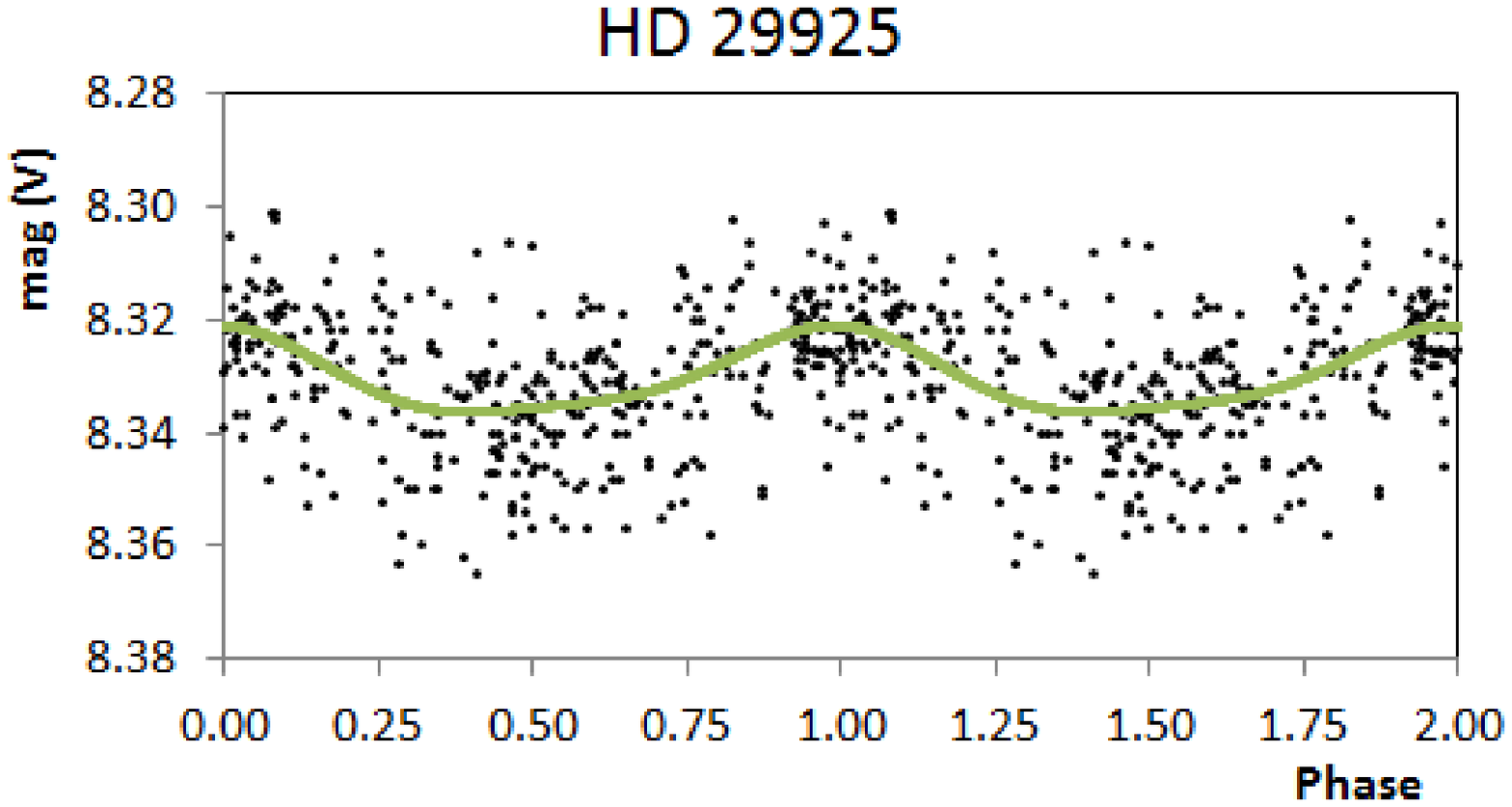}
\includegraphics[width=0.42\textwidth]{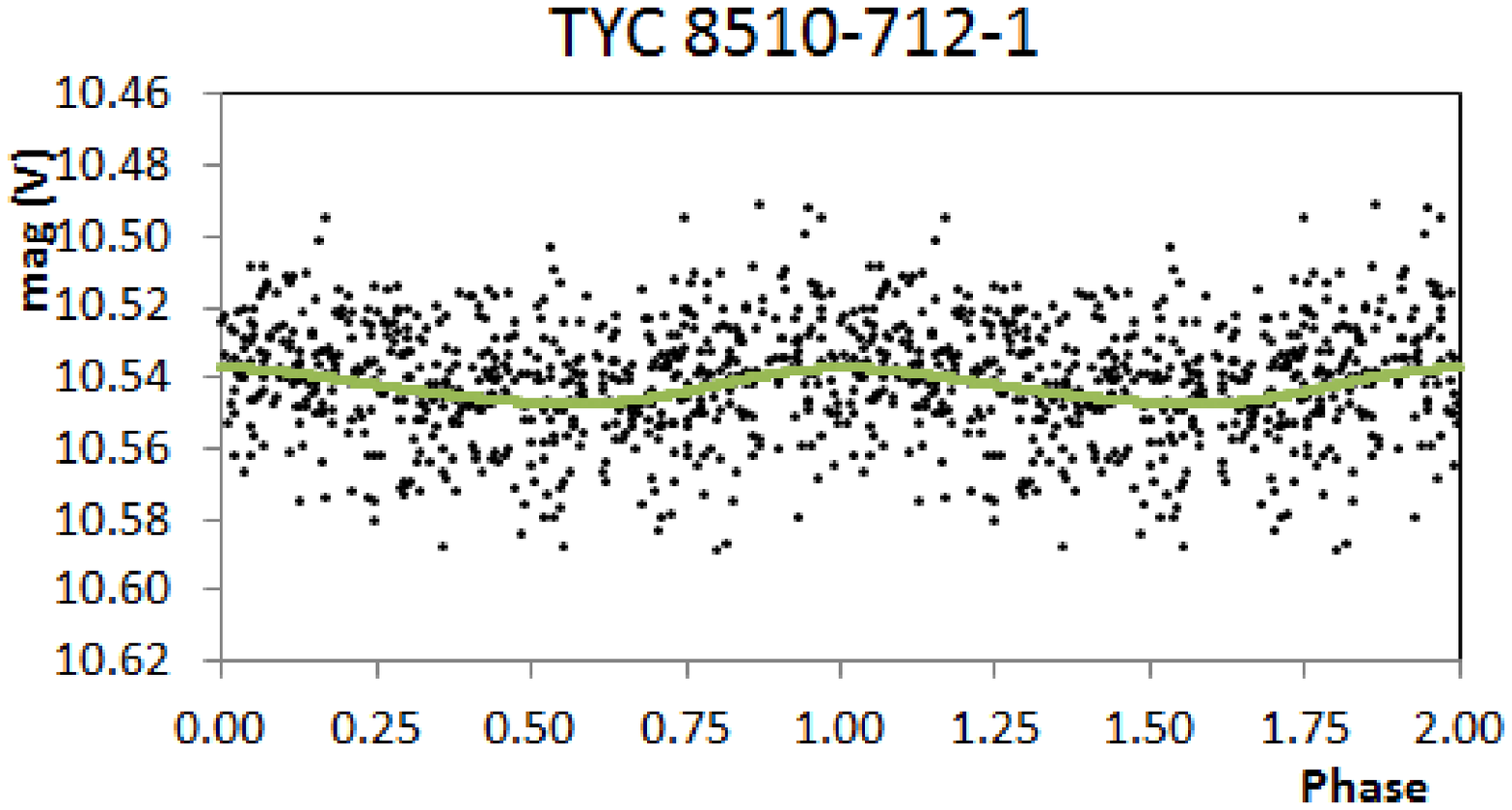}
\includegraphics[width=0.42\textwidth]{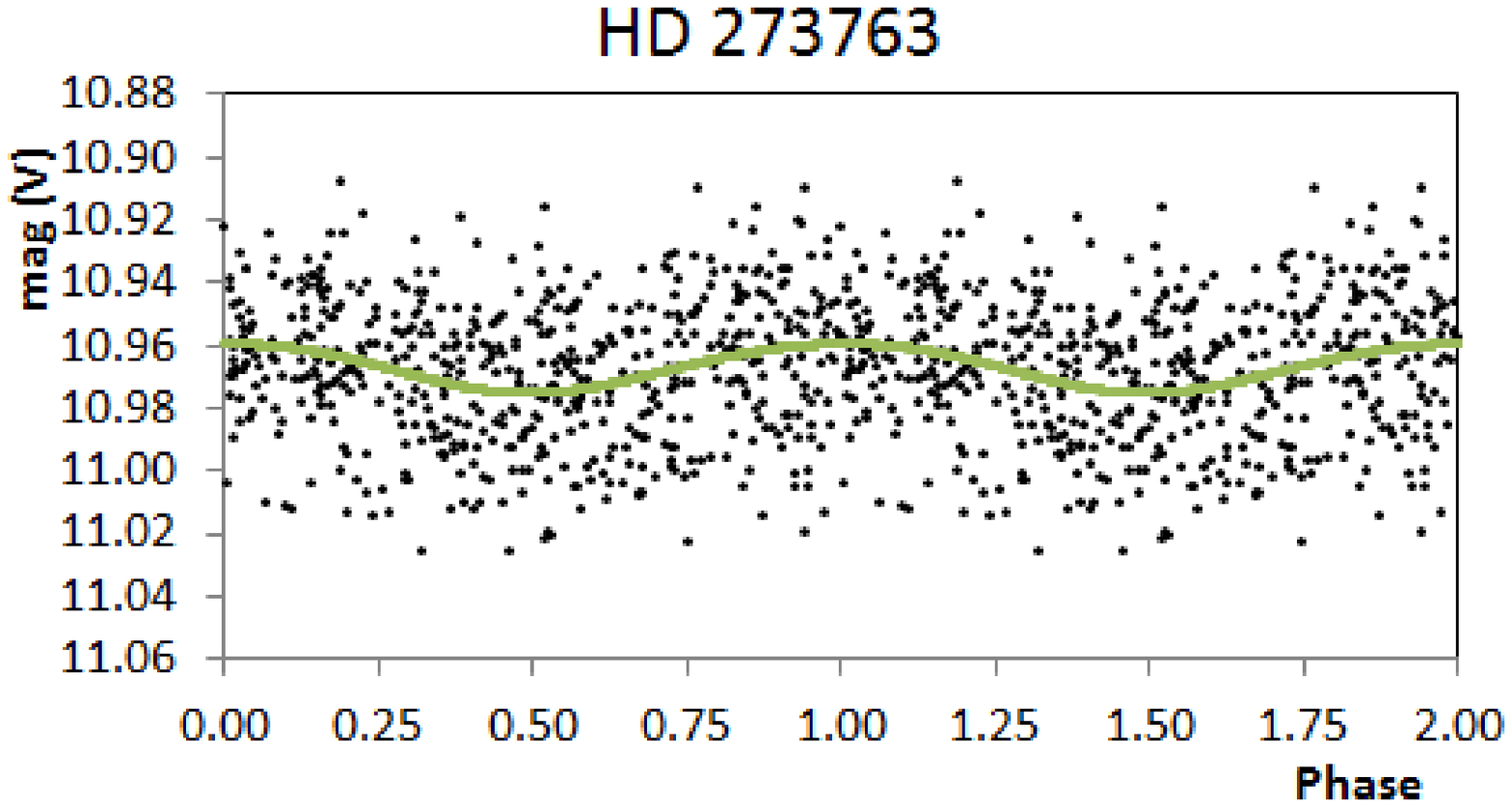}
\includegraphics[width=0.42\textwidth]{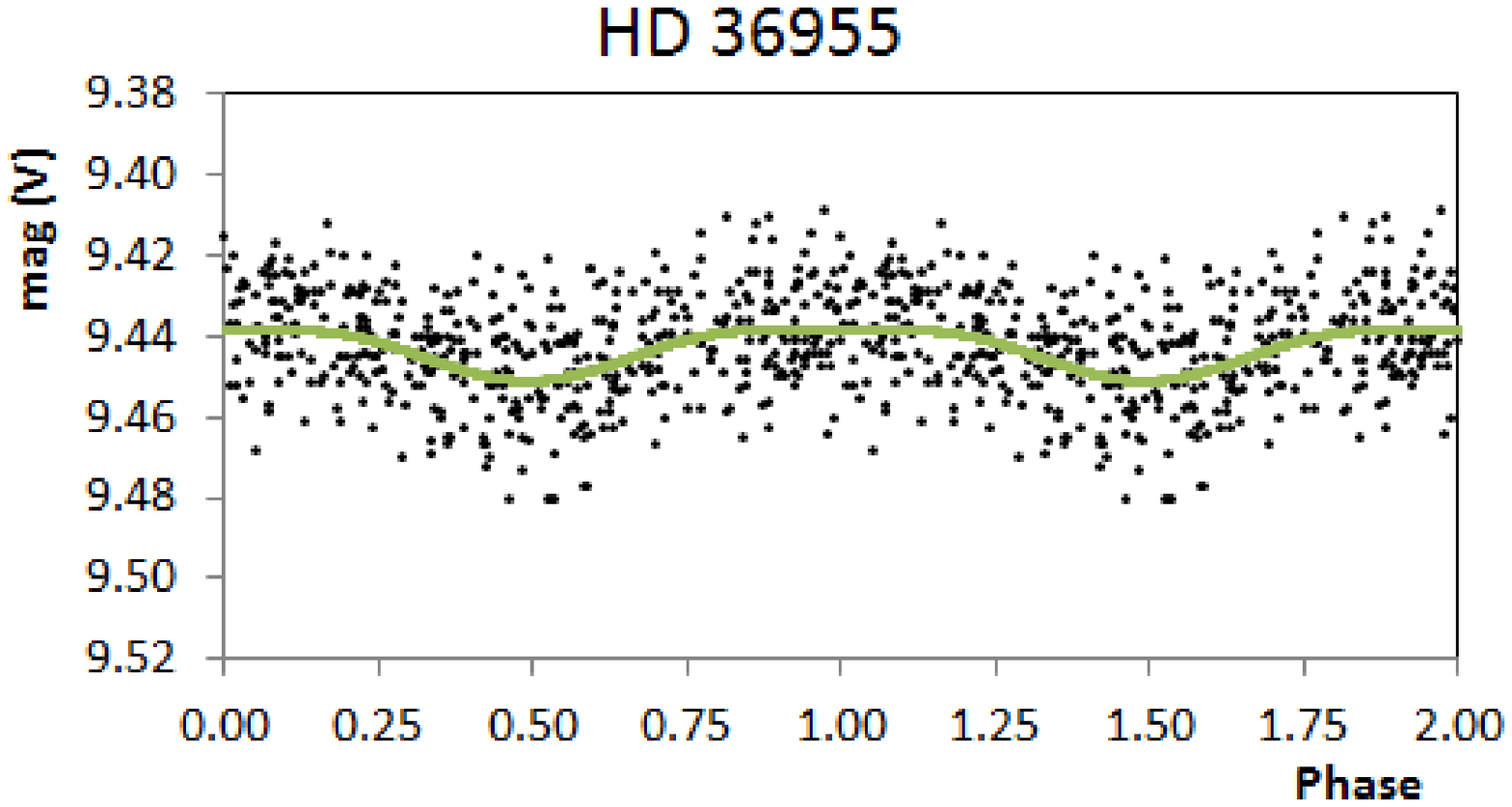}
\includegraphics[width=0.42\textwidth]{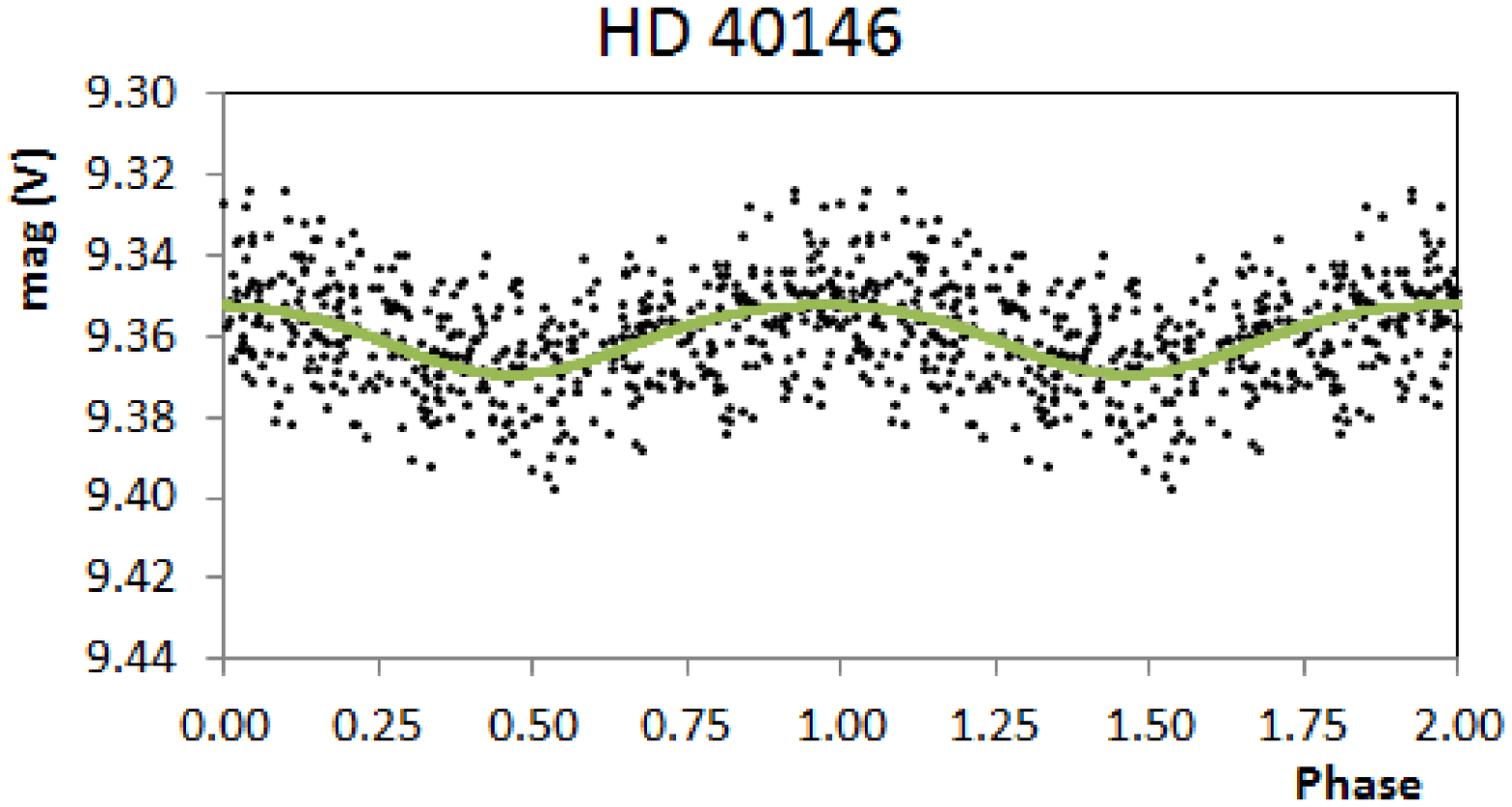}
\includegraphics[width=0.42\textwidth]{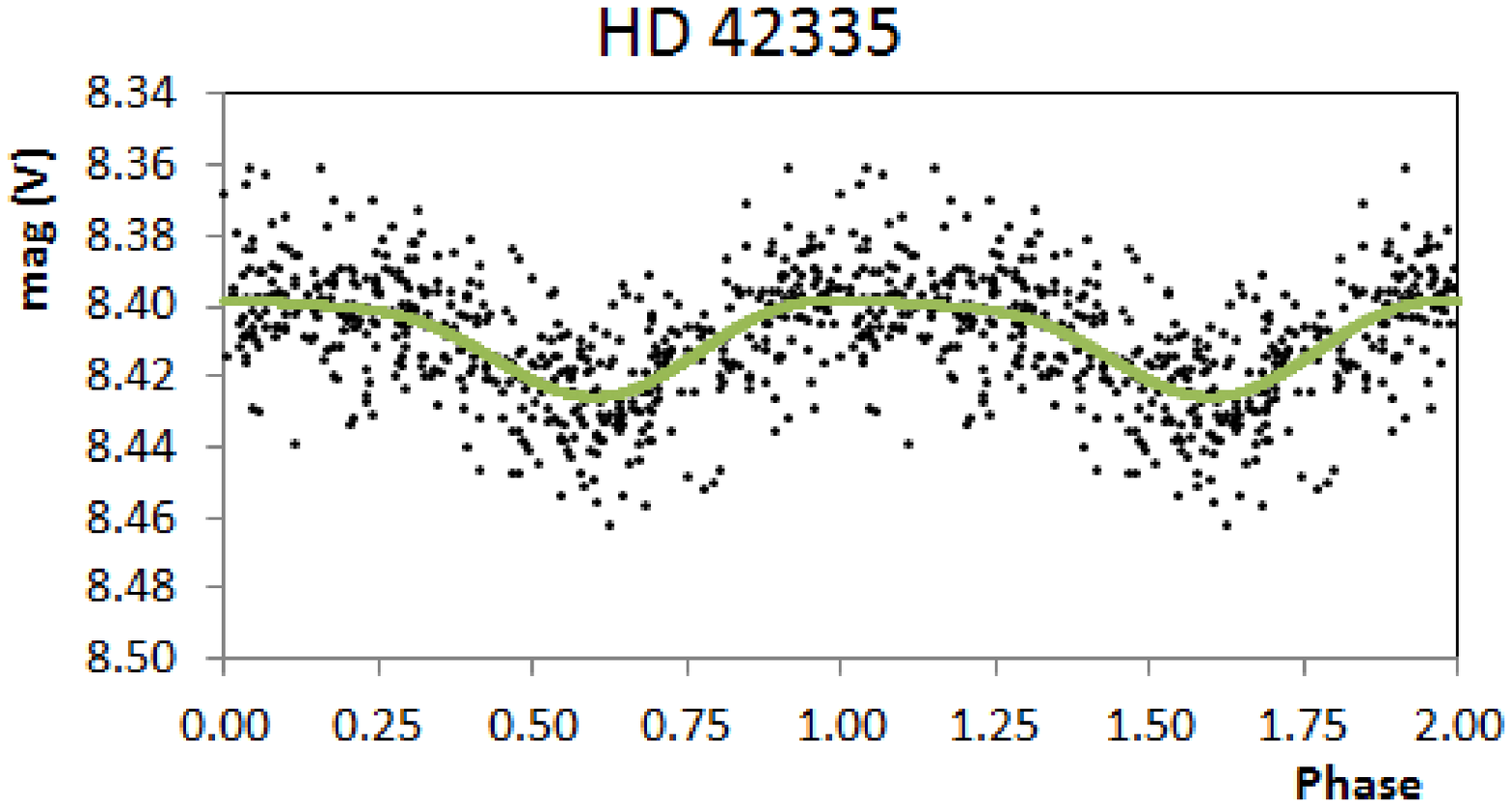}
\caption{Light curves of all stars analyzed with ASAS-3 data, folded with the periods listed in Table \ref{table_masterASASKELT}. The fit curves corresponding to the light curve parameters given in the same table are indicated by the solid lines.} 
 \label{figurepage1}
\end{figure*}


\bsp	
\label{lastpage}
\end{document}